\documentclass[nofootinbib,superscriptaddress,a4paper,twocolumn, floatfix]{revtex4-2}
\usepackage[english]{babel}
\usepackage[utf8]{inputenc}

\usepackage{amsthm}
\usepackage{amssymb}
\usepackage{mathtools}
\usepackage{physics}
\usepackage{xcolor}
\usepackage{graphicx}
\usepackage[left=23mm,right=13mm,top=35mm,columnsep=15pt]{geometry} 

\usepackage{subfigure}
\usepackage{enumitem}

\usepackage{algorithm2e}
\usepackage{algpseudocode}
\makeatletter

\usepackage{glossaries}
\usepackage{siunitx}

\usepackage{adjustbox}
\usepackage{placeins}
\usepackage[T1]{fontenc}
\usepackage{lipsum}
\usepackage{csquotes}
\usepackage[pdftex, pdftitle={Article}, pdfauthor={Author}]{hyperref}
\usepackage{xcolor}
\hypersetup{
    colorlinks,
    linkcolor={red!50!black},
    citecolor={blue!50!black},
    urlcolor={blue!80!black}
}
\usepackage{booktabs}
\usepackage{multirow}
\usepackage{listings}
\usepackage{xcolor}
\definecolor{codegreen}{rgb}{0,0.6,0}
\definecolor{codegray}{rgb}{0.5,0.5,0.5}
\definecolor{codepurple}{rgb}{0.58,0,0.82}
\definecolor{tqblue}{HTML}{08293d}
\definecolor{backcolour}{HTML}{fefdf5}

\lstdefinestyle{mystyle}{
    backgroundcolor=\color{backcolour},   
    commentstyle=\color{codegreen},
    keywordstyle=\color{magenta},
    numberstyle=\tiny\color{codegray},
    stringstyle=\color{codepurple},
    basicstyle=\ttfamily\footnotesize\color{tqblue},
    breakatwhitespace=false,         
    breaklines=true,
    postbreak=\mbox{\textcolor{magenta}{$\hookrightarrow$}\space},                 
    captionpos=b,                    
    keepspaces=true,                 
    numbers=left,                    
    numbersep=5pt,                  
    showspaces=false,                
    showstringspaces=false,
    showtabs=false,                  
    tabsize=2
}

\begin{document}

\title{State Specific Measurement Protocols for the Variational Quantum Eigensolver}

\author{Davide Bincoletto}
\affiliation{{Institute for Computer Science, University of Augsburg, Germany }}

\author{Jakob~S.~Kottmann}
\email[E-mail:]{jakob.kottmann@uni-a.de}
\affiliation{{Institute for Computer Science, University of Augsburg, Germany }}
\affiliation{{Center for Advanced Analytics and Predictive Sciences, University of Augsburg, Germany }}

\date{\today} 

\begin{abstract}
A central roadblock in the realization of variational quantum eigensolvers on quantum hardware is the high overhead associated with measurement repetitions, which hampers the computation of complex problems, such as the simulation of mid- and large-sized molecules. In this work, we propose a novel measurement protocol which relies on computing an approximation of the Hamiltonian expectation value. The method involves measuring cheap grouped operators directly and estimating the residual elements through iterative measurements of new grouped operators in different bases, with the process being truncated at a certain stage. The measured elements comprehend the operators defined by the Hard-Core Bosonic approximation, which encode electron-pair annihilation and creation operators. These can be easily decomposed into three self-commuting groups which can be measured simultaneously. Applied to molecular systems, the method achieves a reduction of $30\%$ to $80\%$ in the number of measurement and gates depth in the measuring circuits compared to state-of-the-art methods. This provides a scalable and cheap measurement protocol, advancing the application of variational approaches for simulating physical systems.

\end{abstract}

\begin{figure*}[ht]
  \centering
	\includegraphics[width=\textwidth]{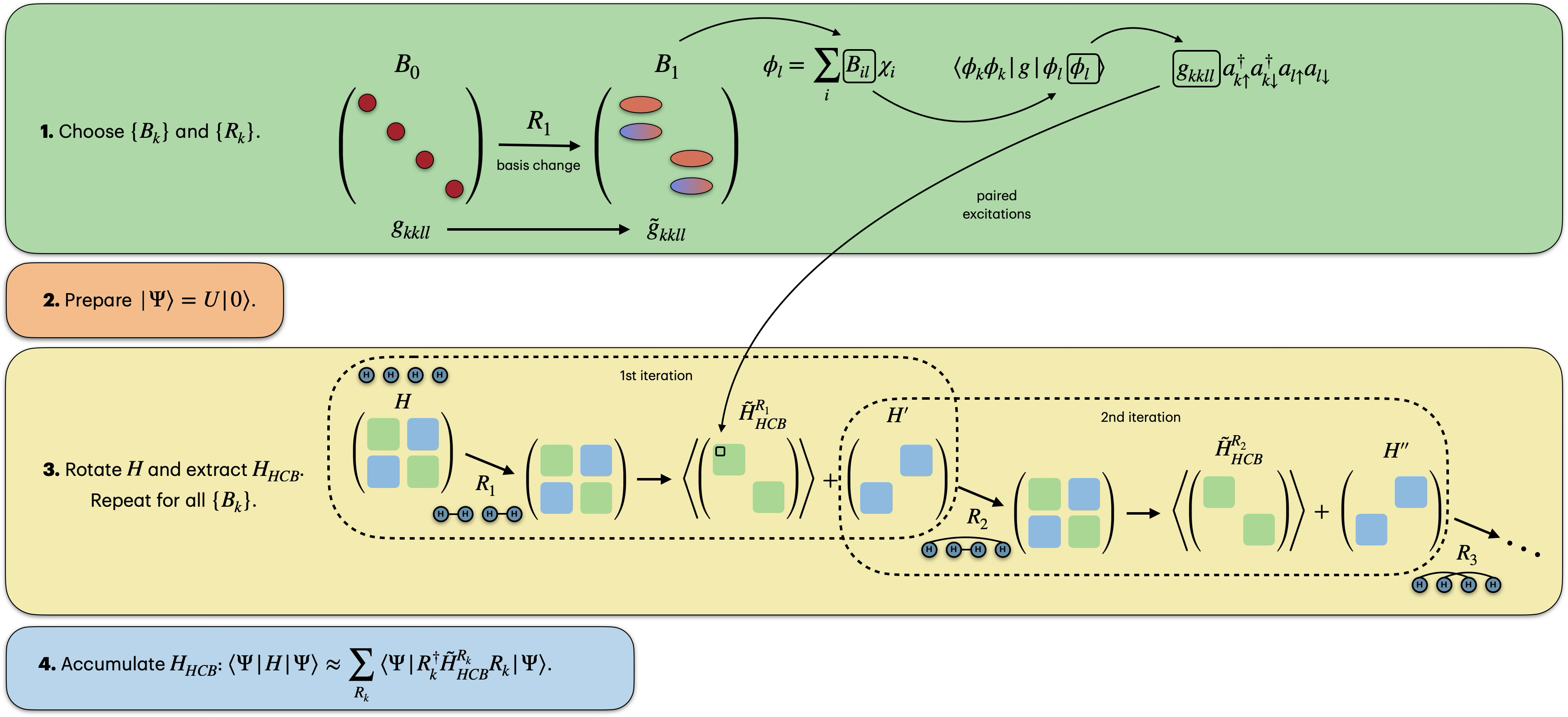}
  \caption{Illustration of the measurement routine used in this article leveraging HCB approximation and basis rotations. The general procedure is applied to hydrogenic systems like H$_4$ (depicted in the Figure), H$_6$ and H$_8$. Here diagonal and off-diagonal matrix elements (green and blue) are used to represent HCB and residual Hamiltonian elements.}
\end{figure*}

\maketitle

\section{Introduction}

The Variational Quantum Eigensolver (VQE) is a class of algorithms, initially designed for Noisy Intermediate-Scale Quantum devices (NISQ). \cite{peruzzo2014variational} They aim to solve the eigenvalue problem in a variational way and are often considered promising candidates for practical quantum advantage. \cite{tillyVariationalQuantumEigensolver2022}

VQEs require lower quantum resources in terms of circuit depth and qubit coherence time than traditional quantum algorithms such as Quantum Phase Estimation. \cite{aspuru2005simulated, cortesAssessing2024} This is due to a hybrid quantum-classical structure. A quantum computer is used to prepare a parametrized quantum state of which an expectation value is measured. A classical computer is then used for updating the parameters in the wavefunction through an optimization algorithm. A relevant aspect of VQE is that the expectation value of an observable can be decomposed and computed as the sum of individual or grouped expectation values. The trade-off of such algorithm is a significant overhead due to measurement repetitions and classical processing. Moreover, the measurement procedure can be affected by noise, which can impact its accuracy and precision. The latter can be attenuated through methods like randomized measurements, quantum detector tomography, and blended scheduling. \cite{korhonenPractical2024}

For the former, consider the electronic structure problem of quantum chemistry, which aims to find the eigestates of many-electron systems. The definition of an electronic Hamiltonian in second quantization, here in non-standard notation, is:
\begin{align}
    H &= H_1 + H_2 \\ &= \sum_{kl} h_{kl} a^{\dagger}_k a_l + \frac{1}{2} \sum_{klmn} g_{klmn} a^{\dagger}_k a^{\dagger}_l a_n a_m \nonumber
\end{align}
and the approximation to the ground-state energy is computed by minimizing the expectation value of the electronic Hamiltonian
\begin{equation}
    E_{VQE} = \min_{\theta} \bra{0} U^{\dagger}(\theta) H U(\theta) \ket{0}
\end{equation}
where $U(\theta)$ is a parametrized quantum circuit and $\ket{0}$ the initial (all-zero) qubit state on the quantum computer. In order to estimate the expectation value of the Hamiltonian, one has to perform measurements on it. For electronic Hamiltonians this reduces to measuring the terms:
\begin{align}
    \langle H \rangle = \sum_{kl} h_{kl} \langle a^\dagger_k a_l \rangle + \frac{1}{2} \sum_{klmn} g_{klmn} \langle a^{\dagger}_k a^{\dagger}_l a_n a_m \rangle
\end{align}
which corresponds to measuring the components of the 1- and 2-body reduced density matrices.
Without further work, the number of measurement groups for each iteration of the VQE optimization process is proportional to the number of Hamiltonian terms, which results in a growth of $\order{N^4}$, with $N$ the number of spin-orbitals in the system. This creates a practical obstacle for applying VQE to big size molecules. \cite{aspuru2005simulated, gonthierMeasurementsRoadblockTerm2022}\\

In order to be processed by a quantum computer, the Hamiltonian has to be mapped via a Fermion-to-qubit encoding, such as Jordan-Wigner transformation \cite{JordanWigner1928, tranter2018comparison}, into a sum of Pauli strings ($N$-fold tensor product of the three Pauli matrices and the unit matrix)
\begin{equation}
    H = \sum_i w_i P_i
\end{equation}
with $P_i$ a Pauli string and $w_i$ the corresponding weight.\\

Since the standard measurement on quantum computers consists of reading out the classical bit values of the qubits (this corresponds to measuring in the $Z$-basis), we need to transform all the other Pauli operators in the Hamiltonian. This means finding a set of unitary operators such that
\begin{equation}
    P_i^{(d)} \equiv P^z_i = U_i P_i U^{\dagger}_i
\end{equation}
where $P_i^{(d)}$ is a diagonal matrix in the form of a tensor product consisting only of Pauli-$Z$ and unit matrices.\\

Grouping methods identify commuting cliques of Pauli strings in order to determine such set of unitary operators efficiently, and thus mitigate the measurement overhead. Due to commutativity those Pauli strings can all be transformed into $P^z_i$ by the same unitary and thus be measured simultaneously. For qubit Hamiltonians one can group Pauli operators into qubit-wise commuting sets (QWC), meaning that each Pauli string commutes with each other, and each Pauli operator, with the same index within strings, commutes with each other. You can see examples of this in Section II from \cite{verteletskyi2020measurement}. Two proposed methods for finding the sets are called Large First (LF) and Recursive Largest First (RLF), and involve solving the minimum clique cover problem. \cite{verteletskyi2020measurement}\\

An alternative approach involves using unitary transformations, such that any group of mutually commuting operators can be transformed into their qubit-wise commuting form. This strategy, referred to as Fully Commuting (FC), is a less restrictive method, because the grouped strings commute with each other in a normal fashion $[P_i,P_j]=0$. \cite{yen2020measuring} The number of measurement groups needed to determine the expectation value of $H$ is influenced by the Pauli strings grouping strategy and the allocation of measurements to each group. Two proposed methods to address it are Sorted Insertion (SI) \cite{crawford2019efficient, bansinghFidelity2022} and Iterative Coefficient Splitting (ICS) \cite{yenDeterministic2023}. These fall under the class of the so-called qubit-algebra-based methods, because they group the operators following the Jordan-Wigner transformation. In contrast, fermionic-algebra-based methods leverage the commutativity properties of the molecular Hamiltonian operators in second quantization. Two notable approaches are Low-rank decomposition (LR) \cite{hugginsEfficient2021, yenCartan2021} and Fluid Fermionic Fragments (F$^3$) \cite{choiFluid2023}.\\

After grouping the Pauli strings into commuting fragments $H_\alpha$ the expectation value of the Hamiltonian can be expressed as
\begin{equation}
    \begin{split}
        \bra{\Psi} H \ket{\Psi} & = \bra{\Psi} \sum_\alpha H_\alpha \ket{\Psi} \\
        & = \sum_{\alpha} \bra{\Psi} U_{\alpha}^{\dagger} U_{\alpha} H_\alpha U^{\dagger}_{\alpha} U_{\alpha} \ket{\Psi} \\
        & = \sum_{\alpha} \bra{\Phi_{\alpha}}  P^z_{\alpha} \ket{\Phi_{\alpha}}
    \end{split}
\end{equation}
where $U_\alpha$ diagonalized the group of commuting Pauli strings summarized into $H_{\alpha}$.\\

QWC grouping uses diagonalizing circuits, i.e., unitary operators, consisting of single-qubit rotations restricted to individual qubits and will thus have shallow depths. However, grouping this way doesn't reduce the number of terms to measure by much. It has been empirically shown that this grouping method will only reduce the prefactor in the scaling $\order{N^4}$ by a factor of three. \cite{gokhale2019minimizing}\\

The advantage of the FC grouping scheme is that it can reduce the number of terms to measure from $\order{N^4}$ to $\order{N^3}$ \cite{gokhale2019n} or even to $\order{N}$ \cite{hugginsEfficient2021}. The trade-off however is the need for entangling gates in the diagonalizing circuits, which can be a concern considering the fidelity of multi-qubit operators with the hardware being used. Recent results have shown however that using FC grouping together with a circuit optimization procedure to reduce CNOT gate counts can still result in fewer overall measurements than the QWC grouping scheme. \cite{bansinghFidelity2022} Furthermore, a study introduced a framework for designing diagonalization circuits, with two-qubit gates adapted the connectivity constraints of modern quantum computing architectures. \cite{millerHardwareTailoredDiagonalizationCircuits2024} \\

Despite the numerous advancements, there is still no method that efficiently minimizes the overhead in a way that can be handled in practice, especially for mid- and big-size molecules. In this paper, we present a method for reducing measurements overhead by exploiting molecular structure. The procedure aims to approximate the expectation value of the Hamiltonian with respect to a specific target state (in this case, the ground-state of the electronic system), thus, it differs from grouping methods that partition the Hamiltonian itself typically in an exact matter. The goal is to heuristically leverage the structure of the quantum state at hand instead of grouping all the terms of the Hamiltonian, which can become computationally expensive. The trade-off is that the process is not exact, but can be tuned accordingly. Throughout the study, we will focus on a single iteration of the VQE algorithm. Our purpose is to obtain a scaling that is lower than previous state-of-the-art methods and a generalization property on multiple molecular systems.\\

The rest of the paper is structured as follows. In Section \ref{Methodology} we introduce the basic principles, present the steps of the method and introduce the scenarios used for the simulations. In Section \ref{Results} we show the results on representative examples. Finally, in Section \ref{Conclusion} we consolidate the insights gained and discuss future outlooks.

\section{Methodology} \label{Methodology}

The general methodology is based on two observations. The first one is that, given a molecule and its corresponding second quantized Hamiltonian, we can define a physically motivated approximation of the operator. We chose the Hard-Core Boson Hamiltonian (HCB), we will define it in Section \ref{HCB}. \cite{elfving2021simulating} It contains valuable information about the molecule and its elements are cheap to store and elaborate on. Therefore, by evaluating the HCB elements we can retrieve an accurate approximation of the expectation value of the original Hamiltonian.

The second observation, defined in Section \ref{Orbital rotation}, is that we can perform some operation that acts as a basis change from the original orbital basis to a new one, i.e., a linear combination of the original orbitals. This suggests that we can always find a new representation of the elements discarded in the first HCB approximation and then apply a new HCB approximation on them. Thus, by reiterating the process multiple times we can get closer to the correct value.

\subsection{Hard-Core Boson Hamiltonian} \label{HCB}

The HCB Hamiltonian is a paired-electron approximation that treats spin-paired-electrons as quasi-particles, occupying the spatial orbitals. On the other hand, this approximation breaks the invariance of the Hamiltonian with respect to orbital rotations, meaning that different choices of spatial orbitals lead to different approximations. The HCB Hamiltonian can be formulated in two ways, where only the second form is suitable for this work:
\begin{enumerate}[label=(\roman*)]
    \item Bosonic Operator Substitution (compressed). \\
    This approach considers only the electron-pair annihilation and creation operators in the original Hamiltonian and discards the others. Then it substitutes these operators with new ones:
    \begin{equation}
        b_i^{(\dagger)} \sim a^{(\dagger)}_{i\uparrow}a^{(\dagger)}_{i\downarrow}
    \end{equation}
    where $b_i^{(\dagger)}$ act as Hard-Core Bosonic operators that represent the creation or annihilation of an electron pair in orbital $i$ and follow bosonic-like commutation rules. \cite{elfving2021simulating} In this form, the wavefunction can only be a Hard-Core Bosonic wavefunction. In this work, we will not pose such restrictions on the electronic wavefunction which is why we can not use this form. \\
    An example usage of this strategy can be seen in \cite{santosHybrid2024}.

    \item Fermionic Operators for Paired Excitations (uncompressed form). \\    
    This approach uses fermionic operators directly to construct the electron-pair interaction terms, restricting them to only Boson-like excitations. 
    This form allows us to remain in the Jordan-Wigner (or any other fermion-to-qubit encoding) picture for the qubit wavefunction and will only modify the Hamiltonians for the measurements. 
\end{enumerate}

The uncompressed form can be applied to the electronic Hamiltonian directly and its terms are:
\begin{enumerate}
    \item Two electrons are destroyed in orbital $k$ and created in orbital $k$. This corresponds to the number operator.
    \begin{align}
        \alpha_{k} = \sum_{\sigma} h_{kk} & a^{\dagger}_{k\sigma} a_{k\sigma} = \sum_{\sigma} h_{kk} n_{k\sigma} \\
        & \hspace{-0.8em} \includegraphics[width=0.07\textwidth]{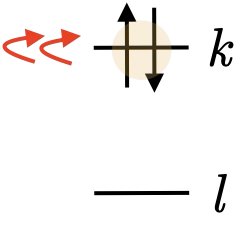} \notag
    \end{align}

    \item Two electrons are destroyed in orbital $l$ and created in orbital $k$.
    \begin{align}
        \beta_{kl} = \sum_{\sigma,\sigma'} & g_{kkll} a^{\dagger}_{k\sigma} a^{\dagger}_{k\sigma'} a_{l\sigma'} a_{l\sigma} \\
      & \includegraphics[width=0.07\textwidth]{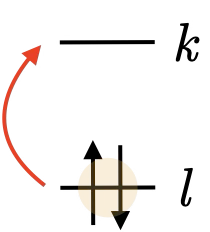} \notag
    \end{align}
    
    \item One electron is destroyed in orbital $l$ and created in the same orbital $l$ and another electron is destroyed in orbital $k$ and created in the same orbital $k$. This corresponds to the number operators of $k$ and $l$.
    \begin{align}
        \gamma_{kl} = \sum_{\sigma,\sigma'} g_{kllk} a^{\dagger}_{k\sigma} a^{\dagger}_{l\sigma'} & a_{l\sigma'} a_{k\sigma} = \sum_{\sigma,\sigma'} g_{kllk} n_{k\sigma} n_{l\sigma'} \\
      & \includegraphics[width=0.07\textwidth]{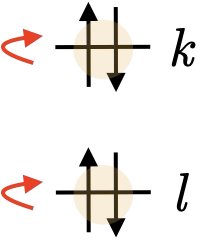} \notag
    \end{align}
    
    \item One electron is destroyed in orbital $l$ and created in orbital $k$ and another electron is destroyed in orbital $k$ and created in orbital $l$.
    \begin{align}
        \delta_{kl} = \sum_{\sigma,\sigma'} & g_{klkl} a^{\dagger}_{k\sigma} a^{\dagger}_{l\sigma'} a_{k\sigma'} a_{l\sigma} \\
      & \includegraphics[width=0.08\textwidth]{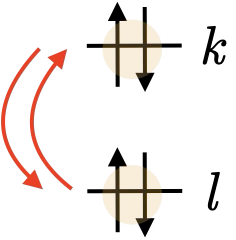} \notag
    \end{align}
\end{enumerate}

In Appendix \ref{Visualization of HCB elements} we present a simple visualization of how the operators act on electron pairs.

Finally, the HCB Hamiltonian has the following expression:
\begin{equation}
    H_{\text{HCB}} = \sum_k \alpha_{k} + \sum_{kl} (\beta_{kl} + \gamma_{kl} + \delta_{kl})
\end{equation}
And for convenience we define the residual Hamiltonian as
\begin{equation}
    H_{\text{res}} = H - H_{\text{HCB}}
\end{equation}
The resulting HCB Hamiltonian, once mapped into Pauli operators, naturally decomposes into three commuting groups: \{$I_0$, $Z_0$, $Z_1$, $Z_0Z_1$, ...\}, \{$Y_0X_1X_2Y_3$, $X_0Y_1Y_2X_3$, $Y_0X_1X_4Y_5$, ...\} and \{$Y_0Y_1X_2X_3$, $X_0X_1Y_2Y_3$, $Y_0Y_1X_4X_5$, ...\}. Namely, $\alpha_{k}$ and $\gamma_{kl}$ parse into the first group, and the combination of $\beta_{kl}$ and $\delta_{kl}$ parses in the second and third groups. This makes it possible to do measurement on multiple elements at the same time, highly reducing the computational overhead.


\subsection{Orbital rotation operation} \label{Orbital rotation}
An orbital basis is a unitary $N\times N$ matrix $B$ operating on the initial set of spatial orbitals. In order to transform, i.e., rotate, the orbital basis into a new one we need to define a proper operation. Although any unitary operator can be applied, only certain preserve the electronic Hamiltonian structure, e.g., the number of Pauli operators or the type of Hamiltonian. An effective 2D rotation, for example, acts as a proper basis change for consecutive orbitals. Thus, in order to rotate any orbital we can use a sequence of effective 2D rotations acting on the atomic orbital space.

To illustrate, this is the matrix representation in the space of two spatial orbitals $p$ and $q$:
\begin{equation}
    R_{\{ p,q \}}(\theta) \equiv R(\theta) = 
    \begin{pmatrix}
    \cos(\theta/2) & \sin(\theta/2)  \\
    -\sin(\theta/2) & \cos(\theta/2)
    \end{pmatrix}
\end{equation}
where $\theta$ is a free parameter. This operation is applied to the molecular integrals in the following way:
\begin{equation}
    \tilde{h}_{kl} = \sum_{xy} R_{kx} R_{ly} h_{xy}\label{eq:one-body-transform}
\end{equation}
\begin{equation}
    \tilde{g}_{klmn} = \sum_{wxyz} R_{kw} R_{lx} R_{my} R_{nz} g_{wxyz}
\end{equation}
and, when done at the same time, defines a global unitary transformation of the Hamiltonian operator. We call this orbital rotation operation.

Such an effective 2D rotation can be also represented as a quantum circuit, given the correspondence between atomic orbital space and qubit space. \cite{kivlichan2018quantum, google2020hartree, kottmannMolecular2023} In fact the unitary operator
\begin{equation}
    U_{R_{ \{ p,q \} }}(\theta) \equiv U_R(\theta) = e^{\frac{\theta}{2}(a^{\dagger}_{p\uparrow} a_{q\uparrow} + a^{\dagger}_{p\downarrow} a_{q\downarrow} - \text{h.c.})},
\end{equation}
which acts on the qubit space, achieves the same result of the orbital rotation operation. \cite{kottmannMolecular2023} Here $p$ and $q$ represent the spatial orbitals affected by the rotation. Thus, analogously to the atomic orbital space, an orbital rotation operation in the qubit space is achieved with a sequence of $U_R(\theta)$.

While the matrix representation is an $N\times N$ operation on the space of of spatial orbitals, with $N$ the number of such orbitals, the circuit representation correspond to a $2^{2N}\times 2^{2N}$ transformation applied on the qubit register.

\subsection{General Procedure}
\label{General Procedure}
The proposed method consists of four steps: a preprocessing phase (steps 1–2), a recursive phase (step 3), and a final phase (step 4). The preprocessing phase is performed once, while the recursive and final phases are executed for each estimation of the expectation value of the whole Hamiltonian.
\begin{enumerate}
    \item \textbf{Choose orbital bases $\mathcal{B}=\left\{ B_k \right\}$.}

    The orbital bases are given as unitary $N\times N$ matrices which operates on the initial set of orbitals, which we will call ``reference orbitals''. Note however, that they do not need to be ``Hartree-Fock'' orbitals, they merely define the reference point for the given matrices. Each matrix in $\mathcal{B}$ is compiled into a orbital rotation operation forming the set $\mathcal{R} = \left\{R_k\right\} = \{ U_{R_k} \}$. \cite{google2020hartree}

    \item \textbf{Choose a quantum circuit to prepare the quantum state of interest.}

    This will provide the wavefunction of interest
    \begin{equation}
        \ket{\Psi} = U \ket{0}
    \end{equation}
    where $U$ is the quantum circuit, and $\ket{0}$ the quantum register. This is the state of which we aim to compute the expectation value.

\item \textbf{Iteratively approximate the expectation value of $H$}
    
    First, $H$ is transformed into
    \begin{equation}
        \tilde{H}_{\text{HCB}}^{R_1} = (R_1 H R_1^{\dagger})_{\text{HCB}}
    \end{equation}
    and
    \begin{equation}
        H' = (R_1 H R_1^{\dagger})_{\text{res}}
    \end{equation}
    To reconstruct $H$ we need to consider both operators, but the former term can be evaluated easily since all the terms can be collected into three commuting groups, as shown before. Therefore, we can isolate it and elaborate on the latter term, which we will call $H'$ for convenience.
    
    $H'$ can be recursively processed in the same way we did for $H$. Each cycle will take a new rotation operation from the set \{$R_1, R_2, R_3, ...$\}, rotate back in the original basis, rotate forward in the new basis and extract HCB and residual Hamiltonian.
    \item \textbf{Accumulate all contributions}
    
    At the end of the procedure we have collected a series of operators that, if not truncated, accumulate to the original Hamiltonian $H$.
    \begin{equation}
        \begin{split}
            \bra{\Psi} H \ket{\Psi} \approx \sum_{R_k \in \mathcal{R}} \bra{\Psi} R_k^{\dagger} \tilde{H}_{\text{HCB}}^{R_k} R_k \ket{\Psi} \\
        \end{split}
    \end{equation}
\end{enumerate}

 The crucial point of the method is that an accurate approximation is bound to a correct choice of the orbital rotations. In the following, we present two typical scenarios for practical applications, which we employed for explicit simulations in \ref{Results}.

\subsection{Scenario I: Leveraging Valence-Bond Structures} \label{Scenario I}
In this scenario we make no assumption on the quantum circuit that produces the state. In order to run the simulations we computed the true ground-state wavefunction of the system through exact diagonalization.

For every system we can define resonance structures or graphs. These are convenient representations for molecules, because we can directly map spatial orbitals into nodes and interactions into edges. These can be used to select the rotation operations to apply in the measurement protocol.

Consider, for example, the molecule H$_4$ arranged in a linear geometry.
\begin{equation*}
    \includegraphics[width=0.15\textwidth]{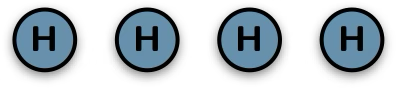}
\end{equation*}

From it we can define two graphs identified by only paired edges, namely $G_1 = \{ \{0,1\},\{2,3\} \}$ and $G_2 = \{ \{0,3\},\{1,2\} \}$:
\begin{equation*}
    \raisebox{0.23cm}{$G_1 = $} \includegraphics[width=0.15\textwidth]{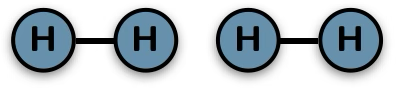}
\end{equation*}
\begin{equation*}
    \raisebox{0.23cm}{$G_2 = $} \includegraphics[width=0.15\textwidth]{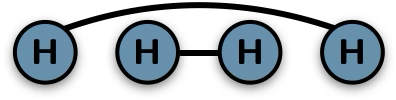}
\end{equation*}

One can use chemistry-inspired heuristics to define the graphs. For example, the pairwise pattern resembles Valence-Bond Theory for chemical bonding construction. \cite{hiberty2007textbook9}

Then, given a set of reference orbitals, we can associate one orbital rotation operation to each graph.

In a minimal STO-3G atomic basis, $G_1$ corresponds to a rotation in a four orbitals space. For $\theta=\frac{\pi}{2}$ $\cos(\frac{\pi}{2}) = \sin(\frac{\pi}{2}) = \frac{1}{\sqrt{2}}$. By arranging the values in rows $\{0,1\},\{2,3\}$, corresponding to the graph nodes, the operation can be represented by the matrix:
\begin{align}
    R_{G_1} \left(\frac{\pi}{2}\right) & = R_{\{ \{0,1\},\{2,3\} \}} \left(\frac{\pi}{2}\right) = \nonumber \\
    & = R_{\{0,1\}} \left(\frac{\pi}{2}\right) \ R_{\{2,3\}} \left(\frac{\pi}{2}\right) = \nonumber \\
    & = \frac{1}{\sqrt{2}}
    \begin{pmatrix}
    1 & 1 & 0 & 0 \\
    -1 & 1 & 0 & 0 \\
    0 & 0 & 1 & 1 \\
    0 & 0 & -1 & 1
    \end{pmatrix}
\end{align}
The coefficients show that the orbitals are now in an equal superposition, thus, we can interpret the first row as a bonding molecular orbital between atomic orbitals 0 and 1 and the second row as an anti-bonding molecular orbital, likewise for the third and fourth rows. Therefore, the rotation operation $R_{G_1}$ is the transformation from the set of ``reference orbitals'' $B_0$, where atomic orbitals have no interaction among them, to the orbital basis $B_1$, where orbital pairs $\{0,1\},\{2,3\}$ are delocalized into bonding and anti-bonding motifs.

Similarly, $G_2$ corresponds to the matrix:
\begin{align}
    R_{G_2} \left(\frac{\pi}{2}\right) & = R_{\{ \{0,3\},\{1,2\} \}} \left(\frac{\pi}{2}\right) = \nonumber \\
    & = R_{\{0,3\}} \left(\frac{\pi}{2}\right) \ R_{\{1,2\}} \left(\frac{\pi}{2}\right) = \nonumber \\
    & = \frac{1}{\sqrt{2}}
    \begin{pmatrix}
    1 & 0 & 0 & 1 \\
    0 & 1 & 1 & 0 \\
    0 & -1 & 1 & 0 \\
    -1 & 0 & 0 & 1
    \end{pmatrix}
\end{align}
The coefficients are the same and so is the bonding anti-bonding motif, but the target orbitals are now $\{0,3\}$ and $\{1,2\}$. This defines the transformation from $B_0$ to $B_2$.

The graph $G_1$ corresponds to $p,q = 0,1$ and $p,q = 2,3$ with $\theta=\frac{\pi}{2}$ and, likewise, the graph $G_2$ corresponds to $p,q = 1,2$ and $p,q = 0,3$ with $\theta=\frac{\pi}{2}$. We will represent such circuits graphically as
\begin{equation}
    \raisebox{1.15cm}{$U_{R_{G_1}} \left(\frac{\pi}{2}\right) \equiv$}  \includegraphics[width=0.055\textwidth]{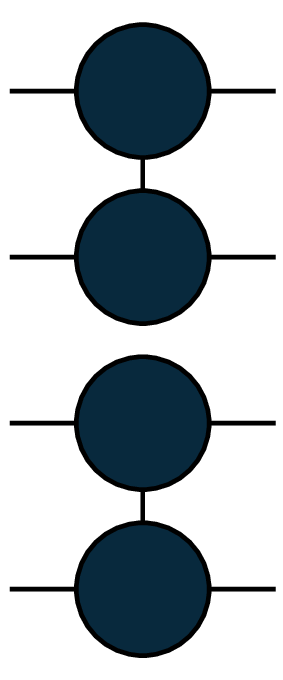},\;\;\;
    \raisebox{1.15cm}{$U_{R_{G_2}} \left(\frac{\pi}{2}\right) \equiv$}  \includegraphics[width=0.08\textwidth]{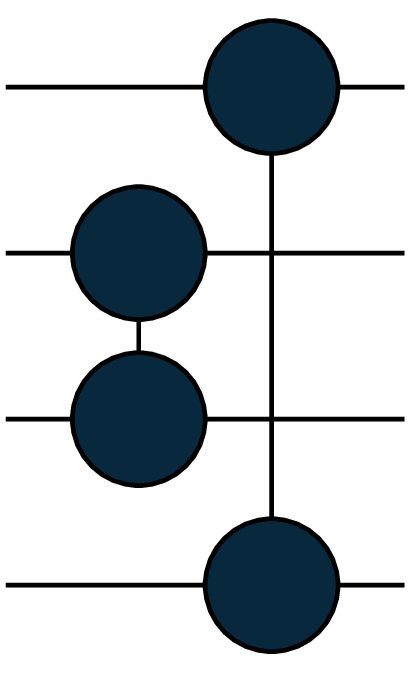}
\end{equation}
where the lines represent spatial orbitals (and therefore 2-qubits in most encodings). The corresponding unitary operators are:
\begin{align}
    U_{R_{G_1}} & = U_{R_{ \{ \{0,1\},\{2,3\} \} }}\left(\frac{\pi}{2}\right) = \nonumber \\
    & = U_{R_{ \{0,1\} }}\left(\frac{\pi}{2}\right) \ U_{R_{ \{2,3\} }}\left(\frac{\pi}{2}\right) = \nonumber \\
    & = e^{\frac{\pi}{4}(a^{\dagger}_{0\uparrow} a_{0\uparrow} + a^{\dagger}_{1\downarrow} a_{1\downarrow} - \text{h.c.})} \ e^{\frac{\pi}{4}(a^{\dagger}_{2\uparrow} a_{2\uparrow} + a^{\dagger}_{3\downarrow} a_{3\downarrow} - \text{h.c.})}
\end{align}
and
\begin{align}
    U_{R_{G_2}} & = U_{R_{ \{ \{0,3\},\{1,2\} \} }}\left(\frac{\pi}{2}\right) = \nonumber \\
    & = U_{R_{ \{0,3\} }}\left(\frac{\pi}{2}\right) \ U_{R_{ \{1,2\} }}\left(\frac{\pi}{2}\right) = \nonumber \\
    & = e^{\frac{\pi}{4}(a^{\dagger}_{0\uparrow} a_{0\uparrow} + a^{\dagger}_{3\downarrow} a_{3\downarrow} - \text{h.c.})} \ e^{\frac{\pi}{4}(a^{\dagger}_{1\uparrow} a_{1\uparrow} + a^{\dagger}_{2\downarrow} a_{2\downarrow} - \text{h.c.})}.
\end{align}

In summary, Scenario I takes a system, defines pairwise graphs, and generates a set of orbital bases, $\mathcal{B}$, or equivalently, a set of orbital rotation operations, $\mathcal{R}$, to be utilized in the measurement procedure.

\subsection{Scenario II: Adapting to Circuit-Designs}
As second explicit scenario we outline a specific quantum circuit that produces the state of interest. As an example one can co-design the ansatz together with the set of rotations. This strategy enables the measurement process to adapt to any specific state produced by the quantum register. Moreover, it takes advantage of the circuit structure for the Hamiltonian evaluations.

Given the input set of orbital bases $\mathcal{B}=\left\{ B_k \right\}$ we defined the set of rotation operations in unitary form, as showed in the previous section. In this instance we built a quantum circuit defined by a sequence of rotations $U_{R_k}$ and double excitations $U_{C_k}$, defined as:
\begin{equation}
    U_{C_{ \{ p,q \} }}(\theta) \equiv U_C(\varphi) = e^{-i\frac{\varphi}{2}(a^{\dagger}_{p\uparrow} a^{\dagger}_{p\downarrow} a_{q\downarrow} a_{q\uparrow} + \text{h.c.})}.
\end{equation}
This Multi-Graph Circuit was defined in \cite{kottmannMolecular2023} and is a cheap and accurate ansatz to compute the ground-state wavefunction. The circuit is constructed as
\begin{equation}
    \begin{split}
        \ket{\Psi} = & \sum_{k>1} U_{R_{G_k}}^{\dagger}(\phi_k) U_{C_{G_k}}(\varphi_k) U_{R_{G_k}}(\phi_k) \\
        & U_{R_{G_1}}^{\dagger}(\phi_1) U_{\text{SPA}}(\varphi_1) \ket{0},
    \end{split}
\end{equation}
where $U_{\text{SPA}}(\varphi)$ corresponds to Eq.(18) of \cite{kottmannMolecular2023} in a minimal basis and $U_{C_{G_k}}(\varphi_k)$ is a tensor product of $U_{C_k}$ gates with a topology, i.e., a set of $\{p,q\}$, defined by the graph $G_k$.

The circuit is made by a sequence of gates which aims at catching all the correlation contributes among the atoms. In order to do that it leverages the graph structure, i.e., the nodes of the graph define the correlated orbitals and the edges define the strength of the interaction.

By this means, the set of rotation operations $\mathcal{R}$, used in the measurement protocol, is exploited by the design of the circuit. The produced state will be an approximation of the true wavefunction, as is typical in VQE algorithms, but we can interpret the rotation operations as existing contributions inside the quantum state, reflecting its underlying structure.

\section{Results} \label{Results}
We tested the method on three molecular systems, H$_4$, H$_6$ and H$_8$, all arranged both on a line with a bond length of 1.5Å and scattered in space. The line configuration is a common benchmark dataset for NISQ algorithms in quantum simulation. For this reason we also considered free geometries which hold no structure. The bond length is chosen to be between the bonding and the dissociation distance, such that the ground-state wavefunction is not trivially simulable.

All the calculations have been carried within the $\textsc{tequila}$ $\textsc{Python}$ package \cite{tequila}. Specifically, the simulations made use of $\textsc{qulacs}$ \cite{qulacs}, the qubit operators elaboration utilized $\textsc{OpenFermion}$ \cite{OpenFermion}, while the integral computations employed the $\textsc{pyscf}$ package \cite{pyscf1} and the exact diagonalization $\textsc{scipy}$ \cite{scipy}. Finally, free geometries have been generated with $\textsc{quanti-gin}$ \cite{steinNylser2025} and circuits depictions are made with $\textsc{qpic}$ \cite{Qpic2025}.

\subsection{Approximation of Hamiltonian expectation value}

For H$_4$ we used a set of orbital rotation operations tailored on the three graphs ($G_1, G_2, G_3$) that can be defined by creating only paired connections between atoms, such as the ones described in Section \ref{Scenario I}. The parameters are always $\theta = \frac{\pi}{2}$.

The set of rotation operations is here shown in quantum circuit representation. As in the previous sections, the wires don't represent individual qubits but spatial orbitals, i.e., pairs of qubits mapped through Jordan-Wigner encoding from a minimal basis set:
\begin{align}
\mathcal{R} = \left\{
    \raisebox{-1.3cm}{\includegraphics[height=0.15\textwidth]{pics/H4_UR1_visual.png}},
    \raisebox{-1.3cm}{\includegraphics[height=0.15\textwidth]{pics/H4_UR2_visual.png}}\label{eq:h4-rotators}, 
    \raisebox{-1.3cm}{\includegraphics[height=0.15\textwidth]{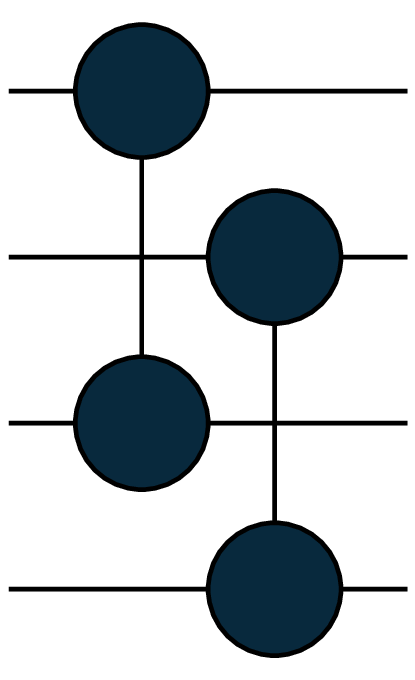}}  \right\}
\end{align}


For Scenario II we used the following circuit from \cite{kottmannMolecular2023}
    \begin{equation}
        \begin{split}
            \ket{\Psi} = & \tilde{U}_{R_{G_2}}^{\dagger}(\phi_2) U_{C_{G_2}}(\varphi_2) \tilde{U}_{R_{G_2}}(\phi_2) \\
            & \tilde{U}_{R_{G_1}}^{\dagger}(\phi_1) U_{\text{SPA}}(\varphi_1) \ket{0}
        \end{split}
    \end{equation}
    where $U_{C_{G_2}}(\varphi)$ corresponds to the spatial orbitals $p,q = 0,3$ and $p,q = 1,2$ and $\varphi$ a free parameter optimized in the preprocessing step and $\tilde{U}_{R_k}=U_{R_k}U_{RR}$ is an extension of the orbital rotation operation which preserves the topology and allows delocalization. The definition of $U_{RR}$ is taken from Eq.(28) of \cite{kottmannMolecular2023} and tailored to each test molecule.

Additionally, we analyzed the H$_4$ system arranged in a square geometry. The $\mathcal{R}$ set is the same as linear H$_4$ and the circuit used as wavefunction is
\begin{equation}
    \begin{split}
        \ket{\Psi} = & \tilde{U}_{R_{G_1}}^{\dagger}(\phi_3) U_{C_{G_1}}(\varphi_3) \tilde{U}_{R_{G_1}}(\phi_3) \\
        & \tilde{U}_{R_{G_2}}^{\dagger}(\phi_2) U_{C_{G_2}}(\varphi_2) \tilde{U}_{R_{G_2}}(\phi_2) \\
        & \tilde{U}_{R_{G_1}}^{\dagger}(\phi_1) U_{\text{SPA}}(\varphi_1) \ket{0}
    \end{split}
\end{equation}
which is an expansion of the one above due to higher accuracy in calculation.

\begin{figure*}
    \centering
    \subfigure[Scenario I]{\includegraphics[width=0.4\textwidth]{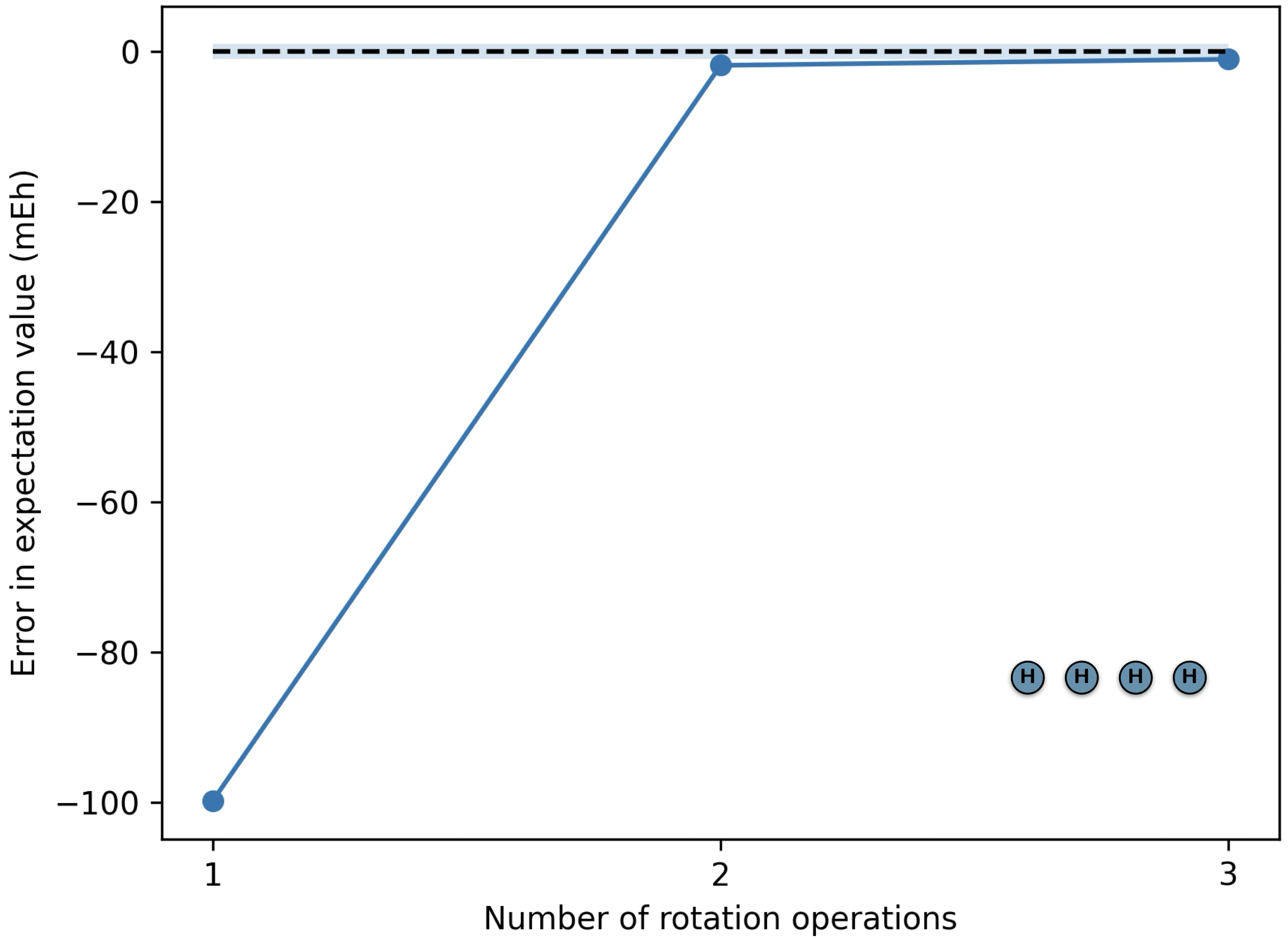}\label{truewfn_linear_H4_result}}
    \subfigure[Scenario II]{\includegraphics[width=0.4\textwidth]{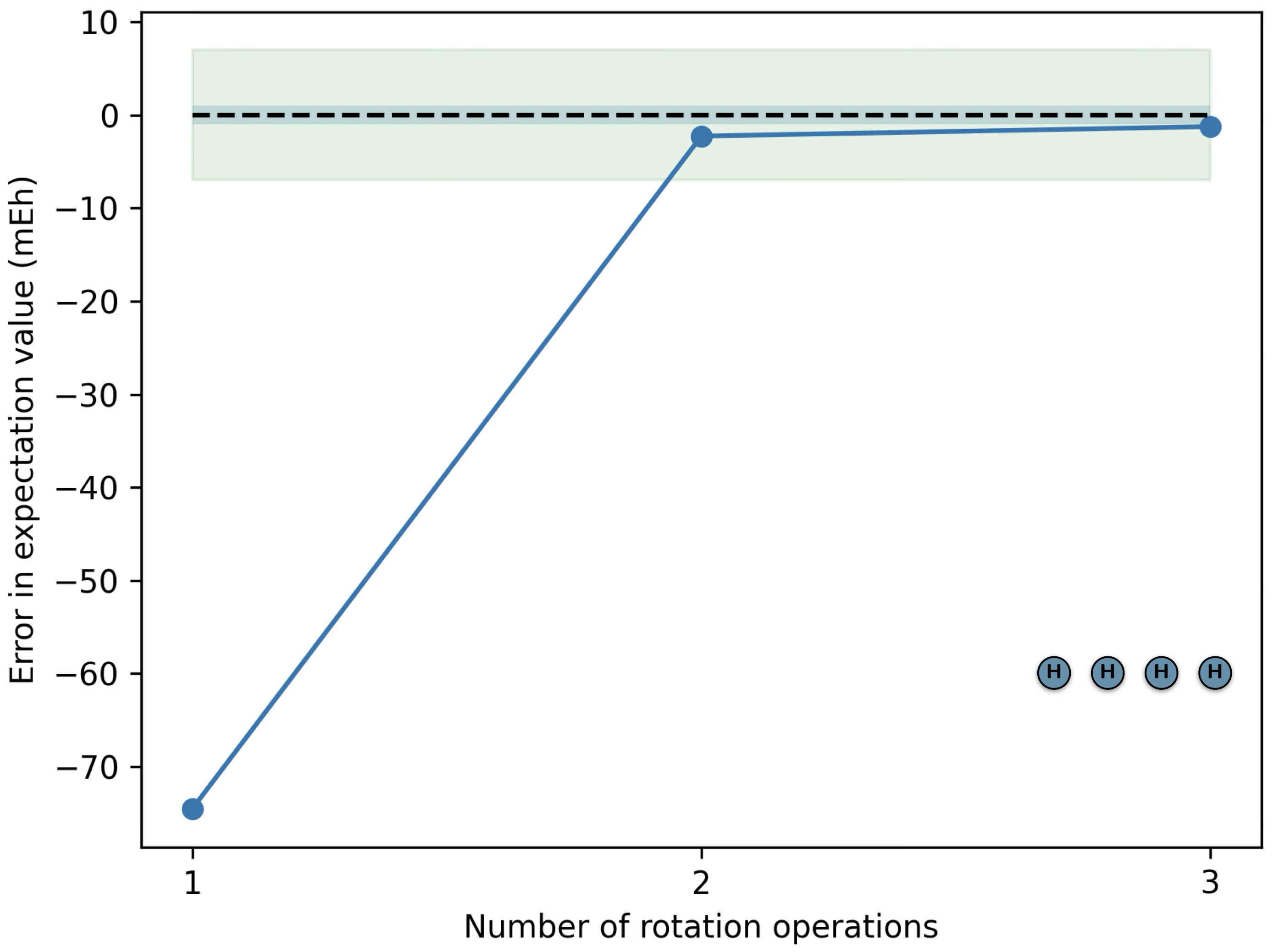}\label{linear_H4_result}} \\
    \subfigure[Scenario I]{\includegraphics[width=0.4\textwidth]{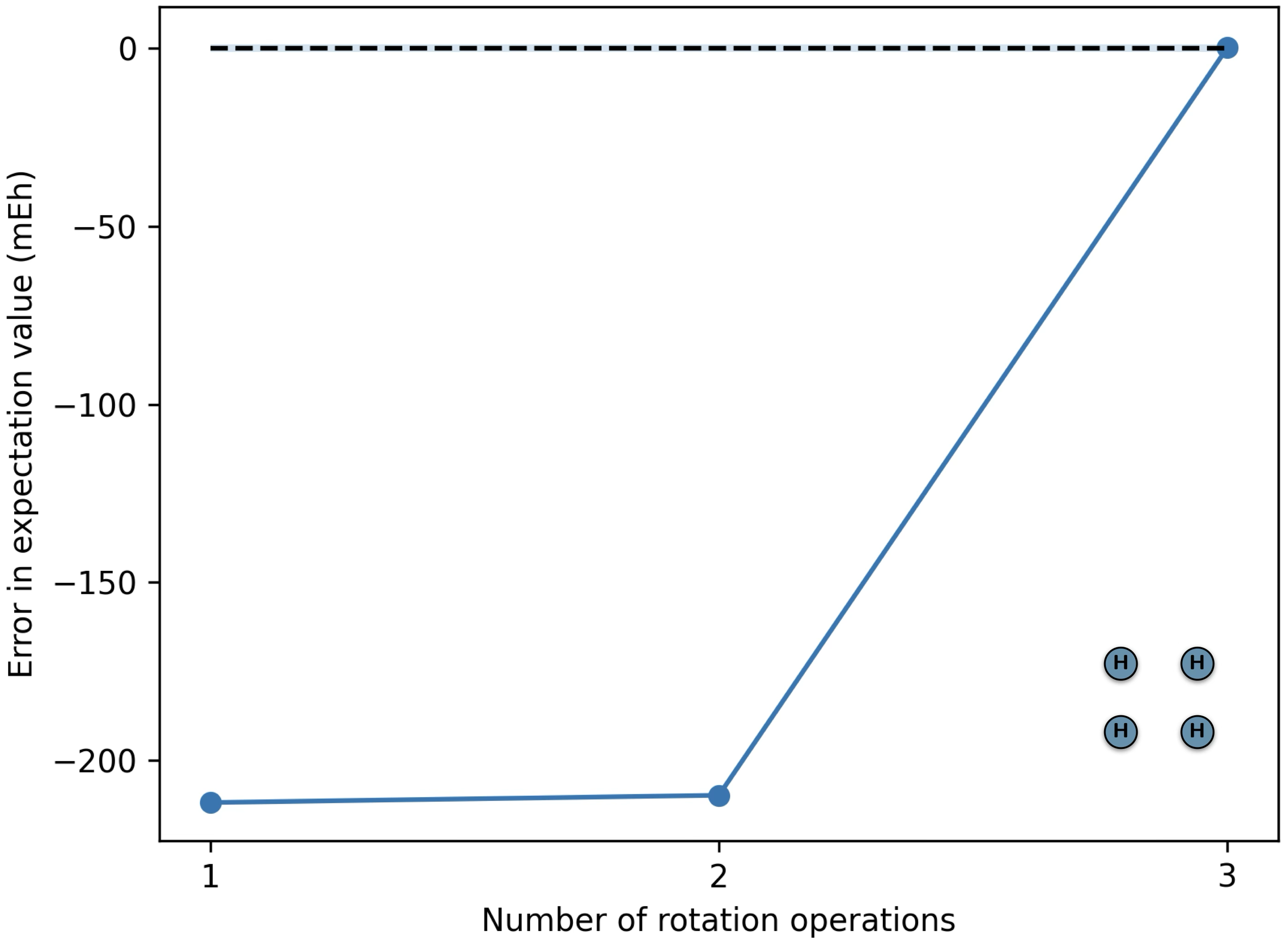}\label{truewfn_square_H4_result}}
    \subfigure[Scenario II]{\includegraphics[width=0.4\textwidth]{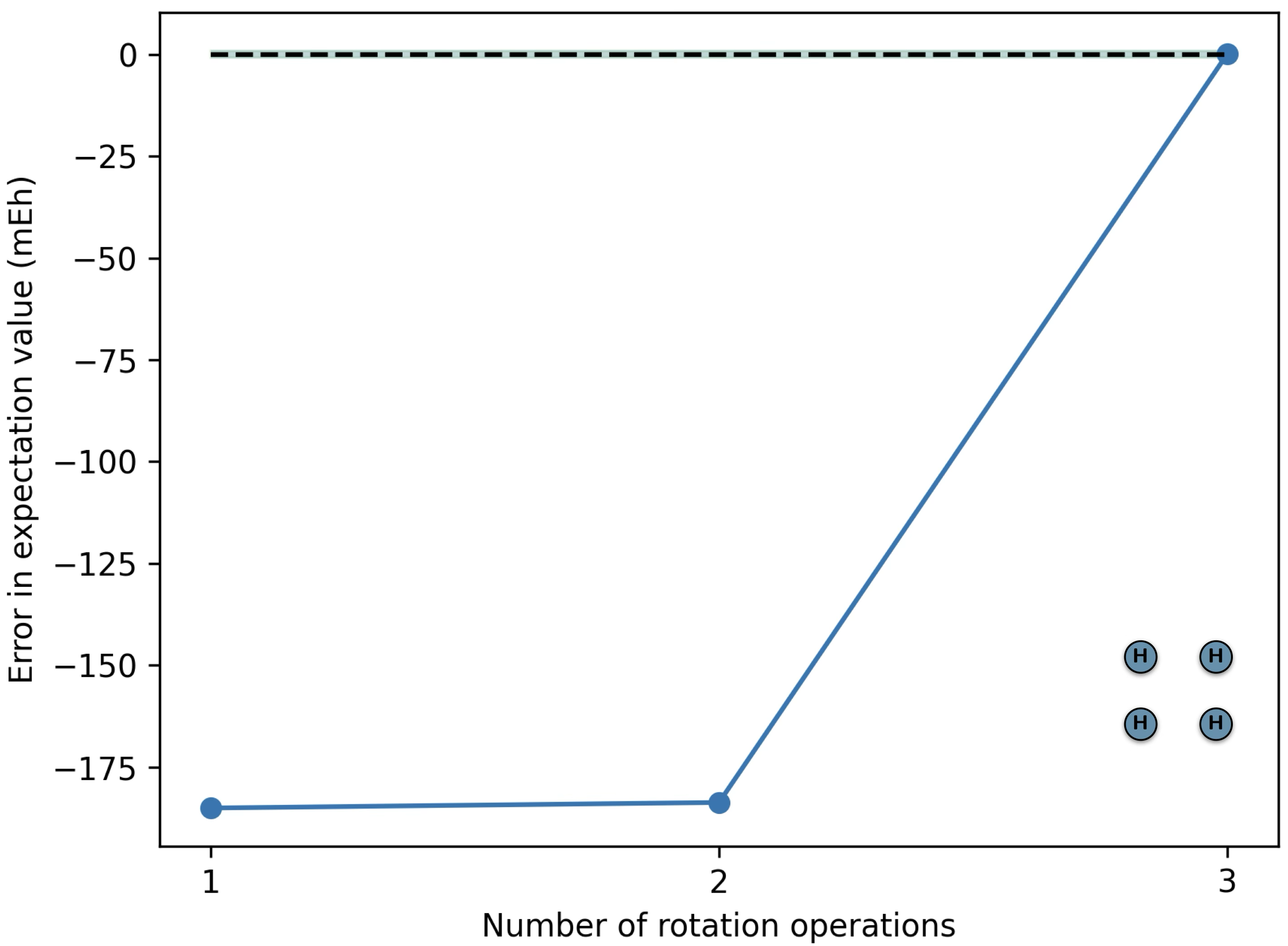}\label{square_H4_result}}
    \caption{Error in approximating the molecular Hamiltonian for (a-b) linear H$_4$ and (c-d) square H$_4$ using the set $\{$$R_{G_1}$, $R_{G_2}$, $R_{G_3}$ $\}$ from Eq.\eqref{eq:h4-rotators}. The blue area is the 1 $\mathrm{mE_h}$ margin of error, which we consider as the desired accuracy, the green area is the error from the chosen circuit ansatz. In Appendix \ref{Close-up visualization} we show a close-up visualization of the results around 1 $\mathrm{mE_h}$.}
    \label{h4_result}
\end{figure*}

Figure \ref{h4_result} displays the results of applying the set of rotation operations on the linear and the square H$_4$ systems and how well it approximates the molecular Hamiltonian $H$ (steps 3-4).

In the same way, for H$_6$ we used a set of graphs ($G_1, ..., G_5$). For Scenario II we used a circuit similar to Eq.(44) from \cite{kottmannMolecular2023}, but with additional transformations, due to higher accuracy in calculation.

Likewise, the set of rotation operations is:
\begin{align}
\mathcal{R} = \left\{
    \raisebox{-1.3cm}{\includegraphics[height=0.15\textwidth]{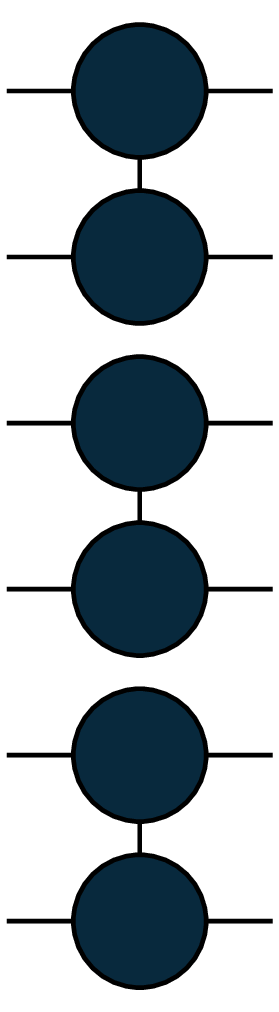}},
    \raisebox{-1.3cm}{\includegraphics[height=0.15\textwidth]{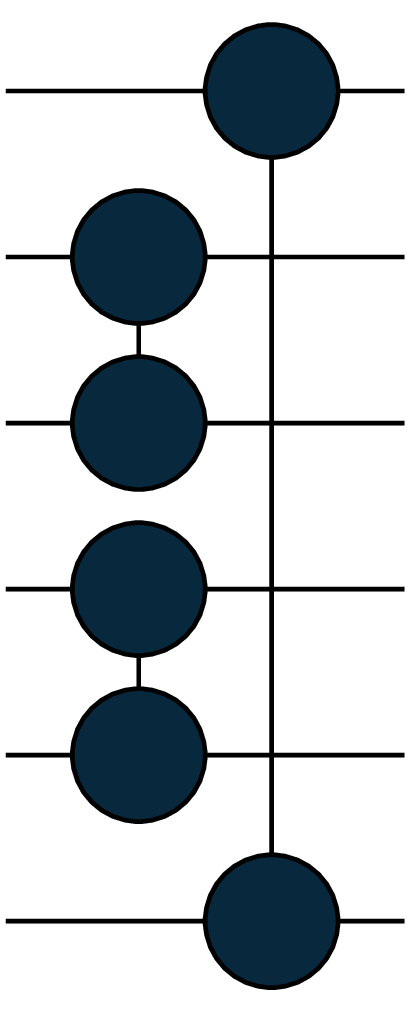}},
    \raisebox{-1.3cm}{\includegraphics[height=0.15\textwidth]{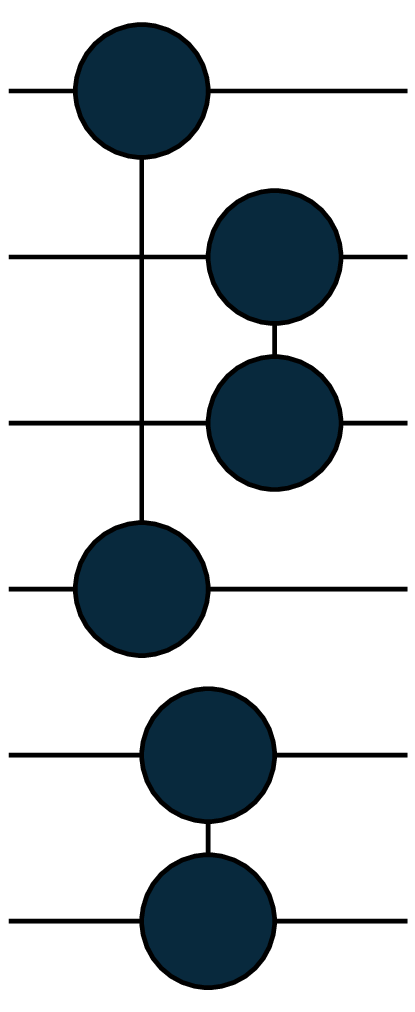}},
    \raisebox{-1.3cm}{\includegraphics[height=0.15\textwidth]{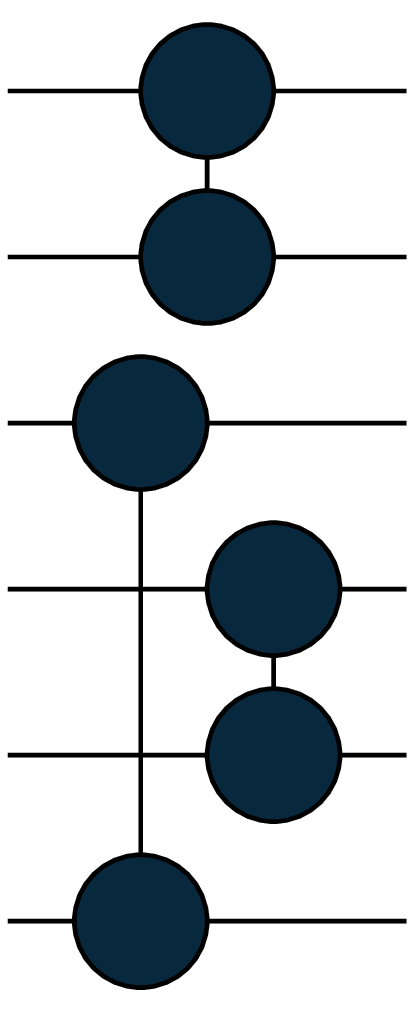}}\label{eq:h6-rotators}, 
    \raisebox{-1.3cm}{\includegraphics[height=0.15\textwidth]{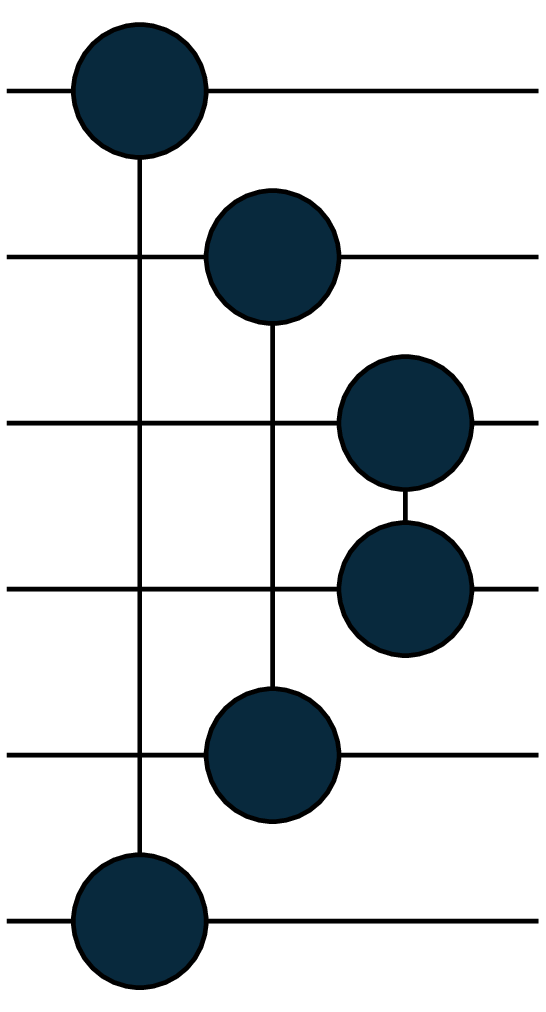}}  \right\}
\end{align}

In addition, we tested H$_6$ arranged in a circular geometry. The set is the same as above and Scenario II circuit is:
\begin{equation}
    \begin{split}
        \ket{\Psi} = & \tilde{U}_{R_{G_4}}^{\dagger}(\phi_4) U_{C_{G_4}}(\varphi_4) \tilde{U}_{R_{G_4}}(\phi_4) \\
        & \tilde{U}_{R_{G_3}}^{\dagger}(\phi_3) U_{C_{G_3}}(\varphi_3) \tilde{U}_{R_{G_3}}(\phi_3) \\
        & \tilde{U}_{R_{G_2}}^{\dagger}(\phi_2) U_{C_{G_2}}(\varphi_2) \tilde{U}_{R_{G_2}}(\phi_2) \\
        & \tilde{U}_{R_{G_1}}^{\dagger}(\phi_1) U_{\text{SPA}}(\varphi_1) \ket{0}
    \end{split}
\end{equation}

\begin{figure*}
    \centering
    \subfigure[Scenario I]{\includegraphics[width=0.4\textwidth]{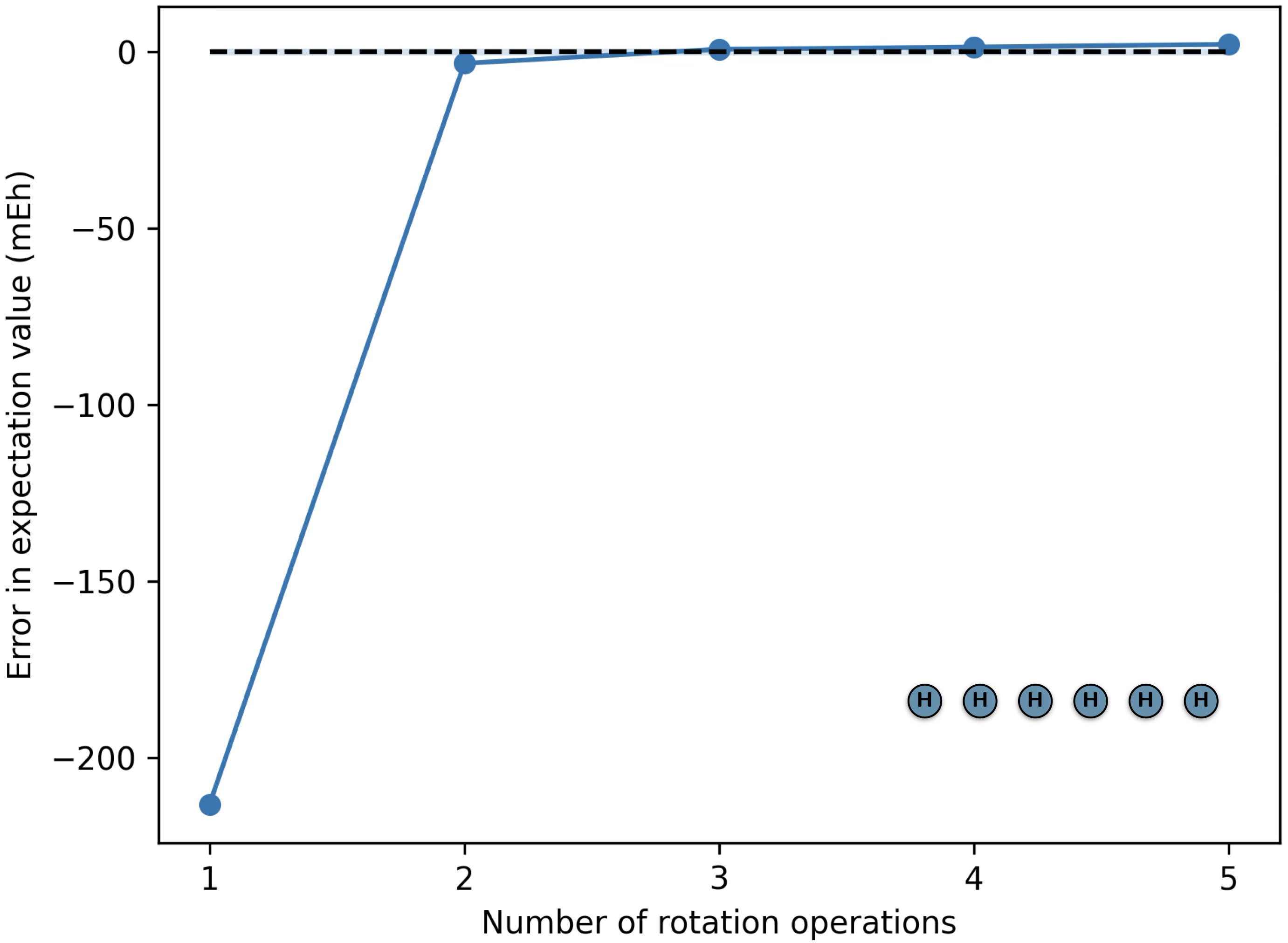}\label{truewfn_linear_H6_result}}
    \subfigure[Scenario II]{\includegraphics[width=0.4\textwidth]{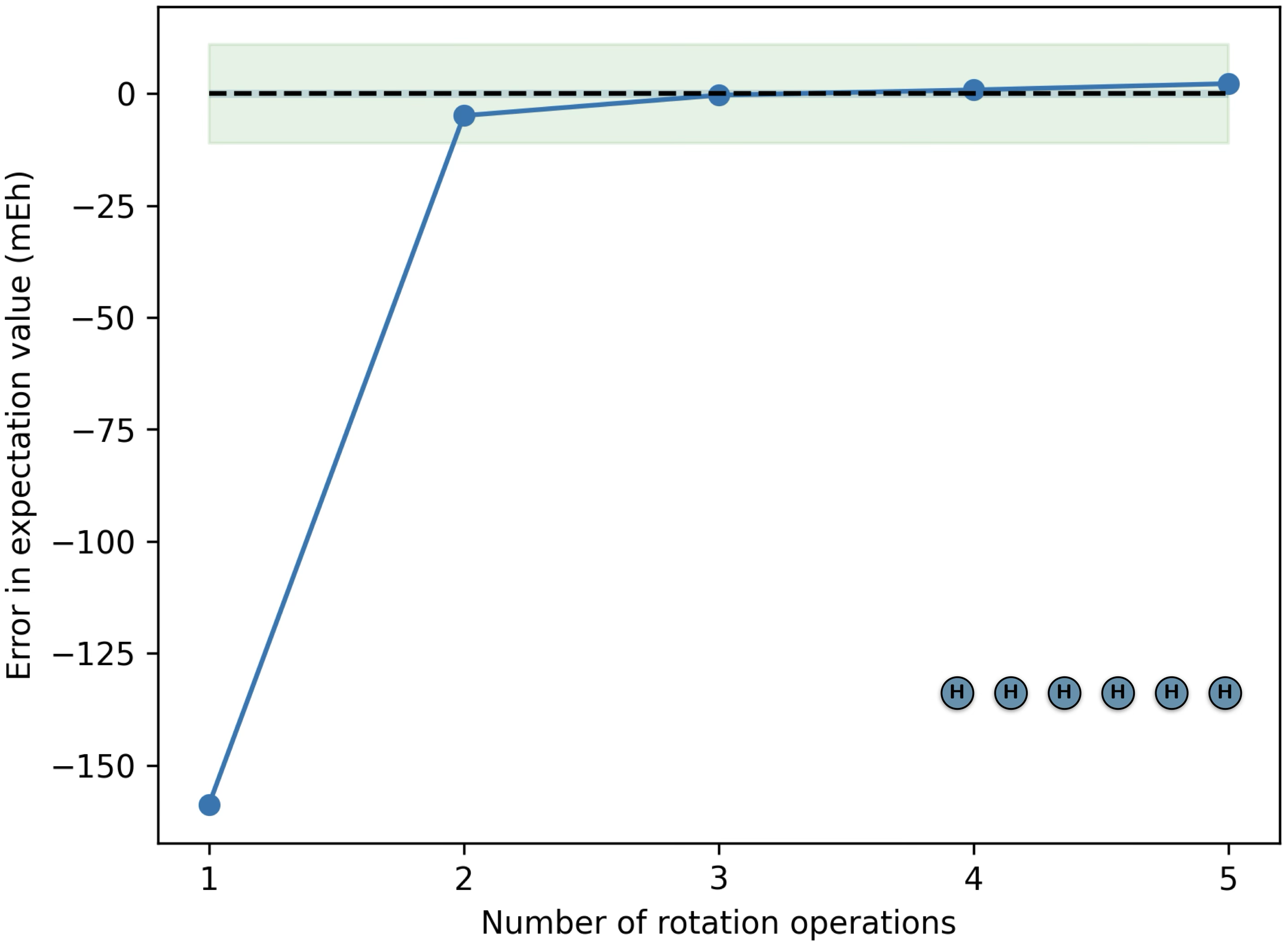}\label{linear_H6_result}} \\
    \subfigure[Scenario I]{\includegraphics[width=0.4\textwidth]{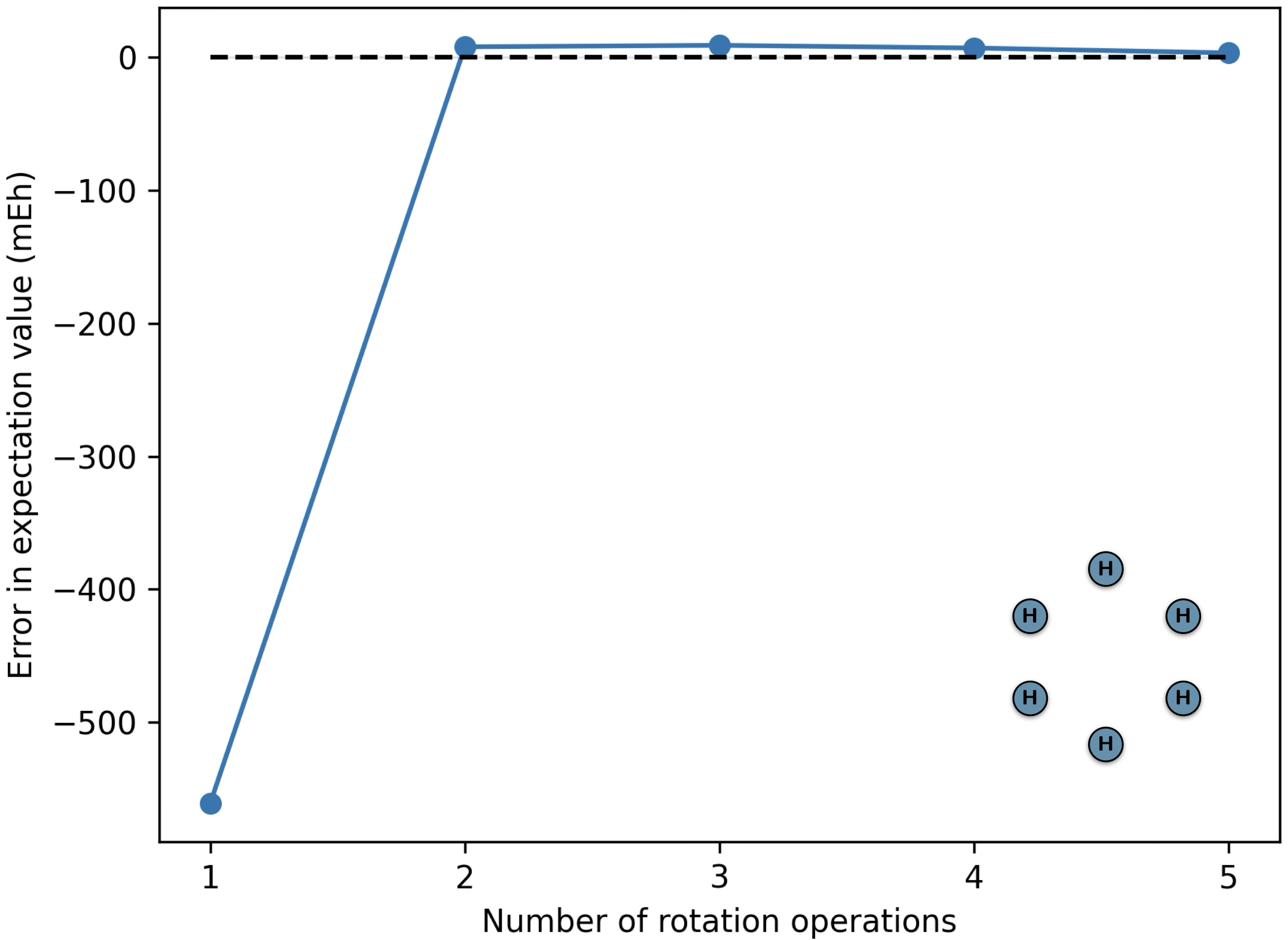}\label{truewfn_circular_H6_result}}
    \subfigure[Scenario II]{\includegraphics[width=0.4\textwidth]{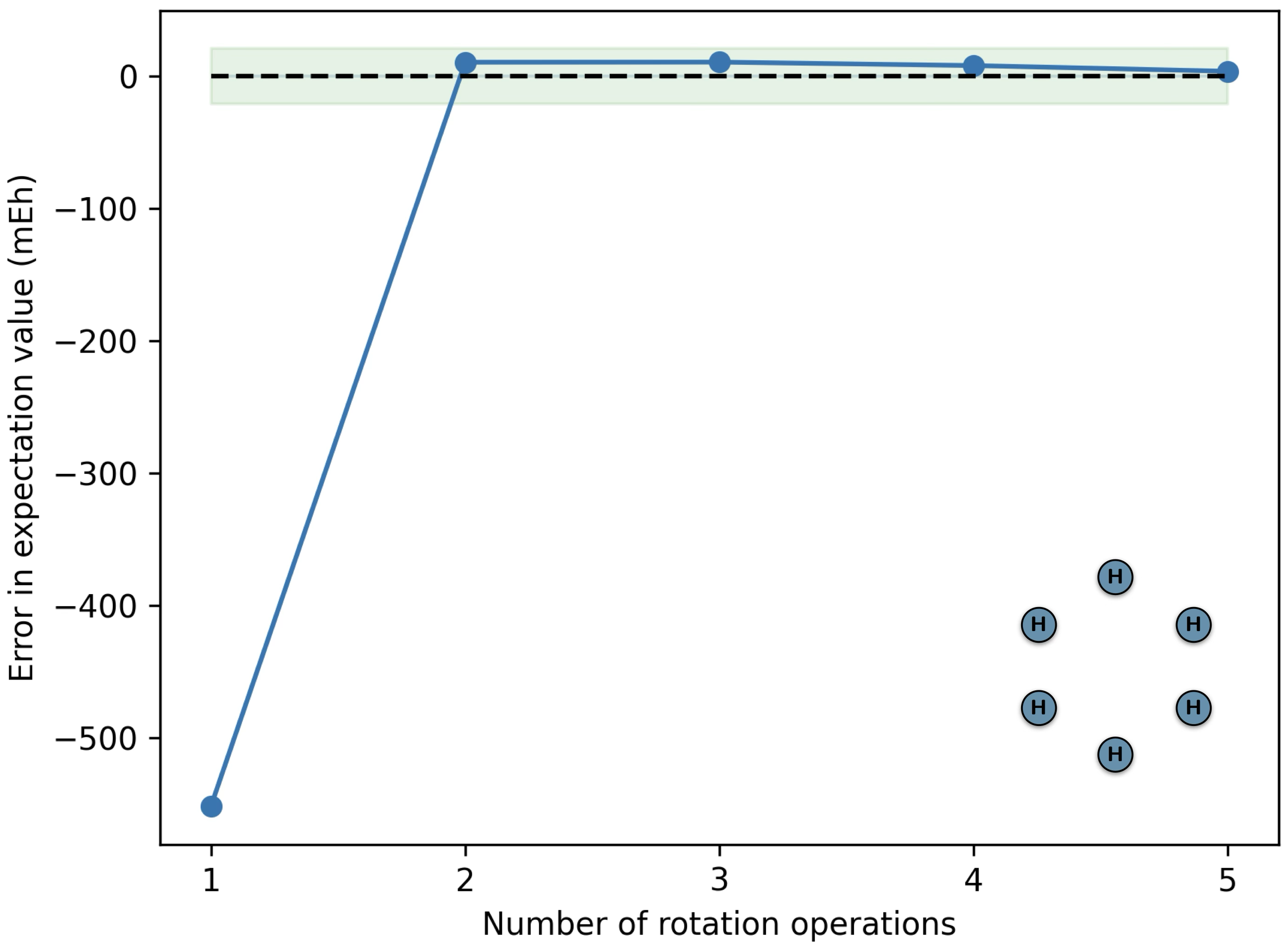}\label{circular_H6_result}}
    \caption{Error in approximating the molecular Hamiltonian for (a-b) linear H$_6$ and (c-d) circular H$_6$ using the set $\{R_{G_1}, ..., R_{G_5}\}$ from Eq.\eqref{eq:h6-rotators}. The blue area is the 1 $\mathrm{mE_h}$ margin of error, which we consider as the desired accuracy, the green area is the error from the chosen circuit ansatz. In Appendix \ref{Close-up visualization} we show a close-up visualization of the results around 1 $\mathrm{mE_h}$.}
    \label{h6_result}
\end{figure*}

Figure \ref{h6_result} shows the results for the linear and circular H$_6$ systems and how well it approximates the molecular Hamiltonian $H$ (steps 3-4).

Furthermore, we tested the approximation for H$_6$ molecules arranged in randomized geometries. As a set of rotation operations we used \{ $R_{G_1}$, ..., $R_{G_6}$ \}, where the graphs are defined by minimum global distance edges. In this case we defined 50 additional unitary transformations randomly generated. Figure \ref{Random_H6_result} shows the result on 100 H$_6$ systems. \\

Finally, regarding H$_8$ we used a set of operations tailored on graphs ($G_1, ..., G_6$). For step 2 we used the same circuit from circular H$_6$, but with the new graphs from H$_8$.

The set of rotation operations is shown here:
\begin{align}
\mathcal{R} = \left\{
    \raisebox{-1.3cm}{\includegraphics[height=0.15\textwidth]{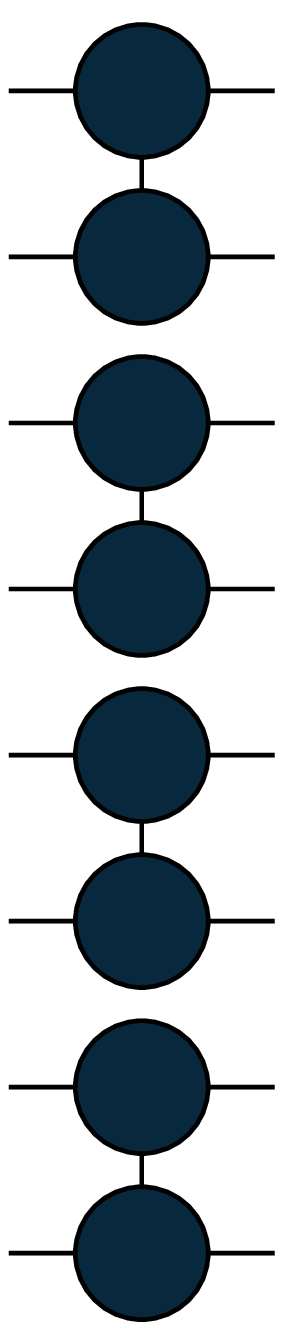}},
    \raisebox{-1.3cm}{\includegraphics[height=0.15\textwidth]{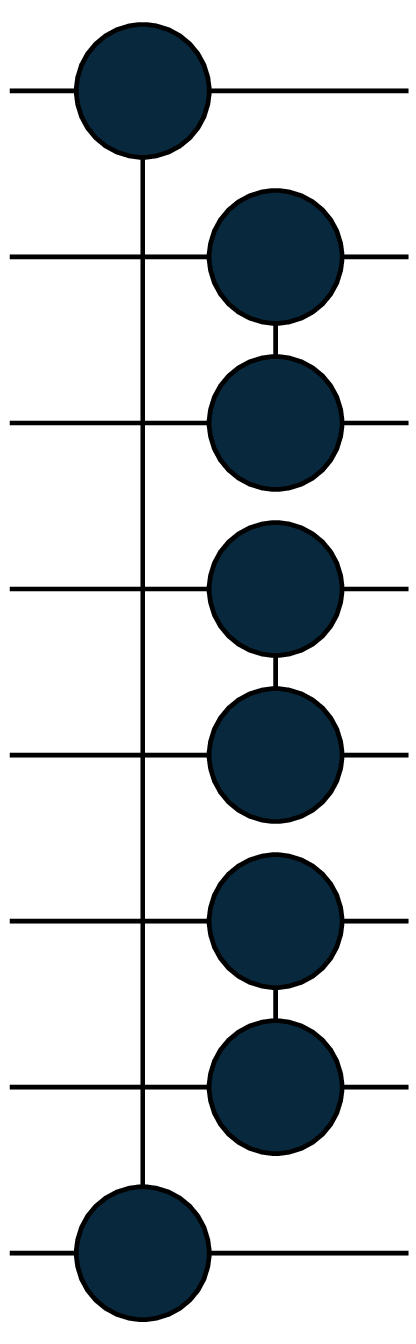}},
    \raisebox{-1.3cm}{\includegraphics[height=0.15\textwidth]{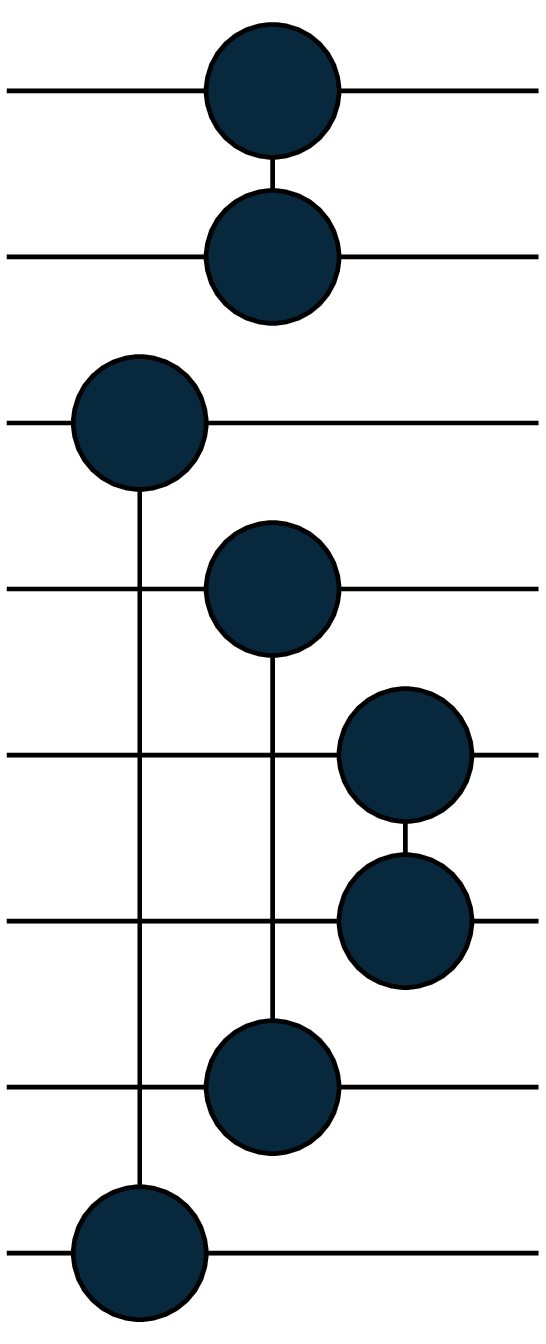}},
    \raisebox{-1.3cm}{\includegraphics[height=0.15\textwidth]{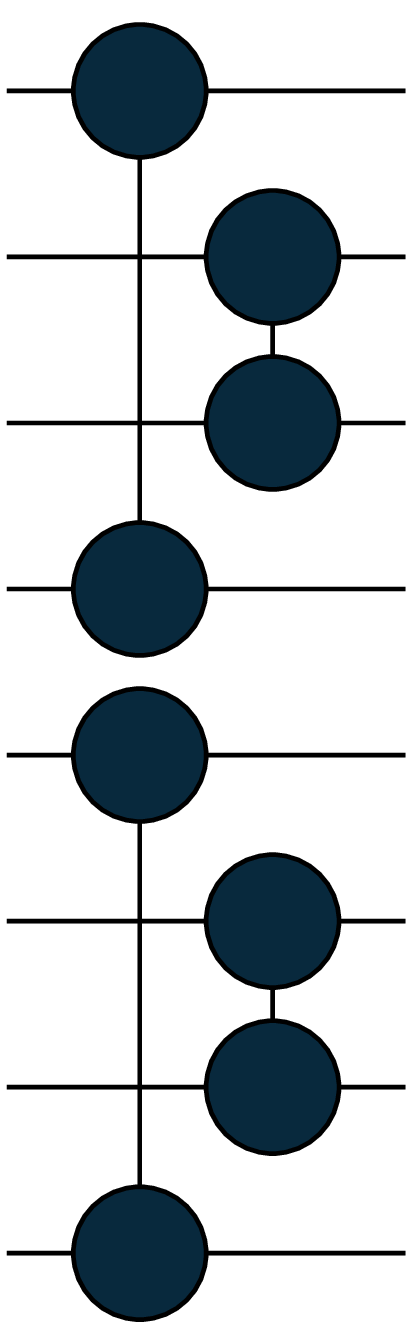}},
    \raisebox{-1.3cm}{\includegraphics[height=0.15\textwidth]{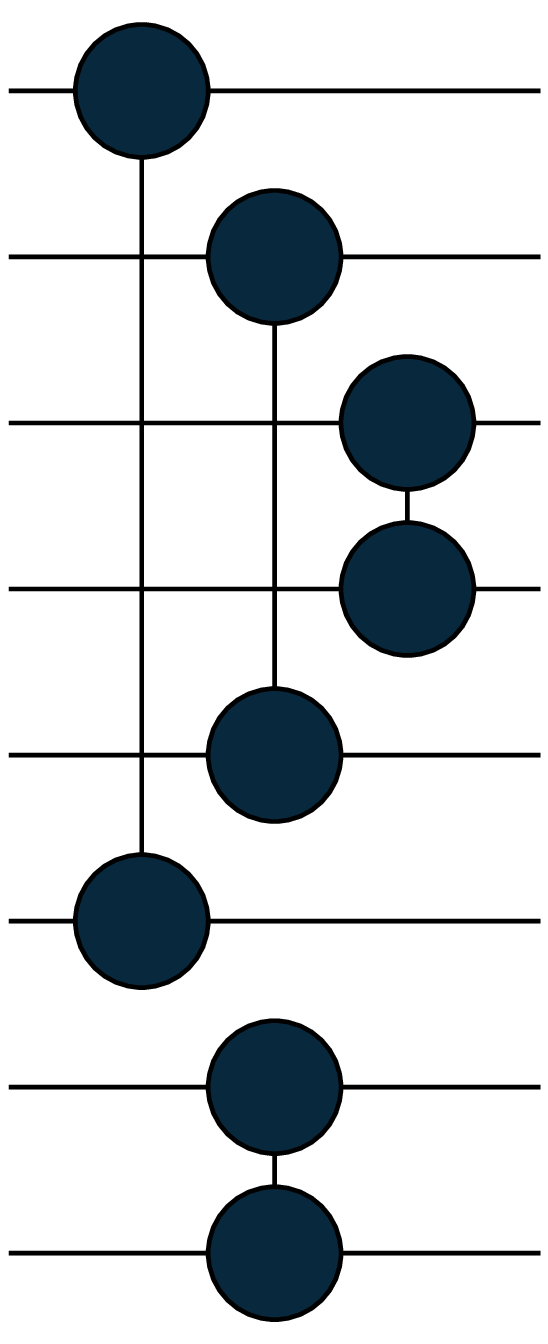}}\label{eq:h8-rotators}, 
    \raisebox{-1.3cm}{\includegraphics[height=0.15\textwidth]{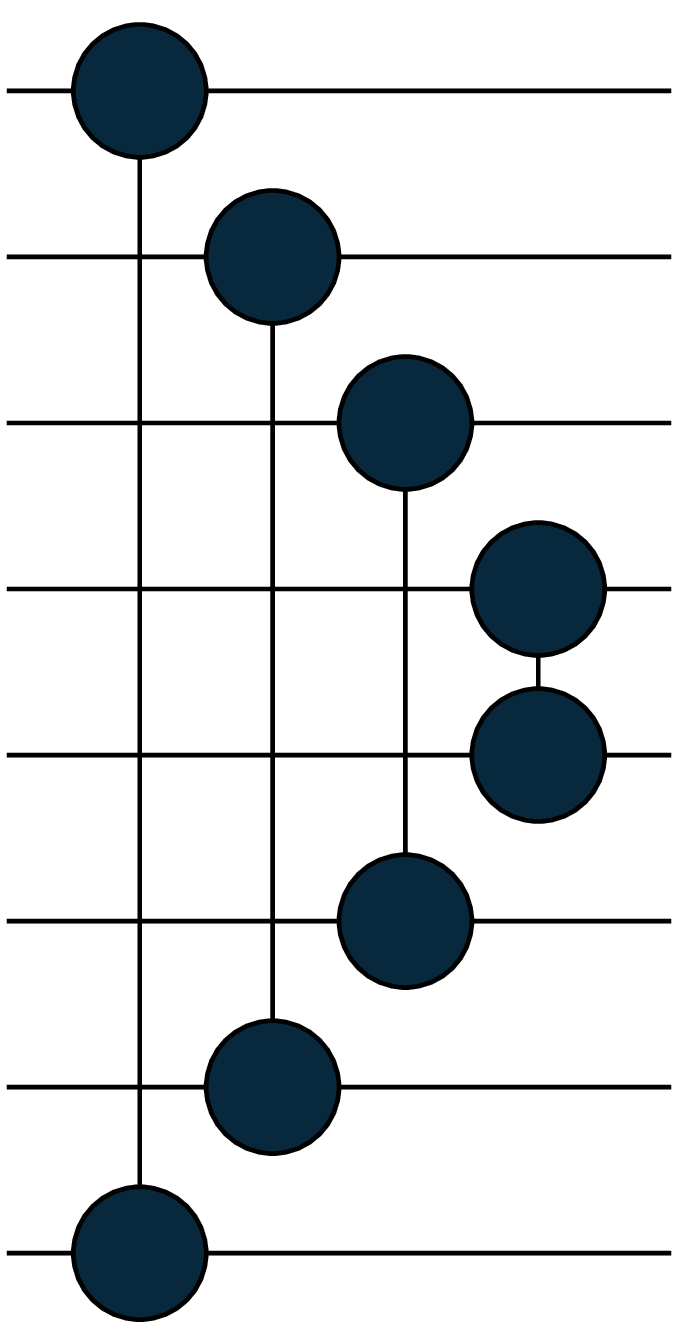}}  \right\}
\end{align}

Figure \ref{h8_result} shows the result for the linear H$_8$ system.

The number of measurement groups (number of expectation values times number of self-commuting groups) needed for all the examples are shown in Figure \ref{Num_of_meas_groups}, with further details in Appendix \ref{Full results tables}.

Figure \ref{Depth_overhead} shows the depth overhead needed for measurements in the three linear examples. Here, the values for Scenario I and II are considered in the Reordered Jordan-Wigner encoding, meaning that the qubits order in the quantum register follows the pattern $\ket{\uparrow\uparrow...\uparrow\downarrow\downarrow...\downarrow}$. This choice leads us to a lower depth overhead by decoupling spin-up and spin-down excitations gates.

\subsection{Number of measurements}
\label{Number of measurements section}

Given $H = \sum_i H_i = \sum_i w_i P_i$, for each Pauli string $P_i$ we estimate the number of measurements as:
\begin{equation}
    M_i = \left( \frac{\abs{w_i}\sqrt{\text{Var}(P_i)}}{\epsilon} \right) ^2 = \left( \frac{\abs{w_i}\sqrt{(1 - \langle P_i \rangle_{\Psi}^2)}}{\epsilon} \right) ^2
\end{equation}
where $\epsilon$ represents the precision and is set to $10^{-3}$. \cite{hugginsEfficient2021}

Then, for each commuting group we only consider the largest value,
\begin{equation}
    M_{\text{group}} = \max_{H_i \in \text{group}} M_i,
\end{equation}
since we can measure all the operators belonging to such group simultaneously. $M_{\text{group}}$ represents the upper bound on our estimated number of measurements.

Finally, we sum together all the contributions from each iteration of our measurement procedure to retrieve the total number of measurements.

Figure \ref{Num_of_meas} shows the number of measurements needed for a complete repetition of the procedure. The full results table is presented in Appendix \ref{Full results tables}.

\subsection{Finite samples simulation}
To validate the consistency of the number of measurements, we evaluated each group using a finite sample size. The number of samples was set to the estimated number of measurements, previously defined as the maximum value among all Pauli strings within the same group. We then repeated this process 100 times and computed the average over all sample simulations for each group. Our assumption is that the final result remains below the previously fixed precision $\epsilon=10^{-3}$.

\begin{figure*}[h]
  \centering
  \subfigure[Linear H$_4$ SI]{\includegraphics[width=0.32\textwidth]{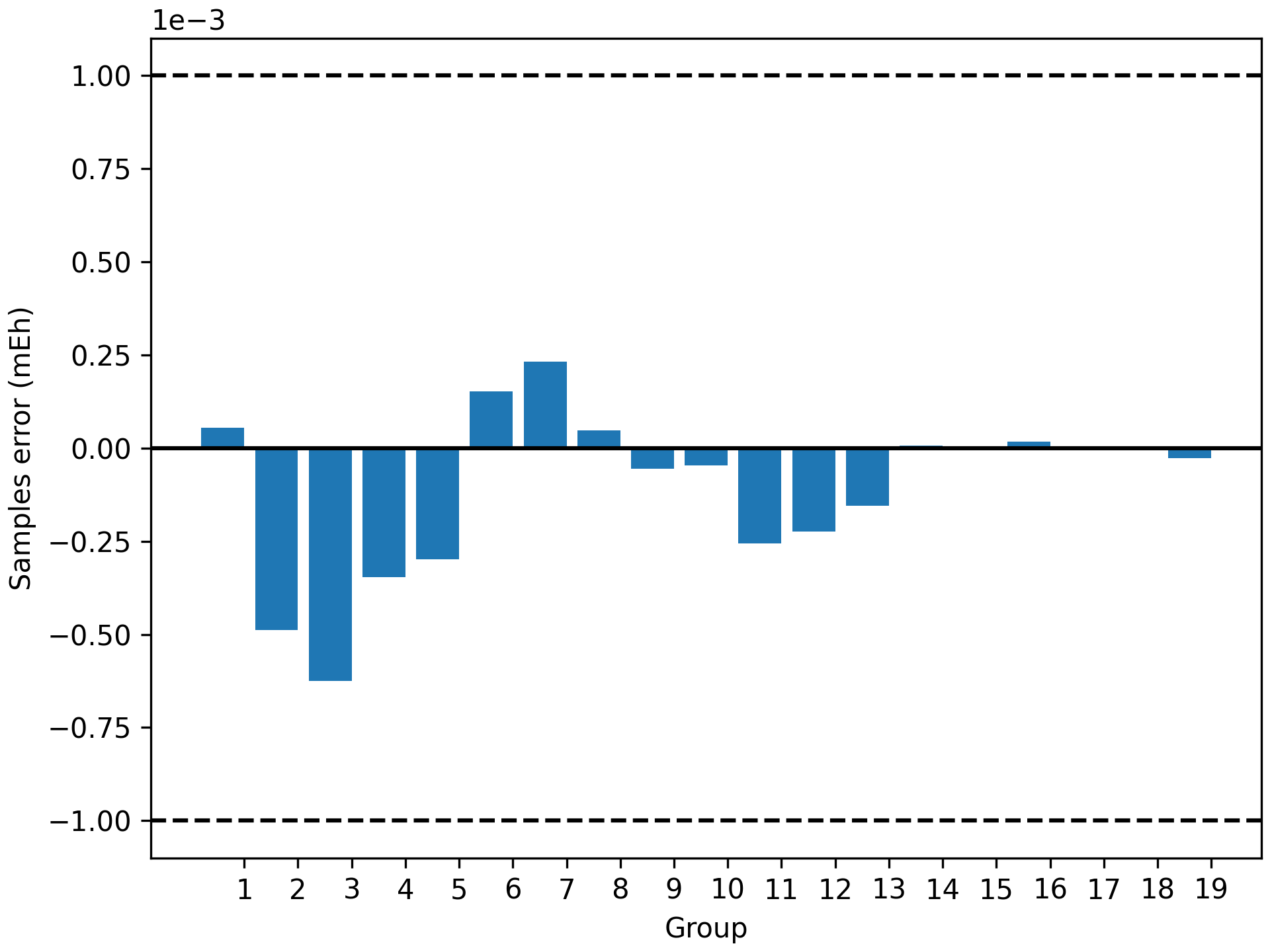}}
  \subfigure[Linear H$_4$ Scenario I]{\includegraphics[width=0.32\textwidth]{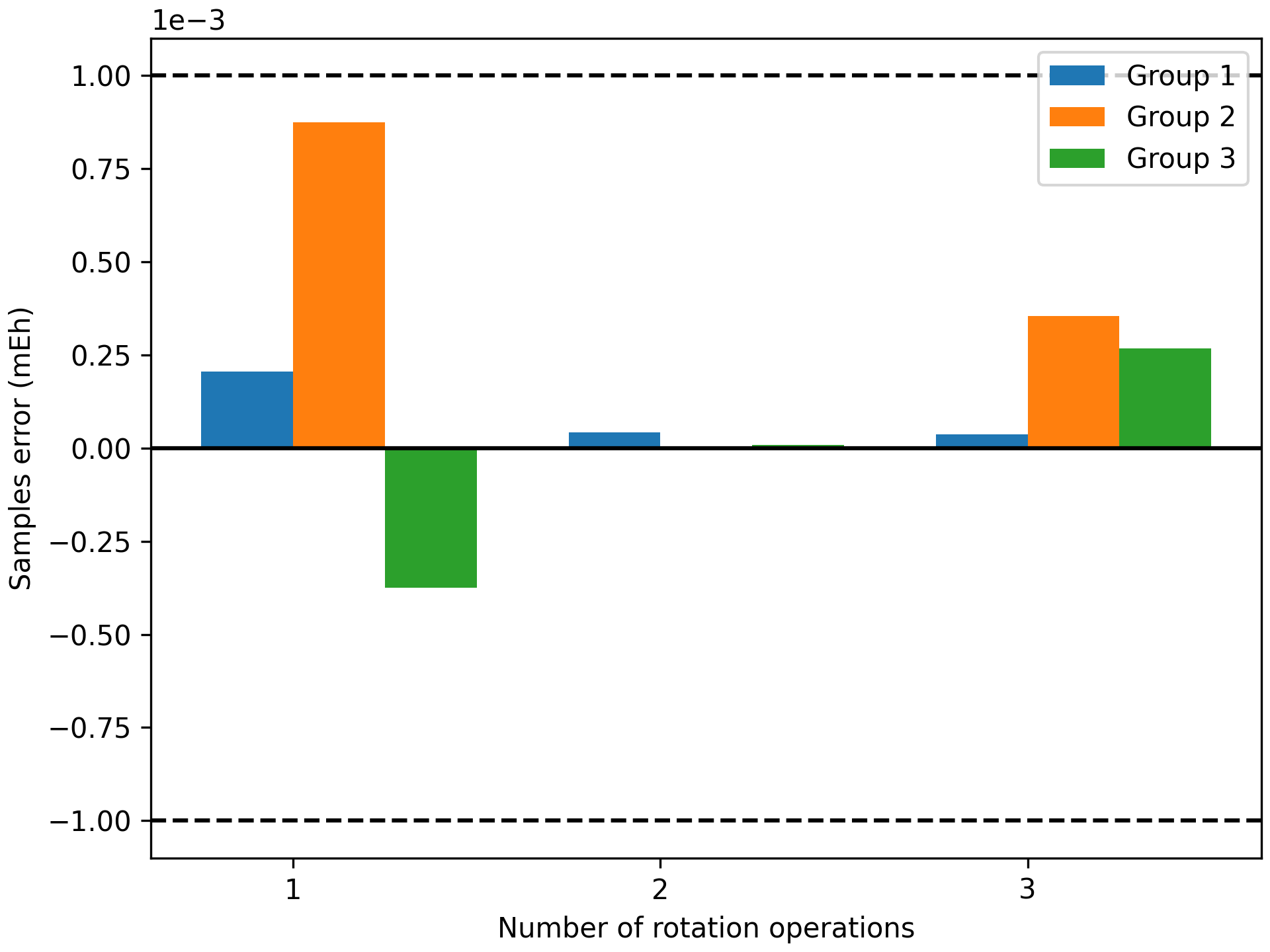}}
  \subfigure[Linear H$_4$ Scenario II]{\includegraphics[width=0.32\textwidth]{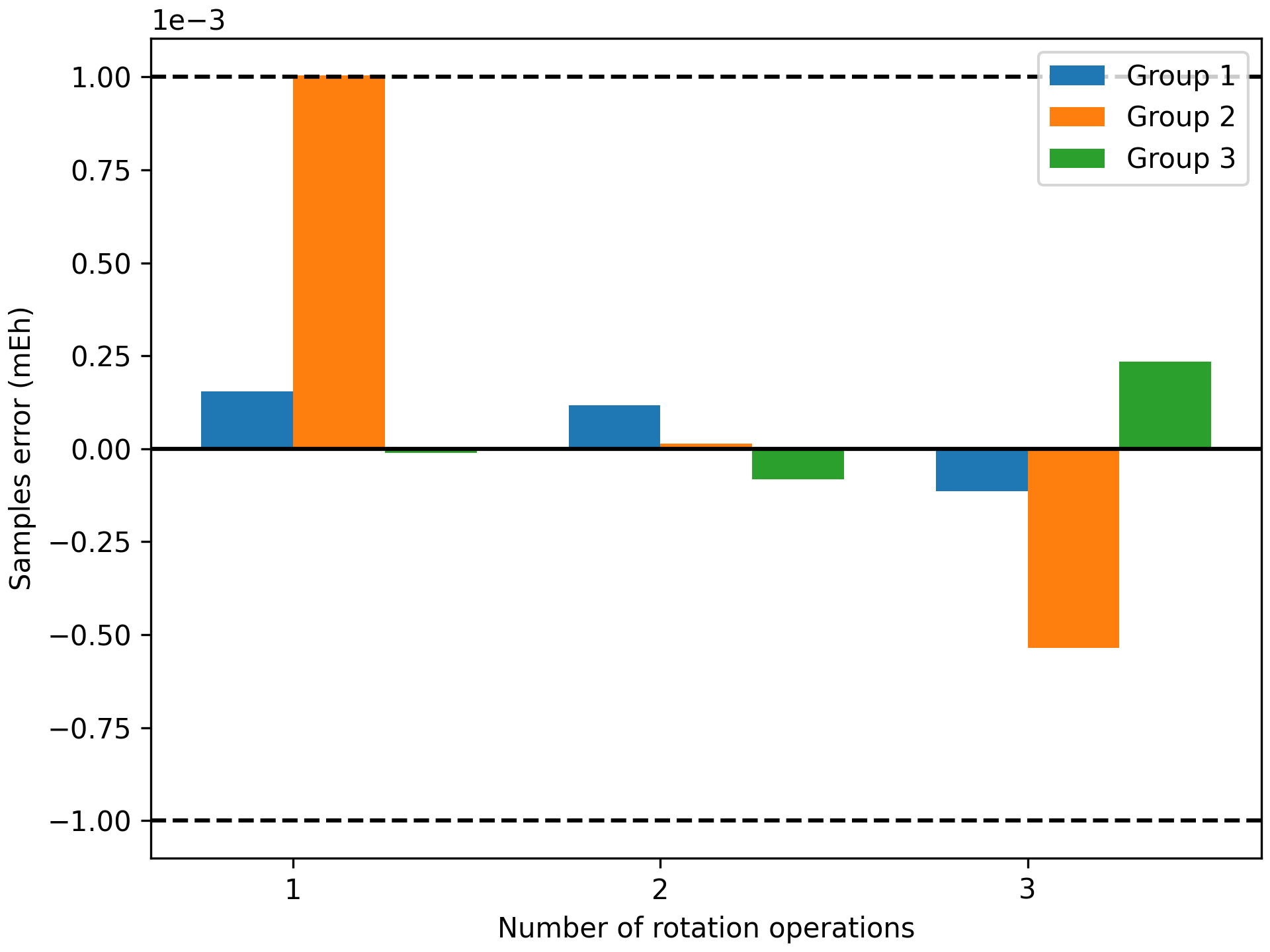}}
  \subfigure[Linear H$_6$ SI]{\includegraphics[width=0.32\textwidth]{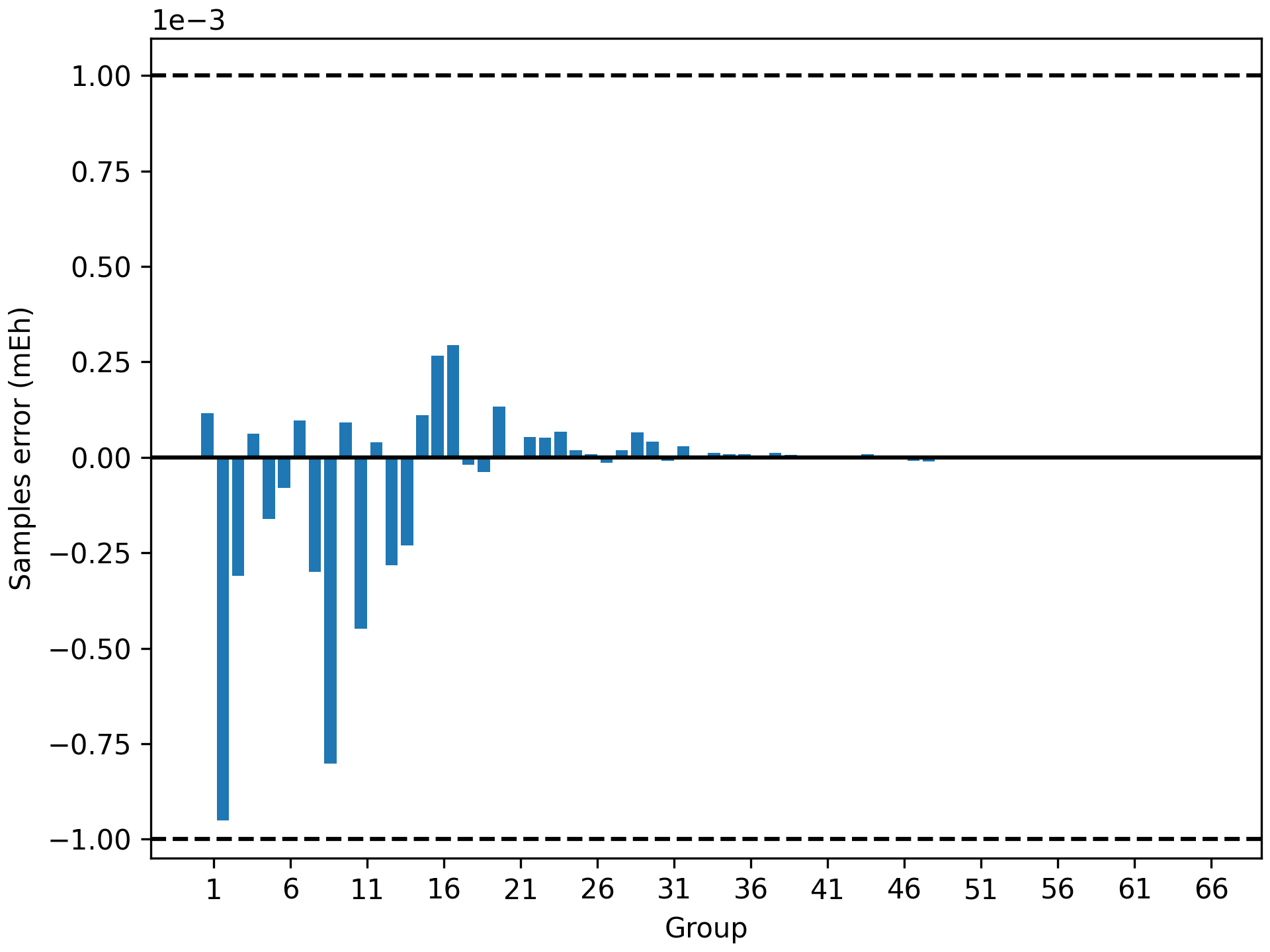}}
  \subfigure[Linear H$_6$ Scenario I]{\includegraphics[width=0.32\textwidth]{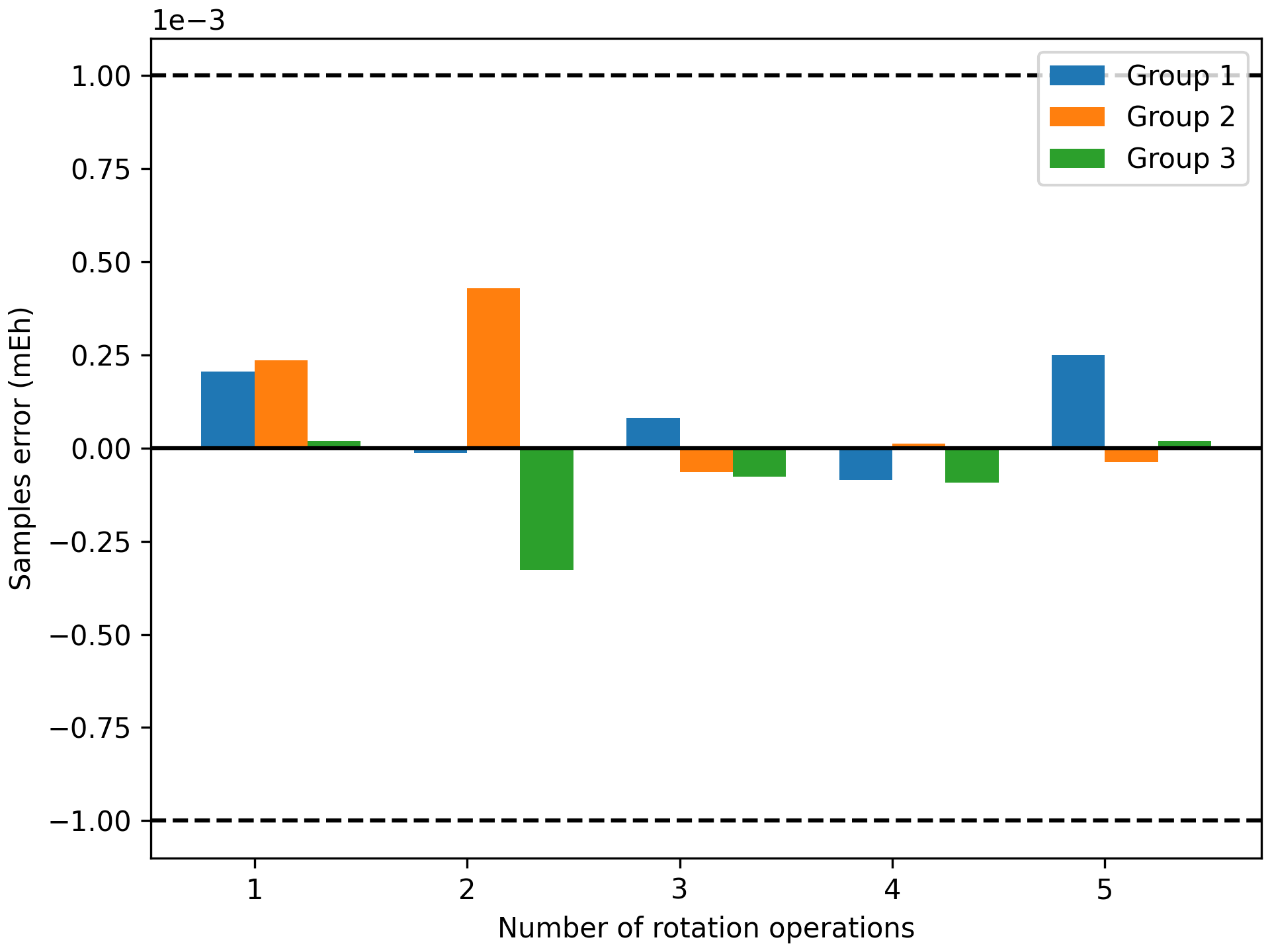}}
  \subfigure[Linear H$_6$ Scenario II]{\includegraphics[width=0.32\textwidth]{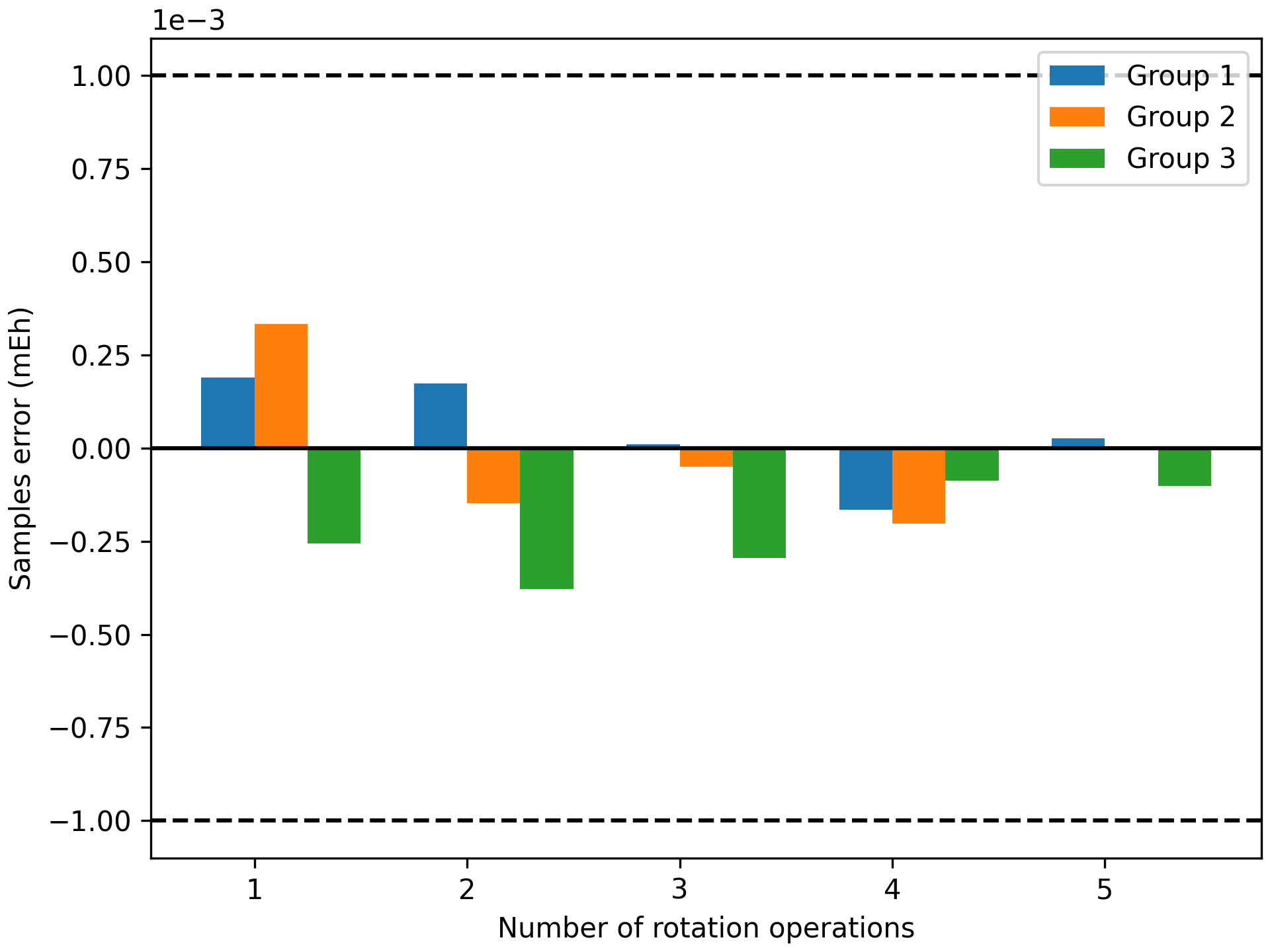}}
  \subfigure[Linear H$_8$ SI]{\includegraphics[width=0.32\textwidth]{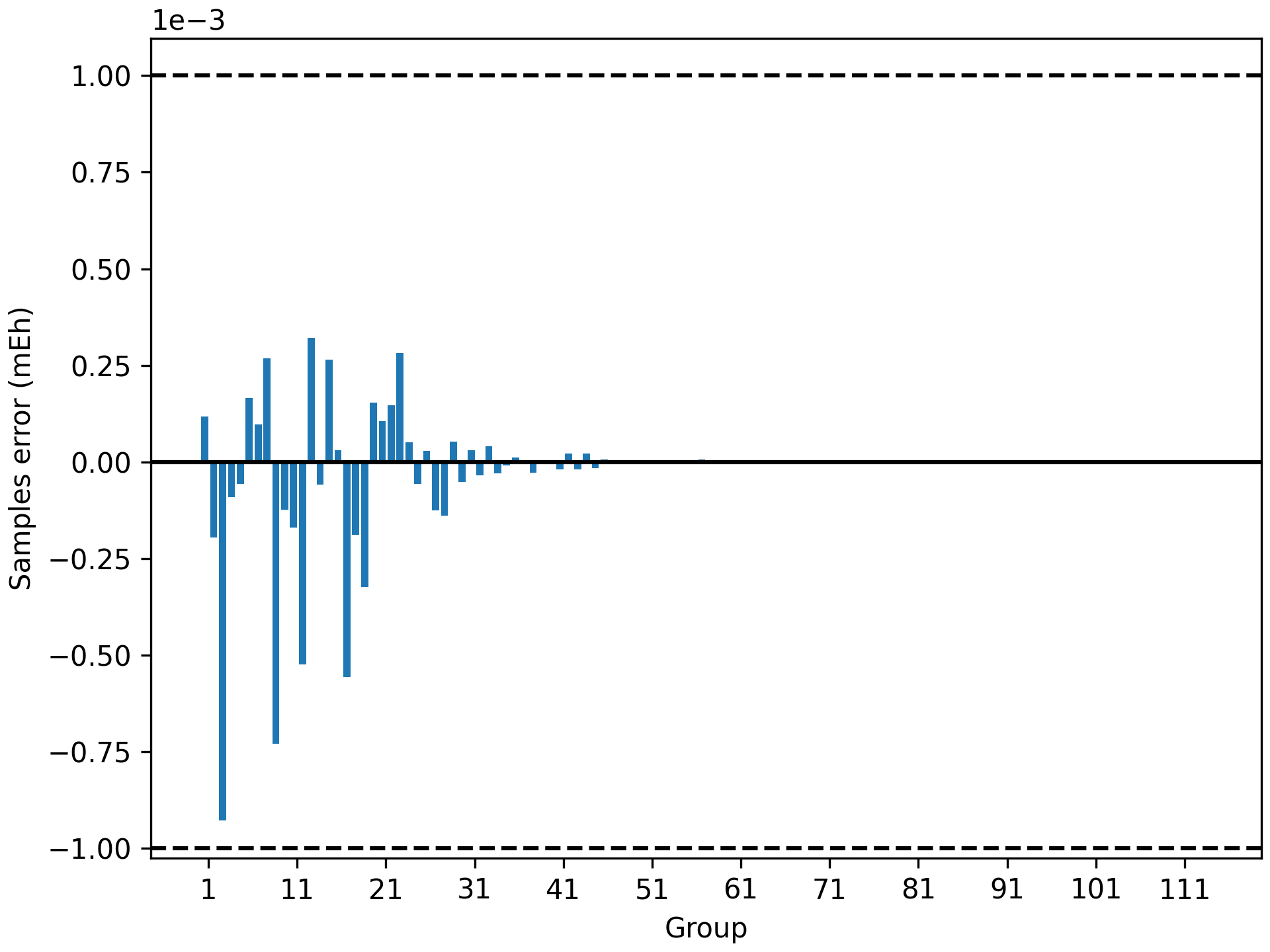}}
  \subfigure[Linear H$_8$ Scenario I]{\includegraphics[width=0.32\textwidth]{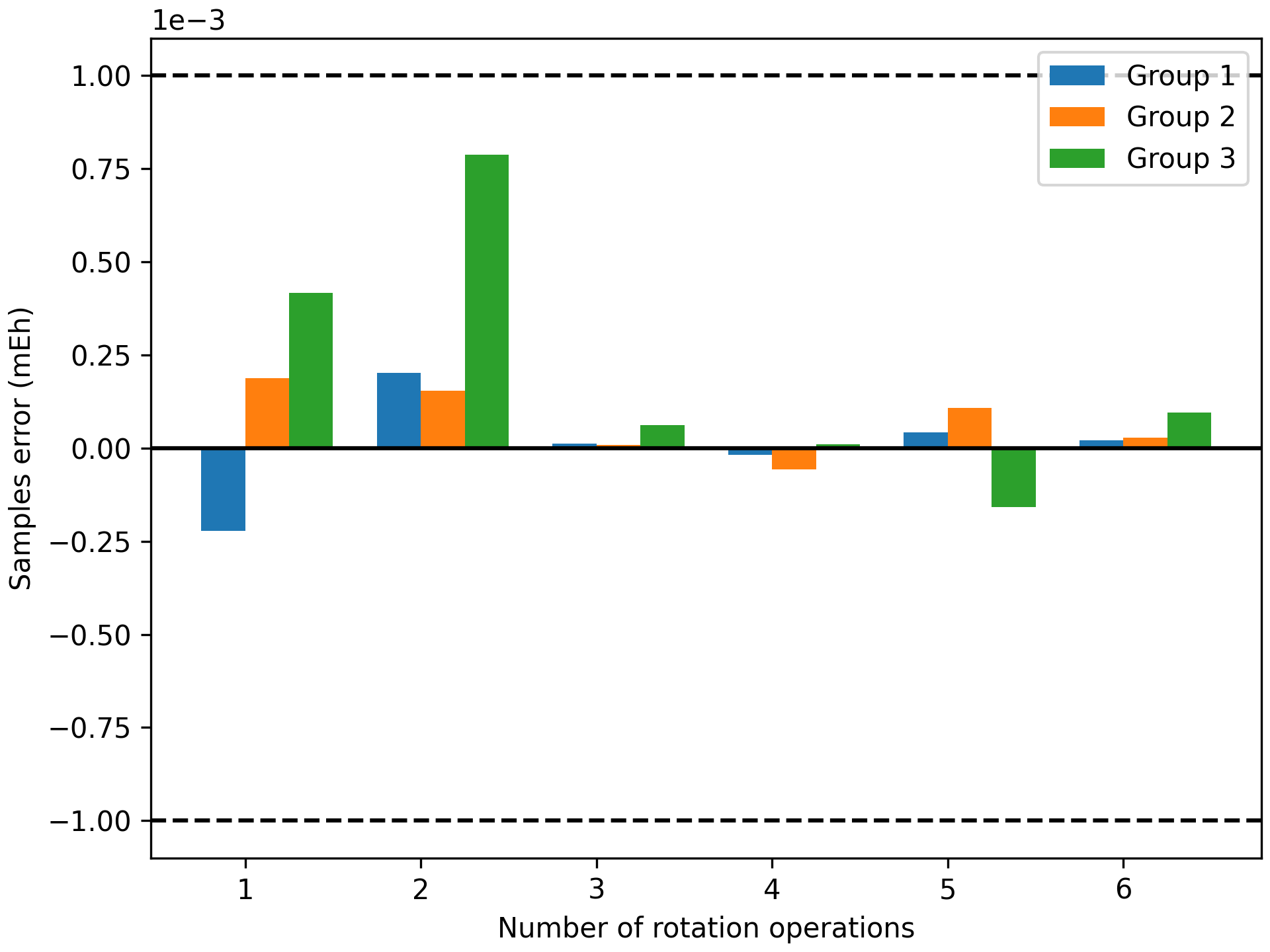}}
  \subfigure[Linear H$_8$ Scenario II]{\includegraphics[width=0.32\textwidth]{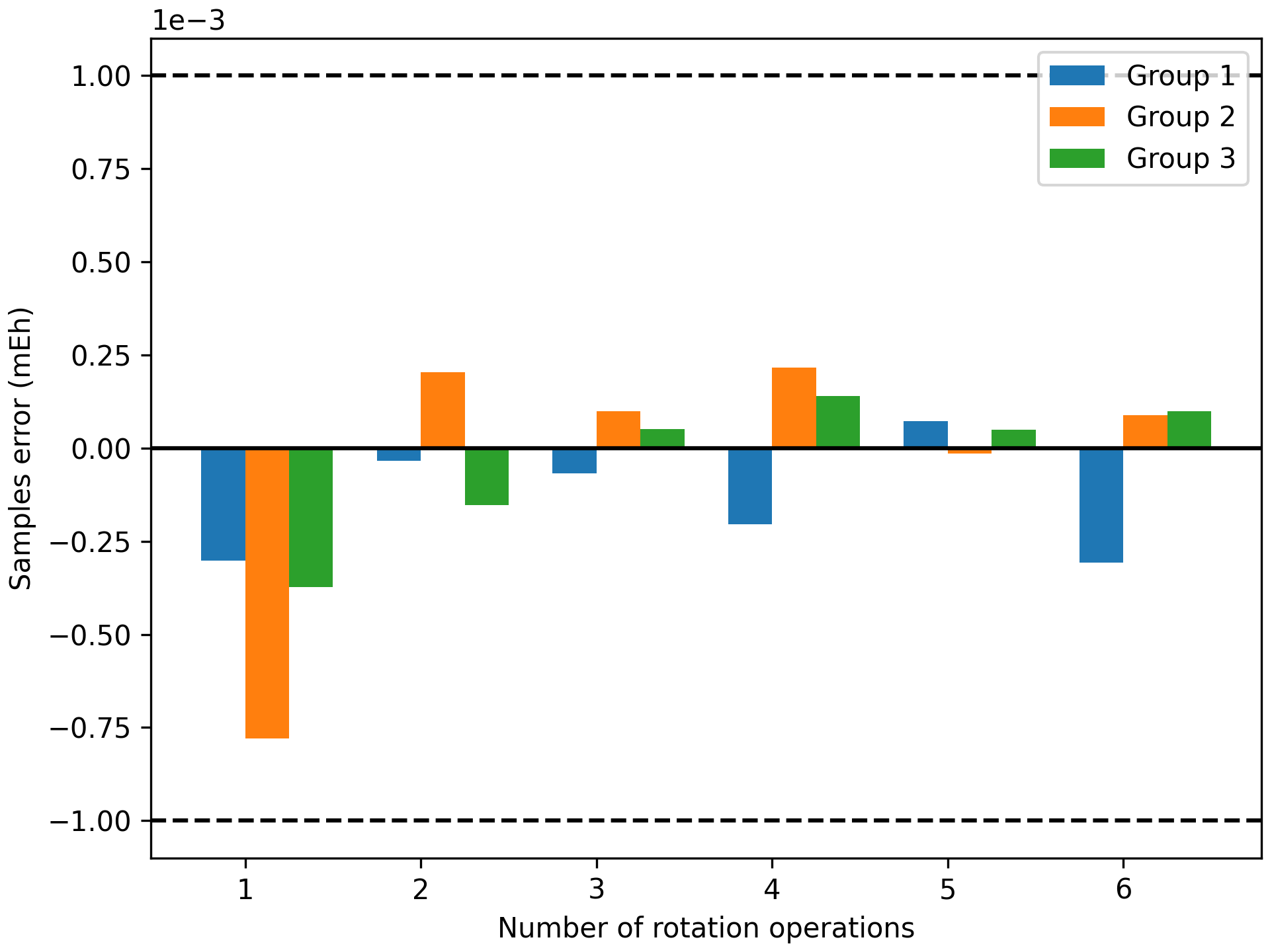}}
  \caption{Finite samples simulation for three molecules: (a-c) linear H$_4$, (d-f) linear H$_6$ and (g-i) linear H$_8$. SI method is used as a benchmark. In Scenario I and II, the rotation operations and the measured groups correspond to those used for the approximation of Hamiltonian expectation values. As mentioned above we set a threshold precision of $\epsilon=10^{-3}$.} \label{Samples}
\end{figure*}

The results are presented in Figure \ref{Samples} together with comparison to the SI method. In Scenario I and II the measurement groups are distributed and evaluated over each rotation operation, whereas for SI they are computed simultaneously. In all considered examples the error never exceeds the precision, thus confirming the consistency of the estimated number of measurements.

\section{Conclusion \& Outlook} \label{Conclusion}
This study introduces a physically motivated methodology for minimizing measurement overhead in the VQE algorithm. The collected data refers to a single optimization iteration. In all cases examined we consistently retrieved accurate approximations for the expectation value of the molecular Hamiltonian, achieving this with a comparably small number of iterations, or rotation operations. These findings show that the HCB elements of an Hamiltonian hold relevant information that can be efficiently leveraged to reduce the computational demand, as we assumed. In fact, the measured operators all respect this structure.

The number of measurement groups is improved by $50\%$ to $80\%$ compared to benchmarks. In the case of less-structured systems, Scenario I shows a large deviation in the distribution, while in Scenario II, $80\%$ of the distribution falls under 10 measurement groups, implying that for most cases only around 10 self-commuting groups need to be measured. All distributions are shown in Appendix \ref{num_of_meas_freeH6}. This behaviour suggests that the co-design of circuit and rotation operations may lead to more expressivity in the HCB Hamiltonian evaluation and thus fewer algorithm iterations needed, though this is not conclusive.

Orbital rotation operations have shown to statistically improve this approximation, even when generated randomly. The graph-based operations, in particular, lead to both accurate and efficient results, although further research is needed to single out the optimal class of operations. This improvement is obtained at a cheap cost in circuit depth overhead, which is lower than state-of-the-art methods.

Notably, the total number of measurements is lower by about $30\%$ to $80\%$ for structured systems, while comparable in magnitude to benchmarks for less-structured examples. These values have been proven consistent from the finite sample simulations. In the latter instance, a possible solution to improving Scenario I and II would be to pick heuristically motivated operations instead of random generated ones. Furthermore, similarly to the measurement groups observation, Scenario II was able to reach a larger improvement compared to Scenario I. This again advocates for a systematic advantage when co-designing circuit and rotation operations. Distributions are shown in Appendix \ref{num_of_meas_freeH6}.

Moreover, the considered systems have been proved to be well described by the graph-based approach introduced in \ref{Scenario I}. \cite{kottmann2022optimized, kottmannMolecular2023} This agreement validates the efficacy of the proposed method, indicating the underlying principles captured by the formalism are fundamental to the structural properties of these systems. Finally, it supports the hypothesis that the method can be systematically generalized to a broader range of systems, potentially extending its utility beyond the originally studied cases. Future heuristics could leverage modern correlation measures \cite{dingConceptOrbitalEntanglement2021, dingQuantumCorrelationsMolecules2022, dingQuantum2023} to capture key interactions and enhance Hamiltonian approximation, or make use of perturbative methods \cite{reascosUniversal2024, diotalleviSymPT2024} to narrow the choice of rotation operations with respect to expressivity and efficiency.

In conclusion, the proposed method represents a promising alternative to the current state-of-the-art techniques in measurement optimization for the VQE algorithm. Although further rigorous testing is required to fully assess its impact on comprehensive variational optimizations, the method shows potential in addressing the challenges associated with simulating physical systems on NISQ devices. By enhancing measurement efficiency, this approach could improve the feasibility and accuracy of quantum simulations, bringing us one step closer to solving larger and more intricate quantum systems — particularly mid- and large-size molecules.

\begin{figure*}[h]
  \centering
  \subfigure[Scenario I]{\includegraphics[width=0.4\textwidth]{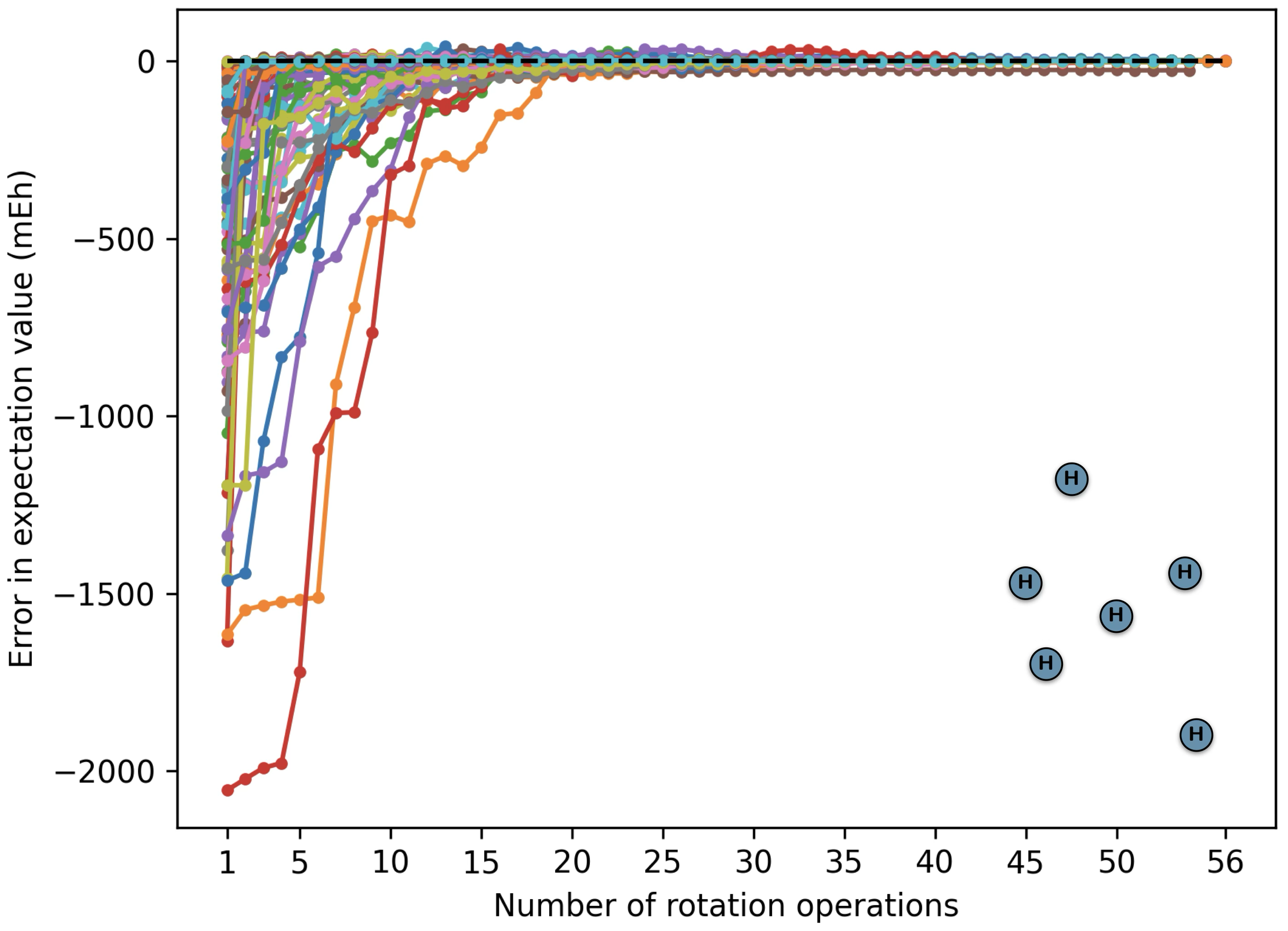}}
  \subfigure[Scenario II]{\includegraphics[width=0.4\textwidth]{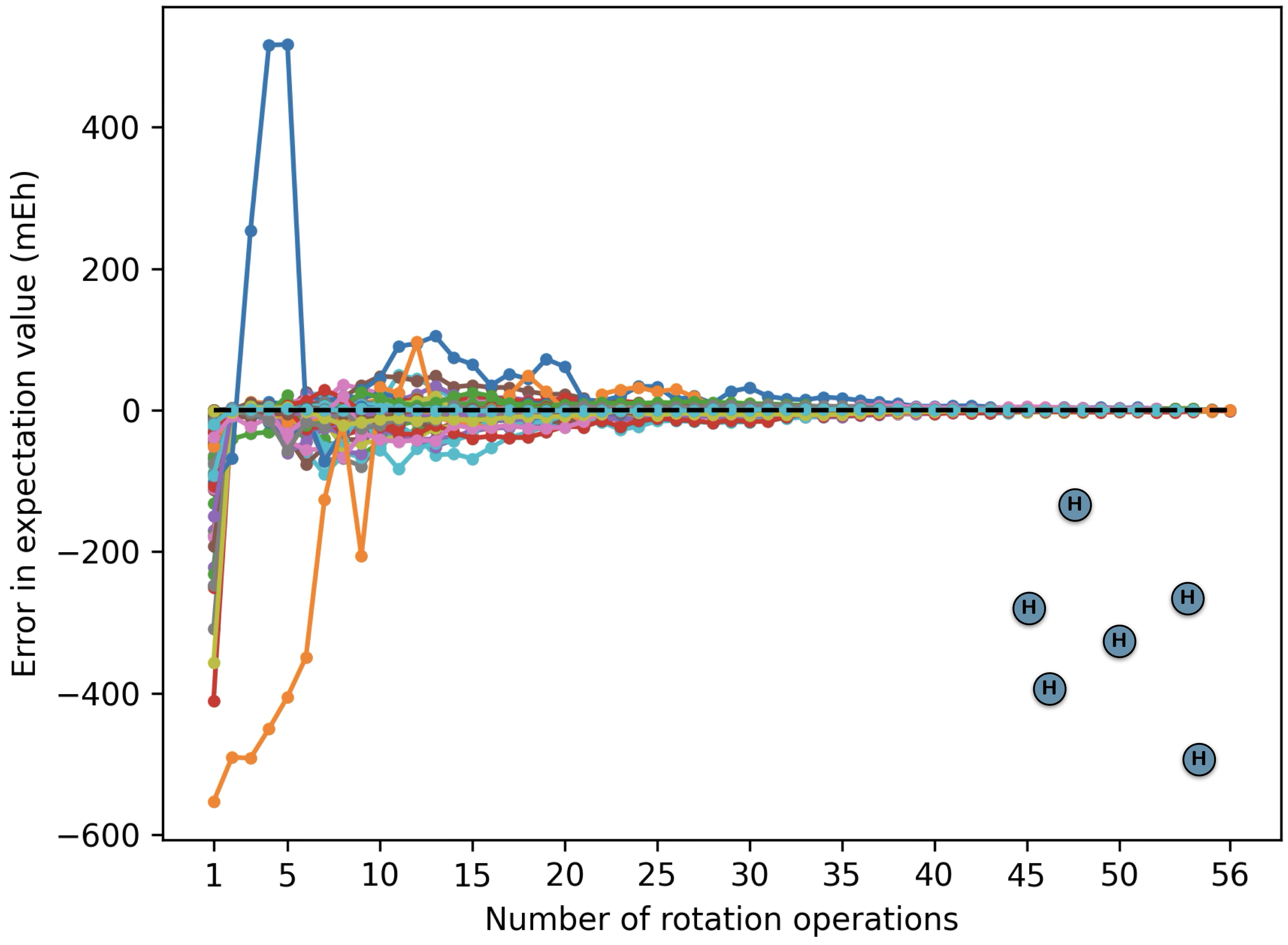}} \\
  \subfigure[Scenario I]{\includegraphics[width=0.4\textwidth]{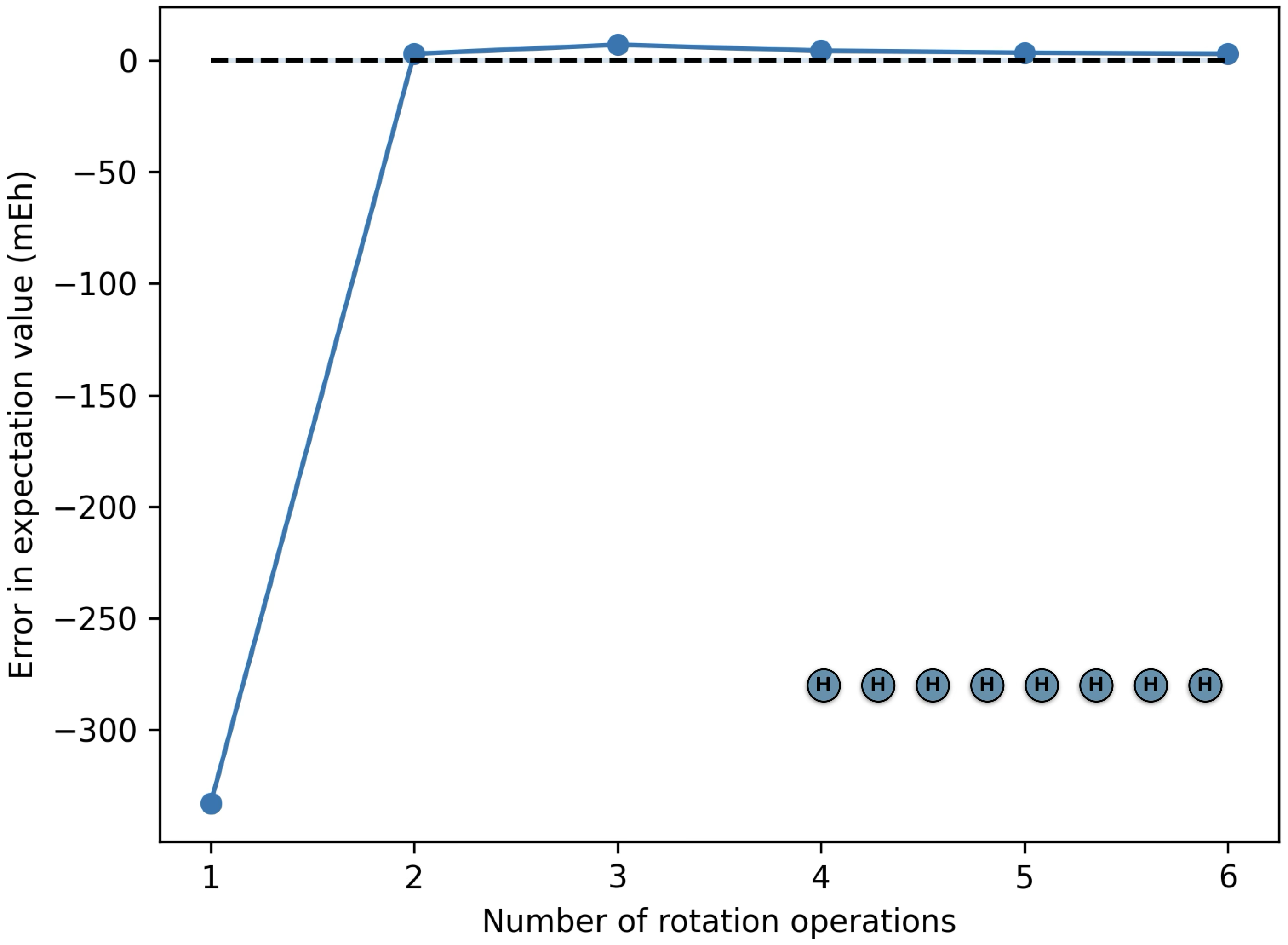}}
  \subfigure[Scenario II]{\includegraphics[width=0.4\textwidth]{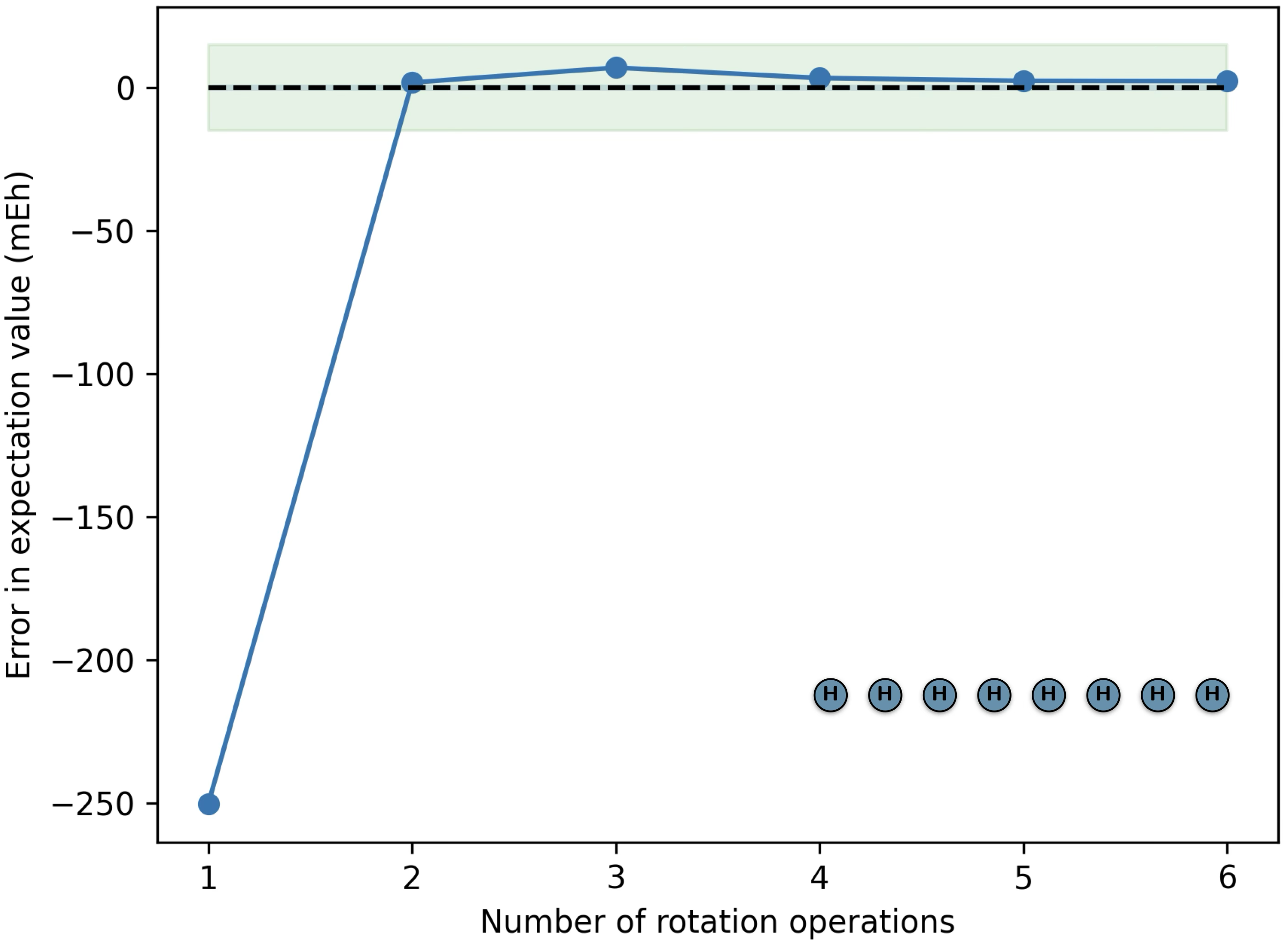}}
  \caption{(a-b)Error in approximating the molecular Hamiltonian for H$_6$ in 100 randomized geometries using the set $\{R_{G_1}, ..., R_{G_5}\}$ from Eq.\eqref{eq:h6-rotators} and 50 random generated unitary transformations. (c-d) Error in approximating the molecular Hamiltonian for linear H$_8$ using the set $\{R_{G_1}, ..., R_{G_6}\}$ from Eq.\eqref{eq:h8-rotators}. The blue area is the 1 $\mathrm{mE_h}$ margin of error, which we consider as the desired accuracy, the green area is the error from the chosen circuit ansatz. In Appendix \ref{Close-up visualization} we show a close-up visualization of the results around 1 $\mathrm{mE_h}$.} \label{Random_H6_result} \label{h8_result}
\end{figure*}

\begin{figure*}[h]
  \centering
  \subfigure[]{\includegraphics[width=0.45\textwidth]{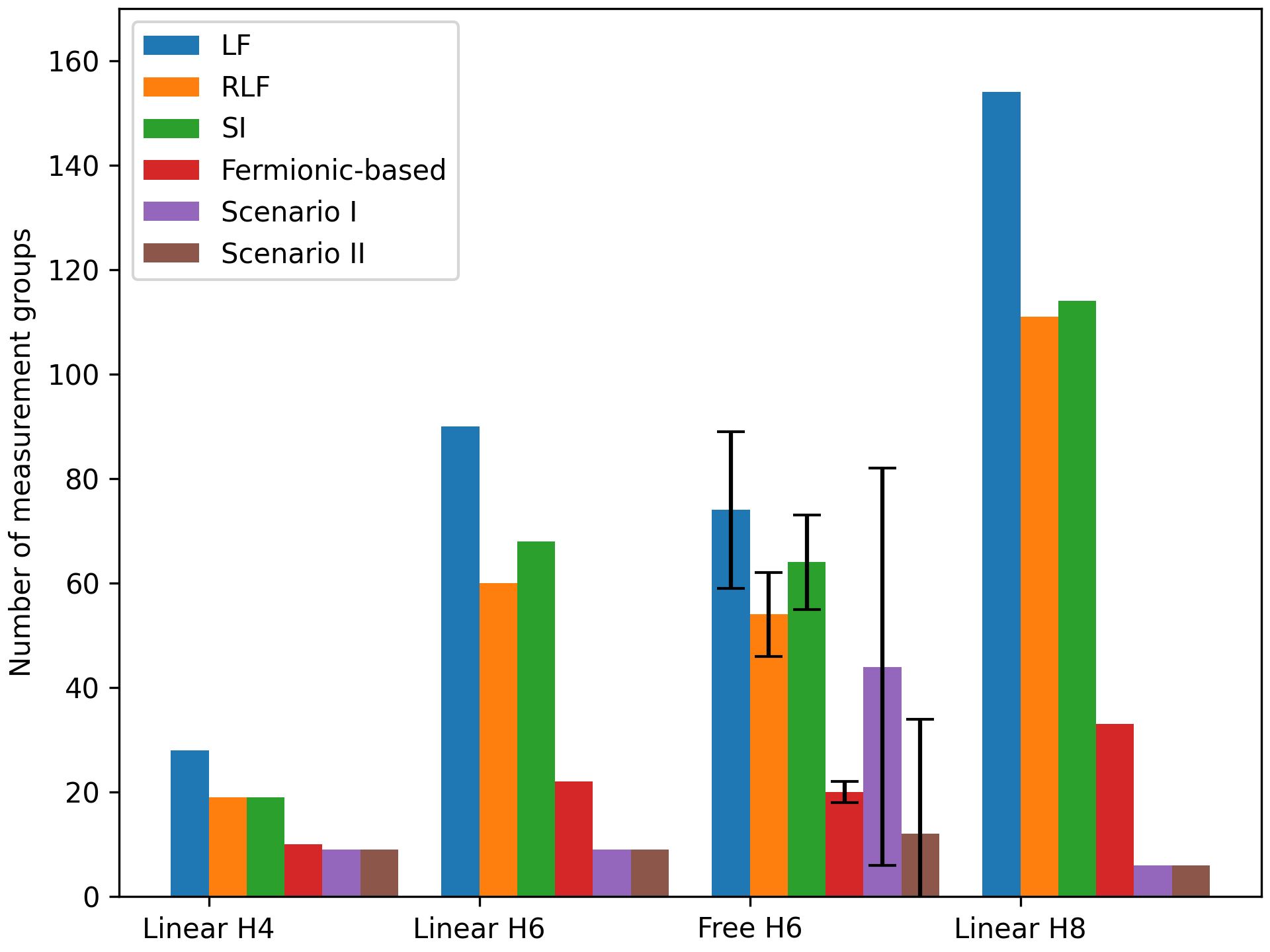} \label{Num_of_meas_groups}}
  \subfigure[]{\includegraphics[width=0.45\textwidth]{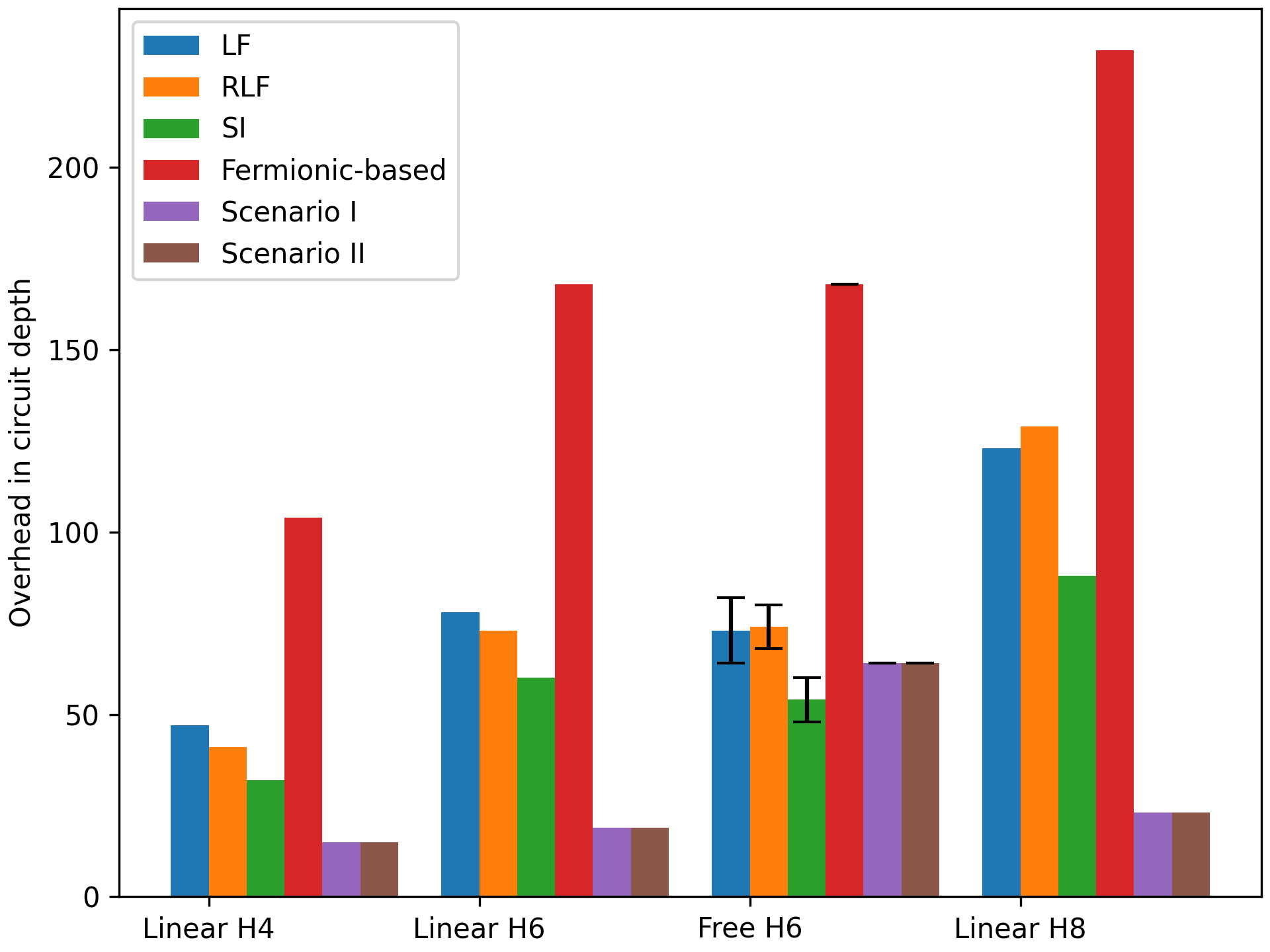} \label{Depth_overhead}}
  \subfigure[]{\includegraphics[width=0.5\textwidth]{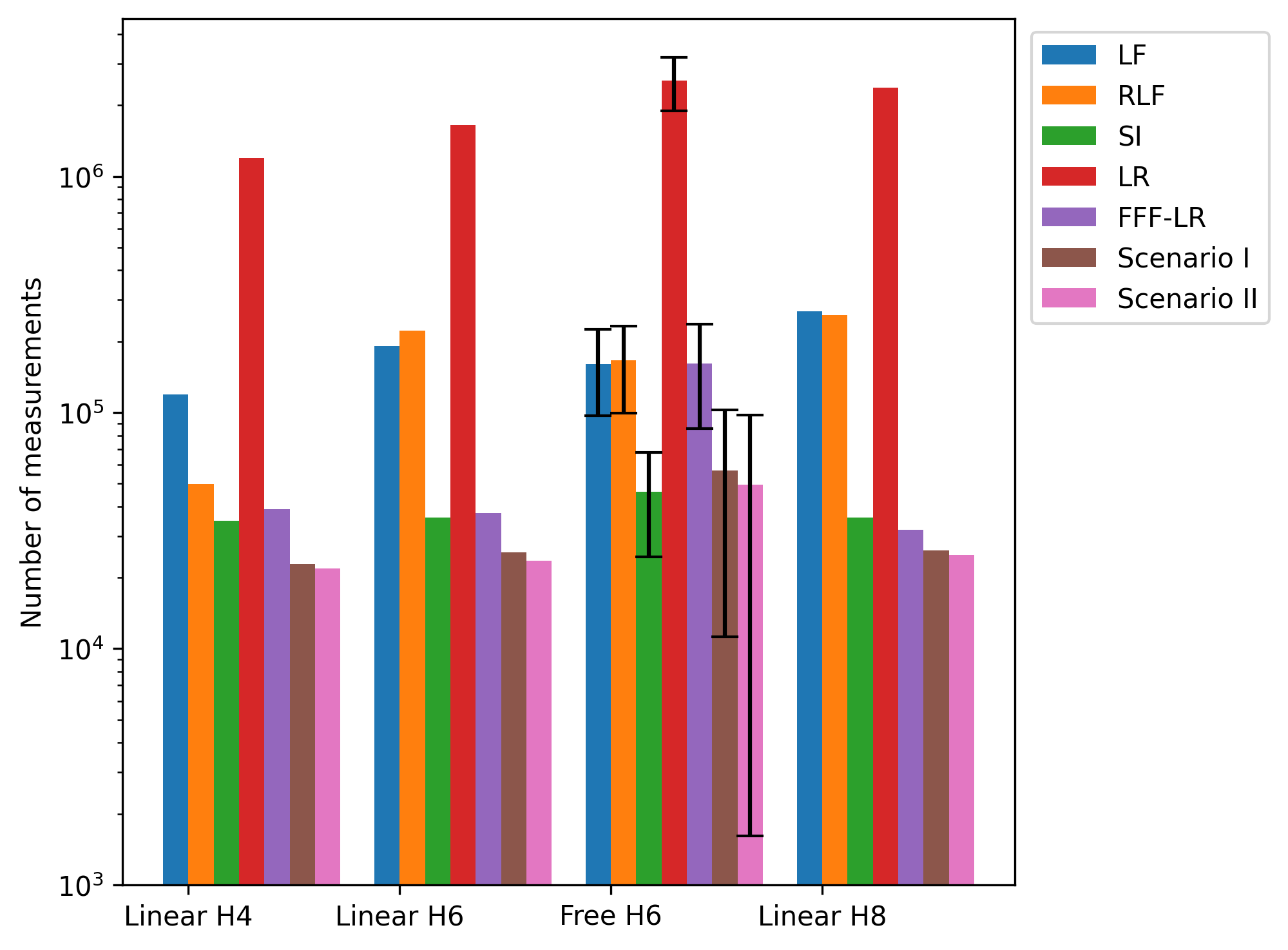} \label{Num_of_meas}}
  \caption{(a) Number of measurement groups needed for different reduction methods. Free H$_6$ refers to randomized molecular geometries, and the values correspond to the mean and standard deviation of the distribution that achieves an error below 2 $\mathrm{mE_h}$. (b) Overhead in circuit depth given by the application of rotation operation on wavefunction in step 4. Here we considered only linear system examples since results are compatible. Values of Scenario I and II are expressed in the Reordered Jordan-Wigner encoding. (c) Number of measurements required to compute all the contributions given by the general procedure, estimated as in Section \ref{Number of measurements section}. This represents the cost for each shot of a VQE algorithm. For acronyms see \hyperref[Glossary]{Glossary}.}
\end{figure*}

\clearpage

\bibliography{main}

\begin{thebibliography}{38}%
\makeatletter
\providecommand \@ifxundefined [1]{%
 \@ifx{#1\undefined}
}%
\providecommand \@ifnum [1]{%
 \ifnum #1\expandafter \@firstoftwo
 \else \expandafter \@secondoftwo
 \fi
}%
\providecommand \@ifx [1]{%
 \ifx #1\expandafter \@firstoftwo
 \else \expandafter \@secondoftwo
 \fi
}%
\providecommand \natexlab [1]{#1}%
\providecommand \enquote  [1]{``#1''}%
\providecommand \bibnamefont  [1]{#1}%
\providecommand \bibfnamefont [1]{#1}%
\providecommand \citenamefont [1]{#1}%
\providecommand \href@noop [0]{\@secondoftwo}%
\providecommand \href [0]{\begingroup \@sanitize@url \@href}%
\providecommand \@href[1]{\@@startlink{#1}\@@href}%
\providecommand \@@href[1]{\endgroup#1\@@endlink}%
\providecommand \@sanitize@url [0]{\catcode `\\12\catcode `\$12\catcode
  `\&12\catcode `\#12\catcode `\^12\catcode `\_12\catcode `\%12\relax}%
\providecommand \@@startlink[1]{}%
\providecommand \@@endlink[0]{}%
\providecommand \url  [0]{\begingroup\@sanitize@url \@url }%
\providecommand \@url [1]{\endgroup\@href {#1}{\urlprefix }}%
\providecommand \urlprefix  [0]{URL }%
\providecommand \Eprint [0]{\href }%
\providecommand \doibase [0]{https://doi.org/}%
\providecommand \selectlanguage [0]{\@gobble}%
\providecommand \bibinfo  [0]{\@secondoftwo}%
\providecommand \bibfield  [0]{\@secondoftwo}%
\providecommand \translation [1]{[#1]}%
\providecommand \BibitemOpen [0]{}%
\providecommand \bibitemStop [0]{}%
\providecommand \bibitemNoStop [0]{.\EOS\space}%
\providecommand \EOS [0]{\spacefactor3000\relax}%
\providecommand \BibitemShut  [1]{\csname bibitem#1\endcsname}%
\let\auto@bib@innerbib\@empty
\bibitem [{\citenamefont {Peruzzo}\ \emph {et~al.}(2014)\citenamefont
  {Peruzzo}, \citenamefont {McClean}, \citenamefont {Shadbolt}, \citenamefont
  {Yung}, \citenamefont {Zhou}, \citenamefont {Love}, \citenamefont
  {{Aspuru-Guzik}},\ and\ \citenamefont {O'brien}}]{peruzzo2014variational}%
  \BibitemOpen
  \bibfield  {author} {\bibinfo {author} {\bibfnamefont {A.}~\bibnamefont
  {Peruzzo}}, \bibinfo {author} {\bibfnamefont {J.}~\bibnamefont {McClean}},
  \bibinfo {author} {\bibfnamefont {P.}~\bibnamefont {Shadbolt}}, \bibinfo
  {author} {\bibfnamefont {M.-H.}\ \bibnamefont {Yung}}, \bibinfo {author}
  {\bibfnamefont {X.-Q.}\ \bibnamefont {Zhou}}, \bibinfo {author}
  {\bibfnamefont {P.~J.}\ \bibnamefont {Love}}, \bibinfo {author}
  {\bibfnamefont {A.}~\bibnamefont {{Aspuru-Guzik}}},\ and\ \bibinfo {author}
  {\bibfnamefont {J.~L.}\ \bibnamefont {O'brien}},\ }\bibfield  {title}
  {\bibinfo {title} {A variational eigenvalue solver on a photonic quantum
  processor},\ }\href {https://doi.org/10.1038/ncomms5213} {\bibfield
  {journal} {\bibinfo  {journal} {Nature Communications}\ }\textbf {\bibinfo
  {volume} {5}},\ \bibinfo {pages} {4213} (\bibinfo {year} {2014})}\BibitemShut
  {NoStop}%
\bibitem [{\citenamefont {Tilly}\ \emph {et~al.}(2022)\citenamefont {Tilly},
  \citenamefont {Chen}, \citenamefont {Cao}, \citenamefont {Picozzi},
  \citenamefont {Setia}, \citenamefont {Li}, \citenamefont {Grant},
  \citenamefont {Wossnig}, \citenamefont {Rungger}, \citenamefont {Booth},\
  and\ \citenamefont {Tennyson}}]{tillyVariationalQuantumEigensolver2022}%
  \BibitemOpen
  \bibfield  {author} {\bibinfo {author} {\bibfnamefont {J.}~\bibnamefont
  {Tilly}}, \bibinfo {author} {\bibfnamefont {H.}~\bibnamefont {Chen}},
  \bibinfo {author} {\bibfnamefont {S.}~\bibnamefont {Cao}}, \bibinfo {author}
  {\bibfnamefont {D.}~\bibnamefont {Picozzi}}, \bibinfo {author} {\bibfnamefont
  {K.}~\bibnamefont {Setia}}, \bibinfo {author} {\bibfnamefont
  {Y.}~\bibnamefont {Li}}, \bibinfo {author} {\bibfnamefont {E.}~\bibnamefont
  {Grant}}, \bibinfo {author} {\bibfnamefont {L.}~\bibnamefont {Wossnig}},
  \bibinfo {author} {\bibfnamefont {I.}~\bibnamefont {Rungger}}, \bibinfo
  {author} {\bibfnamefont {G.~H.}\ \bibnamefont {Booth}},\ and\ \bibinfo
  {author} {\bibfnamefont {J.}~\bibnamefont {Tennyson}},\ }\bibfield  {title}
  {\bibinfo {title} {The {{Variational Quantum Eigensolver}}: {{A}} review of
  methods and best practices},\ }\href
  {https://doi.org/10.1016/j.physrep.2022.08.003} {\bibfield  {journal}
  {\bibinfo  {journal} {Physics Reports}\ }\bibinfo {series} {The {{Variational
  Quantum Eigensolver}}: A Review of Methods and Best Practices},\ \textbf
  {\bibinfo {volume} {986}},\ \bibinfo {pages} {1} (\bibinfo {year}
  {2022})}\BibitemShut {NoStop}%
\bibitem [{\citenamefont {{Aspuru-Guzik}}\ \emph {et~al.}(2005)\citenamefont
  {{Aspuru-Guzik}}, \citenamefont {Dutoi}, \citenamefont {Love},\ and\
  \citenamefont {{Head-Gordon}}}]{aspuru2005simulated}%
  \BibitemOpen
  \bibfield  {author} {\bibinfo {author} {\bibfnamefont {A.}~\bibnamefont
  {{Aspuru-Guzik}}}, \bibinfo {author} {\bibfnamefont {A.~D.}\ \bibnamefont
  {Dutoi}}, \bibinfo {author} {\bibfnamefont {P.~J.}\ \bibnamefont {Love}},\
  and\ \bibinfo {author} {\bibfnamefont {M.}~\bibnamefont {{Head-Gordon}}},\
  }\bibfield  {title} {\bibinfo {title} {Simulated quantum computation of
  molecular energies},\ }\href {https://doi.org/10.1126/science.1113479}
  {\bibfield  {journal} {\bibinfo  {journal} {Science (New York, N.Y.)}\
  }\textbf {\bibinfo {volume} {309}},\ \bibinfo {pages} {1704} (\bibinfo {year}
  {2005})}\BibitemShut {NoStop}%
\bibitem [{\citenamefont {Cortes}\ \emph {et~al.}()\citenamefont {Cortes},
  \citenamefont {Rocca}, \citenamefont {Gonthier}, \citenamefont {Ollitrault},
  \citenamefont {Parrish}, \citenamefont {Anselmetti}, \citenamefont
  {Degroote}, \citenamefont {Moll}, \citenamefont {Santagati},\ and\
  \citenamefont {Streif}}]{cortesAssessing2024}%
  \BibitemOpen
  \bibfield  {author} {\bibinfo {author} {\bibfnamefont {C.~L.}\ \bibnamefont
  {Cortes}}, \bibinfo {author} {\bibfnamefont {D.}~\bibnamefont {Rocca}},
  \bibinfo {author} {\bibfnamefont {J.}~\bibnamefont {Gonthier}}, \bibinfo
  {author} {\bibfnamefont {P.~J.}\ \bibnamefont {Ollitrault}}, \bibinfo
  {author} {\bibfnamefont {R.~M.}\ \bibnamefont {Parrish}}, \bibinfo {author}
  {\bibfnamefont {G.-L.~R.}\ \bibnamefont {Anselmetti}}, \bibinfo {author}
  {\bibfnamefont {M.}~\bibnamefont {Degroote}}, \bibinfo {author}
  {\bibfnamefont {N.}~\bibnamefont {Moll}}, \bibinfo {author} {\bibfnamefont
  {R.}~\bibnamefont {Santagati}},\ and\ \bibinfo {author} {\bibfnamefont
  {M.}~\bibnamefont {Streif}},\ }\bibfield  {title} {\bibinfo {title}
  {Assessing the query complexity limits of quantum phase estimation using
  symmetry aware spectral bounds},\ }\href {http://arxiv.org/abs/2403.04737} {\
  }\Eprint {https://arxiv.org/abs/2403.04737} {2403.04737} \BibitemShut
  {NoStop}%
\bibitem [{\citenamefont {Korhonen}\ \emph {et~al.}()\citenamefont {Korhonen},
  \citenamefont {Vappula}, \citenamefont {Glos}, \citenamefont {Cattaneo},
  \citenamefont {Zimborás}, \citenamefont {Borrelli}, \citenamefont {Rossi},
  \citenamefont {García-Pérez},\ and\ \citenamefont
  {Cavalcanti}}]{korhonenPractical2024}%
  \BibitemOpen
  \bibfield  {author} {\bibinfo {author} {\bibfnamefont {K.}~\bibnamefont
  {Korhonen}}, \bibinfo {author} {\bibfnamefont {H.}~\bibnamefont {Vappula}},
  \bibinfo {author} {\bibfnamefont {A.}~\bibnamefont {Glos}}, \bibinfo {author}
  {\bibfnamefont {M.}~\bibnamefont {Cattaneo}}, \bibinfo {author}
  {\bibfnamefont {Z.}~\bibnamefont {Zimborás}}, \bibinfo {author}
  {\bibfnamefont {E.-M.}\ \bibnamefont {Borrelli}}, \bibinfo {author}
  {\bibfnamefont {M.~A.~C.}\ \bibnamefont {Rossi}}, \bibinfo {author}
  {\bibfnamefont {G.}~\bibnamefont {García-Pérez}},\ and\ \bibinfo {author}
  {\bibfnamefont {D.}~\bibnamefont {Cavalcanti}},\ }\bibfield  {title}
  {\bibinfo {title} {Practical techniques for high precision measurements on
  near-term quantum hardware: A {{Case Study}} in {{Molecular Energy
  Estimation}}}\ }\href {https://doi.org/10.48550/arXiv.2409.02575}
  {10.48550/arXiv.2409.02575},\ \Eprint {https://arxiv.org/abs/2409.02575}
  {2409.02575} \BibitemShut {NoStop}%
\bibitem [{\citenamefont {Gonthier}\ \emph {et~al.}()\citenamefont {Gonthier},
  \citenamefont {Radin}, \citenamefont {Buda}, \citenamefont {Doskocil},
  \citenamefont {Abuan},\ and\ \citenamefont
  {Romero}}]{gonthierMeasurementsRoadblockTerm2022}%
  \BibitemOpen
  \bibfield  {author} {\bibinfo {author} {\bibfnamefont {J.~F.}\ \bibnamefont
  {Gonthier}}, \bibinfo {author} {\bibfnamefont {M.~D.}\ \bibnamefont {Radin}},
  \bibinfo {author} {\bibfnamefont {C.}~\bibnamefont {Buda}}, \bibinfo {author}
  {\bibfnamefont {E.~J.}\ \bibnamefont {Doskocil}}, \bibinfo {author}
  {\bibfnamefont {C.~M.}\ \bibnamefont {Abuan}},\ and\ \bibinfo {author}
  {\bibfnamefont {J.}~\bibnamefont {Romero}},\ }\bibfield  {title} {\bibinfo
  {title} {Measurements as a roadblock to near-term practical quantum advantage
  in chemistry: {{Resource}} analysis},\ }\href
  {https://doi.org/10.1103/PhysRevResearch.4.033154} {\bibfield  {journal}
  {\bibinfo  {journal} {Physical Review Research}\ }\textbf {\bibinfo {volume}
  {4}},\ \bibinfo {pages} {033154}}\BibitemShut {NoStop}%
\bibitem [{\citenamefont {Jordan}\ and\ \citenamefont
  {Wigner}(1928)}]{JordanWigner1928}%
  \BibitemOpen
  \bibfield  {author} {\bibinfo {author} {\bibfnamefont {P.}~\bibnamefont
  {Jordan}}\ and\ \bibinfo {author} {\bibfnamefont {E.}~\bibnamefont
  {Wigner}},\ }\bibfield  {title} {\bibinfo {title} {Über das paulische
  Äquivalenzverbot},\ }\href {https://doi.org/10.1007/BF01331938} {\bibfield
  {journal} {\bibinfo  {journal} {Zeitschrift für Physik}\ }\textbf {\bibinfo
  {volume} {47}},\ \bibinfo {pages} {631} (\bibinfo {year} {1928})}\BibitemShut
  {NoStop}%
\bibitem [{\citenamefont {Tranter}\ \emph {et~al.}(2018)\citenamefont
  {Tranter}, \citenamefont {Love}, \citenamefont {Mintert},\ and\ \citenamefont
  {Coveney}}]{tranter2018comparison}%
  \BibitemOpen
  \bibfield  {author} {\bibinfo {author} {\bibfnamefont {A.}~\bibnamefont
  {Tranter}}, \bibinfo {author} {\bibfnamefont {P.~J.}\ \bibnamefont {Love}},
  \bibinfo {author} {\bibfnamefont {F.}~\bibnamefont {Mintert}},\ and\ \bibinfo
  {author} {\bibfnamefont {P.~V.}\ \bibnamefont {Coveney}},\ }\bibfield
  {title} {\bibinfo {title} {A comparison of the
  {{Bravyi}}\textendash{{Kitaev}} and {{Jordan}}\textendash{{Wigner}}
  transformations for the quantum simulation of quantum chemistry},\ }\href
  {https://doi.org/10.1021/acs.jctc.8b00450} {\bibfield  {journal} {\bibinfo
  {journal} {Journal of Chemical Theory and Computation}\ }\textbf {\bibinfo
  {volume} {14}},\ \bibinfo {pages} {5617} (\bibinfo {year}
  {2018})}\BibitemShut {NoStop}%
\bibitem [{\citenamefont {Verteletskyi}\ \emph {et~al.}(2020)\citenamefont
  {Verteletskyi}, \citenamefont {Yen},\ and\ \citenamefont
  {Izmaylov}}]{verteletskyi2020measurement}%
  \BibitemOpen
  \bibfield  {author} {\bibinfo {author} {\bibfnamefont {V.}~\bibnamefont
  {Verteletskyi}}, \bibinfo {author} {\bibfnamefont {T.-C.}\ \bibnamefont
  {Yen}},\ and\ \bibinfo {author} {\bibfnamefont {A.~F.}\ \bibnamefont
  {Izmaylov}},\ }\bibfield  {title} {\bibinfo {title} {Measurement optimization
  in the variational quantum eigensolver using a minimum clique cover},\ }\href
  {https://doi.org/10.1063/1.5141458} {\bibfield  {journal} {\bibinfo
  {journal} {Journal of Chemical Physics}\ }\textbf {\bibinfo {volume} {152}},\
  \bibinfo {pages} {124114} (\bibinfo {year} {2020})}\BibitemShut {NoStop}%
\bibitem [{\citenamefont {Yen}\ \emph {et~al.}(2020)\citenamefont {Yen},
  \citenamefont {Verteletskyi},\ and\ \citenamefont
  {Izmaylov}}]{yen2020measuring}%
  \BibitemOpen
  \bibfield  {author} {\bibinfo {author} {\bibfnamefont {T.-C.}\ \bibnamefont
  {Yen}}, \bibinfo {author} {\bibfnamefont {V.}~\bibnamefont {Verteletskyi}},\
  and\ \bibinfo {author} {\bibfnamefont {A.~F.}\ \bibnamefont {Izmaylov}},\
  }\bibfield  {title} {\bibinfo {title} {Measuring all compatible operators in
  one series of single-qubit measurements using unitary transformations},\
  }\href {https://doi.org/10.1021/acs.jctc.0c00008} {\bibfield  {journal}
  {\bibinfo  {journal} {Journal of Chemical Theory and Computation}\ }\textbf
  {\bibinfo {volume} {16}},\ \bibinfo {pages} {2400} (\bibinfo {year}
  {2020})}\BibitemShut {NoStop}%
\bibitem [{\citenamefont {Crawford}\ \emph {et~al.}(2019)\citenamefont
  {Crawford}, \citenamefont {{van Straaten}}, \citenamefont {Wang},
  \citenamefont {Parks}, \citenamefont {Campbell},\ and\ \citenamefont
  {Brierley}}]{crawford2019efficient}%
  \BibitemOpen
  \bibfield  {author} {\bibinfo {author} {\bibfnamefont {O.}~\bibnamefont
  {Crawford}}, \bibinfo {author} {\bibfnamefont {B.}~\bibnamefont {{van
  Straaten}}}, \bibinfo {author} {\bibfnamefont {D.}~\bibnamefont {Wang}},
  \bibinfo {author} {\bibfnamefont {T.}~\bibnamefont {Parks}}, \bibinfo
  {author} {\bibfnamefont {E.}~\bibnamefont {Campbell}},\ and\ \bibinfo
  {author} {\bibfnamefont {S.}~\bibnamefont {Brierley}},\ }\bibfield  {title}
  {\bibinfo {title} {Efficient quantum measurement of {{Pauli}} operators in
  the presence of finite sampling error},\ }\href
  {https://arxiv.org/abs/1908.06942} {\bibfield  {journal} {\bibinfo  {journal}
  {arxiv:1908.06942}\ } (\bibinfo {year} {2019})},\ \Eprint
  {https://arxiv.org/abs/1908.06942} {arxiv:1908.06942} \BibitemShut {NoStop}%
\bibitem [{\citenamefont {Bansingh}\ \emph {et~al.}()\citenamefont {Bansingh},
  \citenamefont {Yen}, \citenamefont {Johnson},\ and\ \citenamefont
  {Izmaylov}}]{bansinghFidelity2022}%
  \BibitemOpen
  \bibfield  {author} {\bibinfo {author} {\bibfnamefont {Z.~P.}\ \bibnamefont
  {Bansingh}}, \bibinfo {author} {\bibfnamefont {T.-C.}\ \bibnamefont {Yen}},
  \bibinfo {author} {\bibfnamefont {P.~D.}\ \bibnamefont {Johnson}},\ and\
  \bibinfo {author} {\bibfnamefont {A.~F.}\ \bibnamefont {Izmaylov}},\
  }\bibfield  {title} {\bibinfo {title} {Fidelity overhead for non-local
  measurements in variational quantum algorithms}\ }\href
  {https://doi.org/10.48550/arXiv.2205.07113} {10.48550/arXiv.2205.07113},\
  \Eprint {https://arxiv.org/abs/2205.07113} {2205.07113} \BibitemShut
  {NoStop}%
\bibitem [{\citenamefont {Yen}\ \emph {et~al.}()\citenamefont {Yen},
  \citenamefont {Ganeshram},\ and\ \citenamefont
  {Izmaylov}}]{yenDeterministic2023}%
  \BibitemOpen
  \bibfield  {author} {\bibinfo {author} {\bibfnamefont {T.-C.}\ \bibnamefont
  {Yen}}, \bibinfo {author} {\bibfnamefont {A.}~\bibnamefont {Ganeshram}},\
  and\ \bibinfo {author} {\bibfnamefont {A.~F.}\ \bibnamefont {Izmaylov}},\
  }\bibfield  {title} {\bibinfo {title} {Deterministic improvements of quantum
  measurements with grouping of compatible operators, non-local
  transformations, and covariance estimates},\ }\href
  {https://doi.org/10.1038/s41534-023-00683-y} {\bibfield  {journal} {\bibinfo
  {journal} {npj Quantum Information}\ }\textbf {\bibinfo {volume} {9}},\
  \bibinfo {pages} {1}}\BibitemShut {NoStop}%
\bibitem [{\citenamefont {Huggins}\ \emph {et~al.}()\citenamefont {Huggins},
  \citenamefont {McClean}, \citenamefont {Rubin}, \citenamefont {Jiang},
  \citenamefont {Wiebe}, \citenamefont {Whaley},\ and\ \citenamefont
  {Babbush}}]{hugginsEfficient2021}%
  \BibitemOpen
  \bibfield  {author} {\bibinfo {author} {\bibfnamefont {W.~J.}\ \bibnamefont
  {Huggins}}, \bibinfo {author} {\bibfnamefont {J.~R.}\ \bibnamefont
  {McClean}}, \bibinfo {author} {\bibfnamefont {N.~C.}\ \bibnamefont {Rubin}},
  \bibinfo {author} {\bibfnamefont {Z.}~\bibnamefont {Jiang}}, \bibinfo
  {author} {\bibfnamefont {N.}~\bibnamefont {Wiebe}}, \bibinfo {author}
  {\bibfnamefont {K.~B.}\ \bibnamefont {Whaley}},\ and\ \bibinfo {author}
  {\bibfnamefont {R.}~\bibnamefont {Babbush}},\ }\bibfield  {title} {\bibinfo
  {title} {Efficient and noise resilient measurements for quantum chemistry on
  near-term quantum computers},\ }\href
  {https://doi.org/10.1038/s41534-020-00341-7} {\bibfield  {journal} {\bibinfo
  {journal} {npj Quantum Information}\ }\textbf {\bibinfo {volume} {7}},\
  \bibinfo {pages} {1}}\BibitemShut {NoStop}%
\bibitem [{\citenamefont {Yen}\ and\ \citenamefont
  {Izmaylov}()}]{yenCartan2021}%
  \BibitemOpen
  \bibfield  {author} {\bibinfo {author} {\bibfnamefont {T.-C.}\ \bibnamefont
  {Yen}}\ and\ \bibinfo {author} {\bibfnamefont {A.~F.}\ \bibnamefont
  {Izmaylov}},\ }\bibfield  {title} {\bibinfo {title} {Cartan {{Subalgebra
  Approach}} to {{Efficient Measurements}} of {{Quantum Observables}}},\ }\href
  {https://doi.org/10.1103/PRXQuantum.2.040320} {\bibfield  {journal} {\bibinfo
   {journal} {PRX Quantum}\ }\textbf {\bibinfo {volume} {2}},\ \bibinfo {pages}
  {040320}}\BibitemShut {NoStop}%
\bibitem [{\citenamefont {Choi}\ \emph {et~al.}()\citenamefont {Choi},
  \citenamefont {Loaiza},\ and\ \citenamefont {Izmaylov}}]{choiFluid2023}%
  \BibitemOpen
  \bibfield  {author} {\bibinfo {author} {\bibfnamefont {S.}~\bibnamefont
  {Choi}}, \bibinfo {author} {\bibfnamefont {I.}~\bibnamefont {Loaiza}},\ and\
  \bibinfo {author} {\bibfnamefont {A.~F.}\ \bibnamefont {Izmaylov}},\
  }\bibfield  {title} {\bibinfo {title} {Fluid fermionic fragments for
  optimizing quantum measurements of electronic {{Hamiltonians}} in the
  variational quantum eigensolver},\ }\href
  {https://doi.org/10.22331/q-2023-01-03-889} {\bibfield  {journal} {\bibinfo
  {journal} {Quantum}\ }\textbf {\bibinfo {volume} {7}},\ \bibinfo {pages}
  {889}},\ \Eprint {https://arxiv.org/abs/2208.14490} {2208.14490} \BibitemShut
  {NoStop}%
\bibitem [{\citenamefont {Gokhale}\ \emph {et~al.}(2019)\citenamefont
  {Gokhale}, \citenamefont {Angiuli}, \citenamefont {Ding}, \citenamefont
  {Gui}, \citenamefont {Tomesh}, \citenamefont {Suchara}, \citenamefont
  {Martonosi},\ and\ \citenamefont {Chong}}]{gokhale2019minimizing}%
  \BibitemOpen
  \bibfield  {author} {\bibinfo {author} {\bibfnamefont {P.}~\bibnamefont
  {Gokhale}}, \bibinfo {author} {\bibfnamefont {O.}~\bibnamefont {Angiuli}},
  \bibinfo {author} {\bibfnamefont {Y.}~\bibnamefont {Ding}}, \bibinfo {author}
  {\bibfnamefont {K.}~\bibnamefont {Gui}}, \bibinfo {author} {\bibfnamefont
  {T.}~\bibnamefont {Tomesh}}, \bibinfo {author} {\bibfnamefont
  {M.}~\bibnamefont {Suchara}}, \bibinfo {author} {\bibfnamefont
  {M.}~\bibnamefont {Martonosi}},\ and\ \bibinfo {author} {\bibfnamefont
  {F.~T.}\ \bibnamefont {Chong}},\ }\bibfield  {title} {\bibinfo {title}
  {Minimizing state preparations in variational quantum eigensolver by
  partitioning into commuting families},\ }\href
  {https://arxiv.org/abs/1907.13623} {\bibfield  {journal} {\bibinfo  {journal}
  {arxiv:1907.13623}\ } (\bibinfo {year} {2019})},\ \Eprint
  {https://arxiv.org/abs/1907.13623} {arxiv:1907.13623} \BibitemShut {NoStop}%
\bibitem [{\citenamefont {Gokhale}\ and\ \citenamefont
  {Chong}(2019)}]{gokhale2019n}%
  \BibitemOpen
  \bibfield  {author} {\bibinfo {author} {\bibfnamefont {P.}~\bibnamefont
  {Gokhale}}\ and\ \bibinfo {author} {\bibfnamefont {F.~T.}\ \bibnamefont
  {Chong}},\ }\bibfield  {title} {\bibinfo {title} {O({{N}}{$^3$}) measurement
  cost for variational quantum eigensolver on molecular hamiltonians},\ }\href
  {https://arxiv.org/abs/1908.11857} {\bibfield  {journal} {\bibinfo  {journal}
  {arxiv:1908.11857}\ } (\bibinfo {year} {2019})},\ \Eprint
  {https://arxiv.org/abs/1908.11857} {arxiv:1908.11857} \BibitemShut {NoStop}%
\bibitem [{\citenamefont {Miller}\ \emph {et~al.}()\citenamefont {Miller},
  \citenamefont {Fischer}, \citenamefont {Levi}, \citenamefont {Kuehnke},
  \citenamefont {Sokolov}, \citenamefont {Barkoutsos}, \citenamefont {Eisert},\
  and\ \citenamefont
  {Tavernelli}}]{millerHardwareTailoredDiagonalizationCircuits2024}%
  \BibitemOpen
  \bibfield  {author} {\bibinfo {author} {\bibfnamefont {D.}~\bibnamefont
  {Miller}}, \bibinfo {author} {\bibfnamefont {L.~E.}\ \bibnamefont {Fischer}},
  \bibinfo {author} {\bibfnamefont {K.}~\bibnamefont {Levi}}, \bibinfo {author}
  {\bibfnamefont {E.~J.}\ \bibnamefont {Kuehnke}}, \bibinfo {author}
  {\bibfnamefont {I.~O.}\ \bibnamefont {Sokolov}}, \bibinfo {author}
  {\bibfnamefont {P.~K.}\ \bibnamefont {Barkoutsos}}, \bibinfo {author}
  {\bibfnamefont {J.}~\bibnamefont {Eisert}},\ and\ \bibinfo {author}
  {\bibfnamefont {I.}~\bibnamefont {Tavernelli}},\ }\bibfield  {title}
  {\bibinfo {title} {Hardware-{{Tailored Diagonalization Circuits}}}\ }\href
  {https://doi.org/10.48550/arXiv.2203.03646} {10.48550/arXiv.2203.03646},\
  \Eprint {https://arxiv.org/abs/2203.03646} {2203.03646} \BibitemShut
  {NoStop}%
\bibitem [{\citenamefont {Elfving}\ \emph {et~al.}(2021)\citenamefont
  {Elfving}, \citenamefont {Millaruelo}, \citenamefont {G{\'a}mez},\ and\
  \citenamefont {Gogolin}}]{elfving2021simulating}%
  \BibitemOpen
  \bibfield  {author} {\bibinfo {author} {\bibfnamefont {V.~E.}\ \bibnamefont
  {Elfving}}, \bibinfo {author} {\bibfnamefont {M.}~\bibnamefont {Millaruelo}},
  \bibinfo {author} {\bibfnamefont {J.~A.}\ \bibnamefont {G{\'a}mez}},\ and\
  \bibinfo {author} {\bibfnamefont {C.}~\bibnamefont {Gogolin}},\ }\bibfield
  {title} {\bibinfo {title} {Simulating quantum chemistry in the seniority-zero
  space on qubit-based quantum computers},\ }\href
  {https://doi.org/10.1103/PhysRevA.103.032605} {\bibfield  {journal} {\bibinfo
   {journal} {Physical Review A: Atomic, Molecular, and Optical Physics}\
  }\textbf {\bibinfo {volume} {103}},\ \bibinfo {pages} {032605} (\bibinfo
  {year} {2021})}\BibitemShut {NoStop}%
\bibitem [{\citenamefont {Santos}\ and\ \citenamefont
  {Kottmann}()}]{santosHybrid2024}%
  \BibitemOpen
  \bibfield  {author} {\bibinfo {author} {\bibfnamefont {F.~J. D.~A.}\
  \bibnamefont {Santos}}\ and\ \bibinfo {author} {\bibfnamefont {J.~S.}\
  \bibnamefont {Kottmann}},\ }\bibfield  {title} {\bibinfo {title} {A {{Hybrid
  Qubit Encoding}}: {{Splitting Fock Space}} into {{Fermionic}} and {{Bosonic
  Subspaces}}}\ }\href {https://doi.org/10.48550/arXiv.2411.14096}
  {10.48550/arXiv.2411.14096},\ \Eprint {https://arxiv.org/abs/2411.14096}
  {2411.14096} \BibitemShut {NoStop}%
\bibitem [{\citenamefont {Kivlichan}\ \emph {et~al.}(2018)\citenamefont
  {Kivlichan}, \citenamefont {McClean}, \citenamefont {Wiebe}, \citenamefont
  {Gidney}, \citenamefont {{Aspuru-Guzik}}, \citenamefont {Chan},\ and\
  \citenamefont {Babbush}}]{kivlichan2018quantum}%
  \BibitemOpen
  \bibfield  {author} {\bibinfo {author} {\bibfnamefont {I.~D.}\ \bibnamefont
  {Kivlichan}}, \bibinfo {author} {\bibfnamefont {J.}~\bibnamefont {McClean}},
  \bibinfo {author} {\bibfnamefont {N.}~\bibnamefont {Wiebe}}, \bibinfo
  {author} {\bibfnamefont {C.}~\bibnamefont {Gidney}}, \bibinfo {author}
  {\bibfnamefont {A.}~\bibnamefont {{Aspuru-Guzik}}}, \bibinfo {author}
  {\bibfnamefont {G.~K.-L.}\ \bibnamefont {Chan}},\ and\ \bibinfo {author}
  {\bibfnamefont {R.}~\bibnamefont {Babbush}},\ }\bibfield  {title} {\bibinfo
  {title} {Quantum simulation of electronic structure with linear depth and
  connectivity},\ }\href {https://doi.org/10.1103/PhysRevLett.120.110501}
  {\bibfield  {journal} {\bibinfo  {journal} {Physical Review Letters}\
  }\textbf {\bibinfo {volume} {120}},\ \bibinfo {pages} {110501} (\bibinfo
  {year} {2018})}\BibitemShut {NoStop}%
\bibitem [{\citenamefont {{Google AI Quantum and
  Collaborators}}(2020)}]{google2020hartree}%
  \BibitemOpen
  \bibfield  {author} {\bibinfo {author} {\bibnamefont {{Google AI Quantum and
  Collaborators}}},\ }\bibfield  {title} {\bibinfo {title} {Hartree-{{Fock}} on
  a superconducting qubit quantum computer},\ }\href
  {https://science.sciencemag.org/content/369/6507/1084} {\bibfield  {journal}
  {\bibinfo  {journal} {Science (New York, N.Y.)}\ }\textbf {\bibinfo {volume}
  {369}},\ \bibinfo {pages} {1084} (\bibinfo {year} {2020})}\BibitemShut
  {NoStop}%
\bibitem [{\citenamefont {Kottmann}()}]{kottmannMolecular2023}%
  \BibitemOpen
  \bibfield  {author} {\bibinfo {author} {\bibfnamefont {J.~S.}\ \bibnamefont
  {Kottmann}},\ }\bibfield  {title} {\bibinfo {title} {Molecular {{Quantum
  Circuit Design}}: {{A Graph-Based Approach}}},\ }\href
  {https://doi.org/10.22331/q-2023-08-03-1073} {\bibfield  {journal} {\bibinfo
  {journal} {Quantum}\ }\textbf {\bibinfo {volume} {7}},\ \bibinfo {pages}
  {1073}}\BibitemShut {NoStop}%
\bibitem [{hib(2007)}]{hiberty2007textbook9}%
  \BibitemOpen
  \bibfield  {title} {\bibinfo {title} {Currently available ab initio valence
  bond computational methods and their principles},\ }in\ \href
  {https://doi.org/10.1002/9780470192597.ch9} {\emph {\bibinfo {booktitle} {A
  Chemist's Guide to Valence Bond Theory}}}\ (\bibinfo  {publisher} {{John
  Wiley \& Sons, Ltd}},\ \bibinfo {year} {2007})\ Chap.~\bibinfo {chapter} {9},
  pp.\ \bibinfo {pages} {238--270},\ \Eprint
  {https://arxiv.org/abs/https://onlinelibrary.wiley.com/doi/pdf/10.1002/9780470192597.ch9}
  {https://onlinelibrary.wiley.com/doi/pdf/10.1002/9780470192597.ch9}
  \BibitemShut {NoStop}%
\bibitem [{\citenamefont {Kottmann}\ \emph {et~al.}(2021)\citenamefont
  {Kottmann}, \citenamefont {Alperin-Lea}, \citenamefont {Tamayo-Mendoza},
  \citenamefont {Cervera-Lierta}, \citenamefont {Lavigne}, \citenamefont {Yen},
  \citenamefont {Verteletskyi}, \citenamefont {Schleich}, \citenamefont
  {Anand}, \citenamefont {Degroote}, \citenamefont {Chaney}, \citenamefont
  {Kesibi}, \citenamefont {Curnow}, \citenamefont {Solo}, \citenamefont
  {Tsilimigkounakis}, \citenamefont {Zendejas-Morales}, \citenamefont
  {Izmaylov},\ and\ \citenamefont {Aspuru-Guzik}}]{tequila}%
  \BibitemOpen
  \bibfield  {author} {\bibinfo {author} {\bibfnamefont {J.~S.}\ \bibnamefont
  {Kottmann}}, \bibinfo {author} {\bibfnamefont {S.}~\bibnamefont
  {Alperin-Lea}}, \bibinfo {author} {\bibfnamefont {T.}~\bibnamefont
  {Tamayo-Mendoza}}, \bibinfo {author} {\bibfnamefont {A.}~\bibnamefont
  {Cervera-Lierta}}, \bibinfo {author} {\bibfnamefont {C.}~\bibnamefont
  {Lavigne}}, \bibinfo {author} {\bibfnamefont {T.-C.}\ \bibnamefont {Yen}},
  \bibinfo {author} {\bibfnamefont {V.}~\bibnamefont {Verteletskyi}}, \bibinfo
  {author} {\bibfnamefont {P.}~\bibnamefont {Schleich}}, \bibinfo {author}
  {\bibfnamefont {A.}~\bibnamefont {Anand}}, \bibinfo {author} {\bibfnamefont
  {M.}~\bibnamefont {Degroote}}, \bibinfo {author} {\bibfnamefont
  {S.}~\bibnamefont {Chaney}}, \bibinfo {author} {\bibfnamefont
  {M.}~\bibnamefont {Kesibi}}, \bibinfo {author} {\bibfnamefont {N.~G.}\
  \bibnamefont {Curnow}}, \bibinfo {author} {\bibfnamefont {B.}~\bibnamefont
  {Solo}}, \bibinfo {author} {\bibfnamefont {G.}~\bibnamefont
  {Tsilimigkounakis}}, \bibinfo {author} {\bibfnamefont {C.}~\bibnamefont
  {Zendejas-Morales}}, \bibinfo {author} {\bibfnamefont {A.~F.}\ \bibnamefont
  {Izmaylov}},\ and\ \bibinfo {author} {\bibfnamefont {A.}~\bibnamefont
  {Aspuru-Guzik}},\ }\bibfield  {title} {\bibinfo {title} {{{TEQUILA}}: A
  platform for rapid development of quantum algorithms},\ }\href
  {https://doi.org/10.1088/2058-9565/abe567} {\bibfield  {journal} {\bibinfo
  {journal} {Quantum Science and Technology}\ }\textbf {\bibinfo {volume}
  {6}},\ \bibinfo {pages} {024009} (\bibinfo {year} {2021})}\BibitemShut
  {NoStop}%
\bibitem [{\citenamefont {Suzuki}\ \emph {et~al.}(2021)\citenamefont {Suzuki},
  \citenamefont {Kawase}, \citenamefont {Masumura}, \citenamefont {Hiraga},
  \citenamefont {Nakadai}, \citenamefont {Chen}, \citenamefont {Nakanishi},
  \citenamefont {Mitarai}, \citenamefont {Imai}, \citenamefont {Tamiya},
  \citenamefont {Yamamoto}, \citenamefont {Yan}, \citenamefont {Kawakubo},
  \citenamefont {Nakagawa}, \citenamefont {Ibe}, \citenamefont {Zhang},
  \citenamefont {Yamashita}, \citenamefont {Yoshimura}, \citenamefont
  {Hayashi},\ and\ \citenamefont {Fujii}}]{qulacs}%
  \BibitemOpen
  \bibfield  {author} {\bibinfo {author} {\bibfnamefont {Y.}~\bibnamefont
  {Suzuki}}, \bibinfo {author} {\bibfnamefont {Y.}~\bibnamefont {Kawase}},
  \bibinfo {author} {\bibfnamefont {Y.}~\bibnamefont {Masumura}}, \bibinfo
  {author} {\bibfnamefont {Y.}~\bibnamefont {Hiraga}}, \bibinfo {author}
  {\bibfnamefont {M.}~\bibnamefont {Nakadai}}, \bibinfo {author} {\bibfnamefont
  {J.}~\bibnamefont {Chen}}, \bibinfo {author} {\bibfnamefont {K.~M.}\
  \bibnamefont {Nakanishi}}, \bibinfo {author} {\bibfnamefont {K.}~\bibnamefont
  {Mitarai}}, \bibinfo {author} {\bibfnamefont {R.}~\bibnamefont {Imai}},
  \bibinfo {author} {\bibfnamefont {S.}~\bibnamefont {Tamiya}}, \bibinfo
  {author} {\bibfnamefont {T.}~\bibnamefont {Yamamoto}}, \bibinfo {author}
  {\bibfnamefont {T.}~\bibnamefont {Yan}}, \bibinfo {author} {\bibfnamefont
  {T.}~\bibnamefont {Kawakubo}}, \bibinfo {author} {\bibfnamefont {Y.~O.}\
  \bibnamefont {Nakagawa}}, \bibinfo {author} {\bibfnamefont {Y.}~\bibnamefont
  {Ibe}}, \bibinfo {author} {\bibfnamefont {Y.}~\bibnamefont {Zhang}}, \bibinfo
  {author} {\bibfnamefont {H.}~\bibnamefont {Yamashita}}, \bibinfo {author}
  {\bibfnamefont {H.}~\bibnamefont {Yoshimura}}, \bibinfo {author}
  {\bibfnamefont {A.}~\bibnamefont {Hayashi}},\ and\ \bibinfo {author}
  {\bibfnamefont {K.}~\bibnamefont {Fujii}},\ }\bibfield  {title} {\bibinfo
  {title} {Qulacs: A fast and versatile quantum circuit simulator for research
  purpose},\ }\href {https://doi.org/10.22331/q-2021-10-06-559} {\bibfield
  {journal} {\bibinfo  {journal} {Quantum}\ }\textbf {\bibinfo {volume} {5}},\
  \bibinfo {pages} {559} (\bibinfo {year} {2021})}\BibitemShut {NoStop}%
\bibitem [{\citenamefont {McClean}\ \emph {et~al.}(2020)\citenamefont
  {McClean}, \citenamefont {Rubin}, \citenamefont {Sung}, \citenamefont
  {Kivlichan}, \citenamefont {Bonet-Monroig}, \citenamefont {Cao},
  \citenamefont {Dai}, \citenamefont {Fried}, \citenamefont {Gidney},
  \citenamefont {Gimby}, \citenamefont {Gokhale}, \citenamefont {Häner},
  \citenamefont {Hardikar}, \citenamefont {Havlíček}, \citenamefont
  {Higgott}, \citenamefont {Huang}, \citenamefont {Izaac}, \citenamefont
  {Jiang}, \citenamefont {Liu}, \citenamefont {McArdle}, \citenamefont
  {Neeley}, \citenamefont {O’Brien}, \citenamefont {O’Gorman},
  \citenamefont {Ozfidan}, \citenamefont {Radin}, \citenamefont {Romero},
  \citenamefont {Sawaya}, \citenamefont {Senjean}, \citenamefont {Setia},
  \citenamefont {Sim}, \citenamefont {Steiger}, \citenamefont {Steudtner},
  \citenamefont {Sun}, \citenamefont {Sun}, \citenamefont {Wang}, \citenamefont
  {Zhang},\ and\ \citenamefont {Babbush}}]{OpenFermion}%
  \BibitemOpen
  \bibfield  {author} {\bibinfo {author} {\bibfnamefont {J.~R.}\ \bibnamefont
  {McClean}}, \bibinfo {author} {\bibfnamefont {N.~C.}\ \bibnamefont {Rubin}},
  \bibinfo {author} {\bibfnamefont {K.~J.}\ \bibnamefont {Sung}}, \bibinfo
  {author} {\bibfnamefont {I.~D.}\ \bibnamefont {Kivlichan}}, \bibinfo {author}
  {\bibfnamefont {X.}~\bibnamefont {Bonet-Monroig}}, \bibinfo {author}
  {\bibfnamefont {Y.}~\bibnamefont {Cao}}, \bibinfo {author} {\bibfnamefont
  {C.}~\bibnamefont {Dai}}, \bibinfo {author} {\bibfnamefont {E.~S.}\
  \bibnamefont {Fried}}, \bibinfo {author} {\bibfnamefont {C.}~\bibnamefont
  {Gidney}}, \bibinfo {author} {\bibfnamefont {B.}~\bibnamefont {Gimby}},
  \bibinfo {author} {\bibfnamefont {P.}~\bibnamefont {Gokhale}}, \bibinfo
  {author} {\bibfnamefont {T.}~\bibnamefont {Häner}}, \bibinfo {author}
  {\bibfnamefont {T.}~\bibnamefont {Hardikar}}, \bibinfo {author}
  {\bibfnamefont {V.}~\bibnamefont {Havlíček}}, \bibinfo {author}
  {\bibfnamefont {O.}~\bibnamefont {Higgott}}, \bibinfo {author} {\bibfnamefont
  {C.}~\bibnamefont {Huang}}, \bibinfo {author} {\bibfnamefont
  {J.}~\bibnamefont {Izaac}}, \bibinfo {author} {\bibfnamefont
  {Z.}~\bibnamefont {Jiang}}, \bibinfo {author} {\bibfnamefont
  {X.}~\bibnamefont {Liu}}, \bibinfo {author} {\bibfnamefont {S.}~\bibnamefont
  {McArdle}}, \bibinfo {author} {\bibfnamefont {M.}~\bibnamefont {Neeley}},
  \bibinfo {author} {\bibfnamefont {T.}~\bibnamefont {O’Brien}}, \bibinfo
  {author} {\bibfnamefont {B.}~\bibnamefont {O’Gorman}}, \bibinfo {author}
  {\bibfnamefont {I.}~\bibnamefont {Ozfidan}}, \bibinfo {author} {\bibfnamefont
  {M.~D.}\ \bibnamefont {Radin}}, \bibinfo {author} {\bibfnamefont
  {J.}~\bibnamefont {Romero}}, \bibinfo {author} {\bibfnamefont {N.~P.~D.}\
  \bibnamefont {Sawaya}}, \bibinfo {author} {\bibfnamefont {B.}~\bibnamefont
  {Senjean}}, \bibinfo {author} {\bibfnamefont {K.}~\bibnamefont {Setia}},
  \bibinfo {author} {\bibfnamefont {S.}~\bibnamefont {Sim}}, \bibinfo {author}
  {\bibfnamefont {D.~S.}\ \bibnamefont {Steiger}}, \bibinfo {author}
  {\bibfnamefont {M.}~\bibnamefont {Steudtner}}, \bibinfo {author}
  {\bibfnamefont {Q.}~\bibnamefont {Sun}}, \bibinfo {author} {\bibfnamefont
  {W.}~\bibnamefont {Sun}}, \bibinfo {author} {\bibfnamefont {D.}~\bibnamefont
  {Wang}}, \bibinfo {author} {\bibfnamefont {F.}~\bibnamefont {Zhang}},\ and\
  \bibinfo {author} {\bibfnamefont {R.}~\bibnamefont {Babbush}},\ }\bibfield
  {title} {\bibinfo {title} {{{OpenFermion}}: The electronic structure package
  for quantum computers},\ }\href {https://doi.org/10.1088/2058-9565/ab8ebc}
  {\bibfield  {journal} {\bibinfo  {journal} {Quantum Science and Technology}\
  } (\bibinfo {year} {2020})}\BibitemShut {NoStop}%
\bibitem [{\citenamefont {Sun}\ \emph {et~al.}(2018)\citenamefont {Sun},
  \citenamefont {Berkelbach}, \citenamefont {Blunt}, \citenamefont {Booth},
  \citenamefont {Guo}, \citenamefont {Li}, \citenamefont {Liu}, \citenamefont
  {McClain}, \citenamefont {Sayfutyarova}, \citenamefont {Sharma},
  \citenamefont {Wouters},\ and\ \citenamefont {Chan}}]{pyscf1}%
  \BibitemOpen
  \bibfield  {author} {\bibinfo {author} {\bibfnamefont {Q.}~\bibnamefont
  {Sun}}, \bibinfo {author} {\bibfnamefont {T.~C.}\ \bibnamefont {Berkelbach}},
  \bibinfo {author} {\bibfnamefont {N.~S.}\ \bibnamefont {Blunt}}, \bibinfo
  {author} {\bibfnamefont {G.~H.}\ \bibnamefont {Booth}}, \bibinfo {author}
  {\bibfnamefont {S.}~\bibnamefont {Guo}}, \bibinfo {author} {\bibfnamefont
  {Z.}~\bibnamefont {Li}}, \bibinfo {author} {\bibfnamefont {J.}~\bibnamefont
  {Liu}}, \bibinfo {author} {\bibfnamefont {J.~D.}\ \bibnamefont {McClain}},
  \bibinfo {author} {\bibfnamefont {E.~R.}\ \bibnamefont {Sayfutyarova}},
  \bibinfo {author} {\bibfnamefont {S.}~\bibnamefont {Sharma}}, \bibinfo
  {author} {\bibfnamefont {S.}~\bibnamefont {Wouters}},\ and\ \bibinfo {author}
  {\bibfnamefont {G.~K.-L.}\ \bibnamefont {Chan}},\ }\bibfield  {title}
  {\bibinfo {title} {{{PySCF}}: The {{Python-based}} simulations of chemistry
  framework},\ }\href {https://doi.org/10.1002/wcms.1340} {\bibfield  {journal}
  {\bibinfo  {journal} {Wiley Interdiscip. Rev. Comput. Mol. Sci.}\ }\textbf
  {\bibinfo {volume} {8}},\ \bibinfo {pages} {e1340} (\bibinfo {year}
  {2018})}\BibitemShut {NoStop}%
\bibitem [{\citenamefont {Virtanen}\ \emph {et~al.}(2020)\citenamefont
  {Virtanen}, \citenamefont {Gommers}, \citenamefont {Oliphant}, \citenamefont
  {Haberland}, \citenamefont {Reddy}, \citenamefont {Cournapeau}, \citenamefont
  {Burovski}, \citenamefont {Peterson}, \citenamefont {Weckesser},
  \citenamefont {Bright}, \citenamefont {{van der Walt}}, \citenamefont
  {Brett}, \citenamefont {Wilson}, \citenamefont {Jarrod~Millman},
  \citenamefont {Mayorov}, \citenamefont {Nelson}, \citenamefont {Jones},
  \citenamefont {Kern}, \citenamefont {Larson}, \citenamefont {Carey},
  \citenamefont {Polat}, \citenamefont {Feng}, \citenamefont {Moore},
  \citenamefont {{Vand erPlas}}, \citenamefont {Laxalde}, \citenamefont
  {Perktold}, \citenamefont {Cimrman}, \citenamefont {Henriksen}, \citenamefont
  {Quintero}, \citenamefont {Harris}, \citenamefont {Archibald}, \citenamefont
  {Ribeiro}, \citenamefont {Pedregosa}, \citenamefont {{van Mulbregt}},\ and\
  \citenamefont {Contributors}}]{scipy}%
  \BibitemOpen
  \bibfield  {author} {\bibinfo {author} {\bibfnamefont {P.}~\bibnamefont
  {Virtanen}}, \bibinfo {author} {\bibfnamefont {R.}~\bibnamefont {Gommers}},
  \bibinfo {author} {\bibfnamefont {T.~E.}\ \bibnamefont {Oliphant}}, \bibinfo
  {author} {\bibfnamefont {M.}~\bibnamefont {Haberland}}, \bibinfo {author}
  {\bibfnamefont {T.}~\bibnamefont {Reddy}}, \bibinfo {author} {\bibfnamefont
  {D.}~\bibnamefont {Cournapeau}}, \bibinfo {author} {\bibfnamefont
  {E.}~\bibnamefont {Burovski}}, \bibinfo {author} {\bibfnamefont
  {P.}~\bibnamefont {Peterson}}, \bibinfo {author} {\bibfnamefont
  {W.}~\bibnamefont {Weckesser}}, \bibinfo {author} {\bibfnamefont
  {J.}~\bibnamefont {Bright}}, \bibinfo {author} {\bibfnamefont {S.~J.}\
  \bibnamefont {{van der Walt}}}, \bibinfo {author} {\bibfnamefont
  {M.}~\bibnamefont {Brett}}, \bibinfo {author} {\bibfnamefont
  {J.}~\bibnamefont {Wilson}}, \bibinfo {author} {\bibfnamefont
  {K.}~\bibnamefont {Jarrod~Millman}}, \bibinfo {author} {\bibfnamefont
  {N.}~\bibnamefont {Mayorov}}, \bibinfo {author} {\bibfnamefont {A.~R.~J.}\
  \bibnamefont {Nelson}}, \bibinfo {author} {\bibfnamefont {E.}~\bibnamefont
  {Jones}}, \bibinfo {author} {\bibfnamefont {R.}~\bibnamefont {Kern}},
  \bibinfo {author} {\bibfnamefont {E.}~\bibnamefont {Larson}}, \bibinfo
  {author} {\bibfnamefont {{\relax CJ}.}~\bibnamefont {Carey}}, \bibinfo
  {author} {\bibfnamefont {{\.I}.}~\bibnamefont {Polat}}, \bibinfo {author}
  {\bibfnamefont {Y.}~\bibnamefont {Feng}}, \bibinfo {author} {\bibfnamefont
  {E.~W.}\ \bibnamefont {Moore}}, \bibinfo {author} {\bibfnamefont
  {J.}~\bibnamefont {{Vand erPlas}}}, \bibinfo {author} {\bibfnamefont
  {D.}~\bibnamefont {Laxalde}}, \bibinfo {author} {\bibfnamefont
  {J.}~\bibnamefont {Perktold}}, \bibinfo {author} {\bibfnamefont
  {R.}~\bibnamefont {Cimrman}}, \bibinfo {author} {\bibfnamefont
  {I.}~\bibnamefont {Henriksen}}, \bibinfo {author} {\bibfnamefont {E.~A.}\
  \bibnamefont {Quintero}}, \bibinfo {author} {\bibfnamefont {C.~R.}\
  \bibnamefont {Harris}}, \bibinfo {author} {\bibfnamefont {A.~M.}\
  \bibnamefont {Archibald}}, \bibinfo {author} {\bibfnamefont {A.~H.}\
  \bibnamefont {Ribeiro}}, \bibinfo {author} {\bibfnamefont {F.}~\bibnamefont
  {Pedregosa}}, \bibinfo {author} {\bibfnamefont {P.}~\bibnamefont {{van
  Mulbregt}}},\ and\ \bibinfo {author} {\bibfnamefont {S.~.~.}\ \bibnamefont
  {Contributors}},\ }\bibfield  {title} {\bibinfo {title} {{{SciPy}} 1.0:
  {{Fundamental}} algorithms for scientific computing in python},\ }\href
  {https://doi.org/10.1038/s41592-019-0686-2} {\bibfield  {journal} {\bibinfo
  {journal} {Nature Methods}\ }\textbf {\bibinfo {volume} {17}},\ \bibinfo
  {pages} {261} (\bibinfo {year} {2020})}\BibitemShut {NoStop}%
\bibitem [{\citenamefont {Stein}(2024)}]{steinNylser2025}%
  \BibitemOpen
  \bibfield  {author} {\bibinfo {author} {\bibfnamefont {K.}~\bibnamefont
  {Stein}},\ }\href {https://github.com/nylser/quanti-gin} {\bibinfo {title}
  {Nylser/quanti-gin}} (\bibinfo {year} {2024})\BibitemShut {NoStop}%
\bibitem [{\citenamefont {G.~Draper}(2025)}]{Qpic2025}%
  \BibitemOpen
  \bibfield  {author} {\bibinfo {author} {\bibfnamefont {S.}~\bibnamefont
  {G.~Draper}, \bibfnamefont {Thomas \and A.~Kutin}},\ }\href
  {https://github.com/qpic} {\bibinfo {title} {Qpic/qpic}},\ \bibinfo
  {howpublished} {qpic} (\bibinfo {year} {2025})\BibitemShut {NoStop}%
\bibitem [{\citenamefont {Kottmann}\ and\ \citenamefont
  {{Aspuru-Guzik}}(2022)}]{kottmann2022optimized}%
  \BibitemOpen
  \bibfield  {author} {\bibinfo {author} {\bibfnamefont {J.~S.}\ \bibnamefont
  {Kottmann}}\ and\ \bibinfo {author} {\bibfnamefont {A.}~\bibnamefont
  {{Aspuru-Guzik}}},\ }\bibfield  {title} {\bibinfo {title} {Optimized
  low-depth quantum circuits for molecular electronic structure using a
  separable-pair approximation},\ }\href
  {https://doi.org/10.1103/PhysRevA.105.032449} {\bibfield  {journal} {\bibinfo
   {journal} {Physical Review A}\ }\textbf {\bibinfo {volume} {105}},\ \bibinfo
  {pages} {032449} (\bibinfo {year} {2022})}\BibitemShut {NoStop}%
\bibitem [{\citenamefont {Ding}\ \emph {et~al.}(2021)\citenamefont {Ding},
  \citenamefont {Mardazad}, \citenamefont {Das}, \citenamefont {Szalay},
  \citenamefont {Schollw{\"o}ck}, \citenamefont {Zimbor{\'a}s},\ and\
  \citenamefont {Schilling}}]{dingConceptOrbitalEntanglement2021}%
  \BibitemOpen
  \bibfield  {author} {\bibinfo {author} {\bibfnamefont {L.}~\bibnamefont
  {Ding}}, \bibinfo {author} {\bibfnamefont {S.}~\bibnamefont {Mardazad}},
  \bibinfo {author} {\bibfnamefont {S.}~\bibnamefont {Das}}, \bibinfo {author}
  {\bibfnamefont {S.}~\bibnamefont {Szalay}}, \bibinfo {author} {\bibfnamefont
  {U.}~\bibnamefont {Schollw{\"o}ck}}, \bibinfo {author} {\bibfnamefont
  {Z.}~\bibnamefont {Zimbor{\'a}s}},\ and\ \bibinfo {author} {\bibfnamefont
  {C.}~\bibnamefont {Schilling}},\ }\bibfield  {title} {\bibinfo {title}
  {Concept of {{Orbital Entanglement}} and {{Correlation}} in {{Quantum
  Chemistry}}},\ }\href {https://doi.org/10.1021/acs.jctc.0c00559} {\bibfield
  {journal} {\bibinfo  {journal} {Journal of Chemical Theory and Computation}\
  }\textbf {\bibinfo {volume} {17}},\ \bibinfo {pages} {79} (\bibinfo {year}
  {2021})}\BibitemShut {NoStop}%
\bibitem [{\citenamefont {Ding}\ \emph {et~al.}(2022)\citenamefont {Ding},
  \citenamefont {Knecht}, \citenamefont {Zimbor{\'a}s},\ and\ \citenamefont
  {Schilling}}]{dingQuantumCorrelationsMolecules2022}%
  \BibitemOpen
  \bibfield  {author} {\bibinfo {author} {\bibfnamefont {L.}~\bibnamefont
  {Ding}}, \bibinfo {author} {\bibfnamefont {S.}~\bibnamefont {Knecht}},
  \bibinfo {author} {\bibfnamefont {Z.}~\bibnamefont {Zimbor{\'a}s}},\ and\
  \bibinfo {author} {\bibfnamefont {C.}~\bibnamefont {Schilling}},\ }\bibfield
  {title} {\bibinfo {title} {Quantum correlations in molecules: From quantum
  resourcing to chemical bonding},\ }\href@noop {} {\bibfield  {journal}
  {\bibinfo  {journal} {Quantum Science and Technology}\ }\textbf {\bibinfo
  {volume} {8}},\ \bibinfo {pages} {015015} (\bibinfo {year}
  {2022})}\BibitemShut {NoStop}%
\bibitem [{\citenamefont {Ding}\ \emph {et~al.}(2023)\citenamefont {Ding},
  \citenamefont {Knecht},\ and\ \citenamefont {Schilling}}]{dingQuantum2023}%
  \BibitemOpen
  \bibfield  {author} {\bibinfo {author} {\bibfnamefont {L.}~\bibnamefont
  {Ding}}, \bibinfo {author} {\bibfnamefont {S.}~\bibnamefont {Knecht}},\ and\
  \bibinfo {author} {\bibfnamefont {C.}~\bibnamefont {Schilling}},\ }\bibfield
  {title} {\bibinfo {title} {Quantum {{Information-Assisted Complete Active
  Space Optimization}} ({{QICAS}})},\ }\href
  {https://doi.org/10.1021/acs.jpclett.3c02536} {\bibfield  {journal} {\bibinfo
   {journal} {The Journal of Physical Chemistry Letters}\ }\textbf {\bibinfo
  {volume} {14}},\ \bibinfo {pages} {11022} (\bibinfo {year}
  {2023})}\BibitemShut {NoStop}%
\bibitem [{\citenamefont {Reascos}\ \emph {et~al.}(2024)\citenamefont
  {Reascos}, \citenamefont {Diotallevi},\ and\ \citenamefont
  {Benito}}]{reascosUniversal2024}%
  \BibitemOpen
  \bibfield  {author} {\bibinfo {author} {\bibfnamefont {L.}~\bibnamefont
  {Reascos}}, \bibinfo {author} {\bibfnamefont {G.~F.}\ \bibnamefont
  {Diotallevi}},\ and\ \bibinfo {author} {\bibfnamefont {M.}~\bibnamefont
  {Benito}},\ }\bibfield  {title} {\bibinfo {title} {Universal solution to the
  {{Schrieffer-Wolff Transformation Generator}}},\ }\bibfield  {journal}
  {\bibinfo  {journal} {quant-ph}\ }\href
  {https://doi.org/10.48550/arXiv.2411.11535} {10.48550/arXiv.2411.11535}
  (\bibinfo {year} {2024}),\ \Eprint {https://arxiv.org/abs/2411.11535}
  {arXiv:2411.11535} \BibitemShut {NoStop}%
\bibitem [{\citenamefont {Diotallevi}\ \emph {et~al.}(2024)\citenamefont
  {Diotallevi}, \citenamefont {Reascos},\ and\ \citenamefont
  {Benito}}]{diotalleviSymPT2024}%
  \BibitemOpen
  \bibfield  {author} {\bibinfo {author} {\bibfnamefont {G.~F.}\ \bibnamefont
  {Diotallevi}}, \bibinfo {author} {\bibfnamefont {L.}~\bibnamefont
  {Reascos}},\ and\ \bibinfo {author} {\bibfnamefont {M.}~\bibnamefont
  {Benito}},\ }\bibfield  {title} {\bibinfo {title} {{{SymPT}}: A comprehensive
  tool for automating effective {{Hamiltonian}} derivations},\ }\bibfield
  {journal} {\bibinfo  {journal} {quant-ph}\ }\href
  {https://doi.org/10.48550/arXiv.2412.10240} {10.48550/arXiv.2412.10240}
  (\bibinfo {year} {2024}),\ \Eprint {https://arxiv.org/abs/2412.10240}
  {arXiv:2412.10240} \BibitemShut {NoStop}%
\end{thebibliography}%

\clearpage

\appendix
\section*{Appendix}

\subsection{Visualization of HCB elements} \label{Visualization of HCB elements}
\begin{figure*}
    \centering
    \subfigure[$A_{kk}$]{\includegraphics[width=0.26\textwidth]{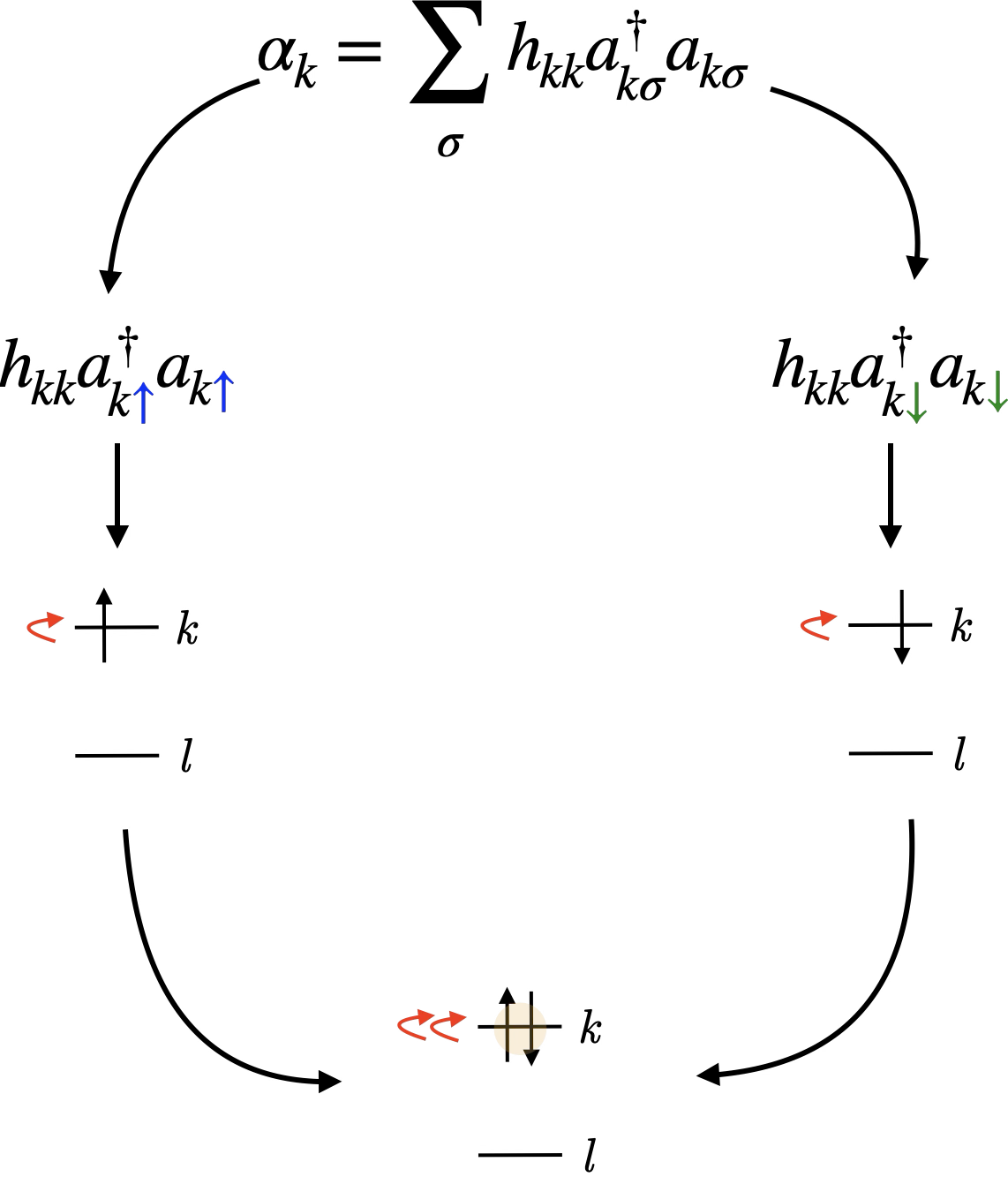}}
    \subfigure[$B_{kl}$]{\includegraphics[width=0.76\textwidth]{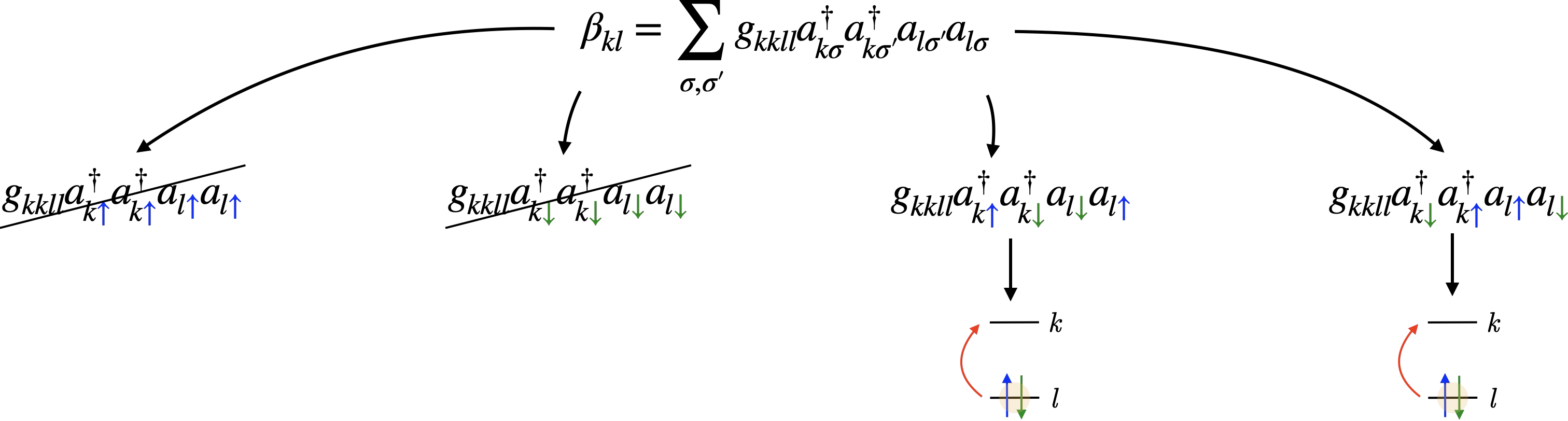}}
    \subfigure[$C_{kl}$]{\includegraphics[width=0.76\textwidth]{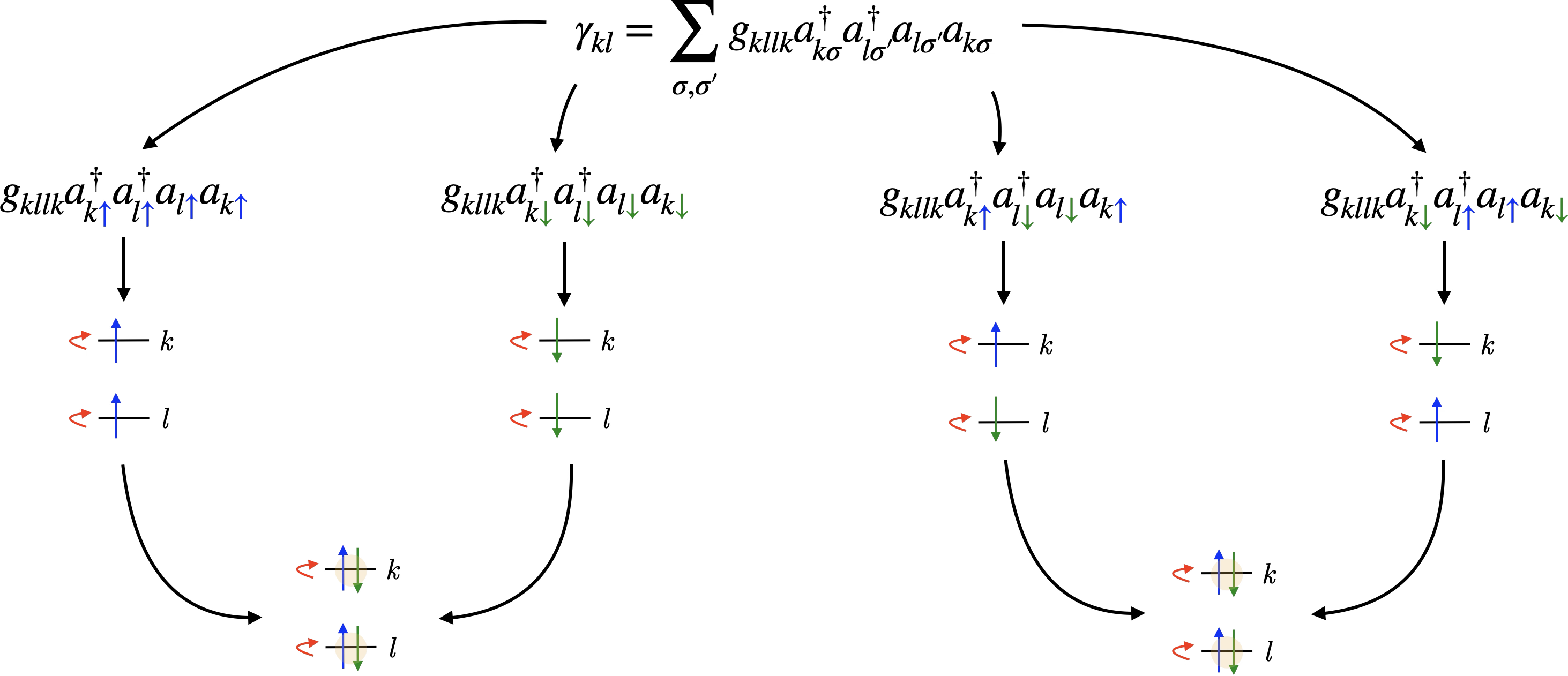}}
    \subfigure[$D_{kl}$]{\includegraphics[width=0.76\textwidth]{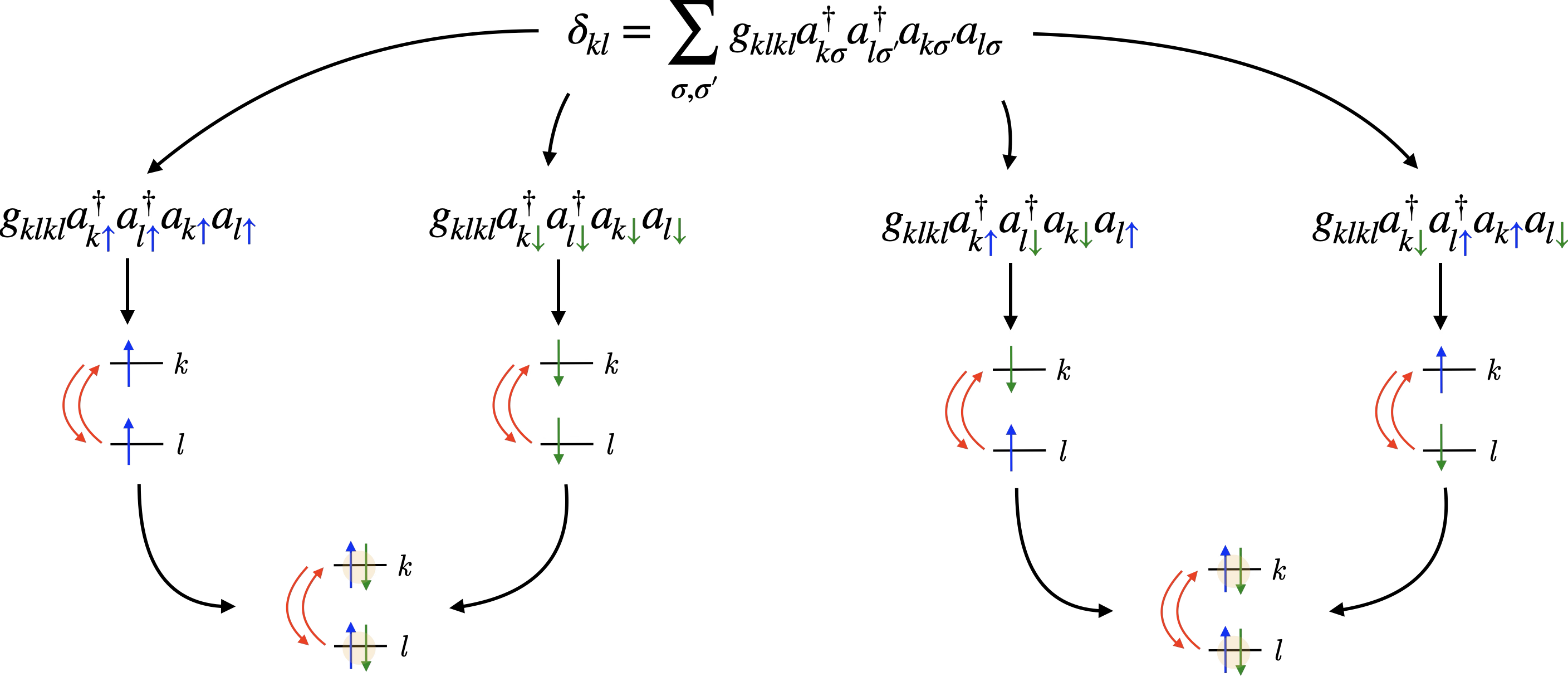}}
    \caption{Visualization of HCB elements. Each operator is expanded in all the possible spin combinations. These are then visually represented to explicit the paired-electrons, or quasi-bosonic particles, interpretation. The first and second terms stemming from $\beta_{kl}$ with all coherent spins are not allowed due to fermionic operators commutativity properties.}
    \label{visual_hcb}
\end{figure*}

The operators $\alpha_{k}$, $\beta_{kl}$, $\gamma_{kl}$ and $\delta_{kl}$ presented in Section \ref{HCB} account for multiple creation and destruction fermionic operators with all possible spin combinations. In Figure \ref{visual_hcb} we display a visual representation of all the terms that one needs to take into consideration and why we can interpret them as paired-electrons or quasi-bosonic particle creation and destruction operators. In each of the four operators we end up with two terms which provide the same contribution. The $\frac{1}{2}$ coefficient in the Hamiltonian definition takes care of this repetition.

\subsection{Full results tables} \label{Full results tables}
Tables \ref{measurement_groups} and \ref{number_of_measurements} show all the numerical results displayed in Figures \ref{Num_of_meas_groups} and \ref{Num_of_meas}.

\begin{table*}[h]
\centering
    \caption{Full results number of measurement groups needed for different reduction methods from \ref{Num_of_meas_groups}. In Scenario I and II we considered the number of steps that achieved the lowest error. Free H$_6$ refers to randomized molecule geometries and the values correspond to mean and standard deviation of the sample.}
    \begin{tabular}{ccccccc}
        \toprule
        \midrule
        Method & Linear H$_4$ & Square H$_4$ & Linear H$_6$ & Circular H$_6$ & Free H$_6$ & Linear H$_8$ \\
        \midrule
        Original $H$ & 361 & 357 & 1623 & 1795 & 1382$\pm$265 & 3985 \\
        \midrule
        \multicolumn{7}{c}{Pauli-grouping} \\
        \midrule
        \hyperlink{LF}{LF} & 28 & 21 & 90 & 91 & 74$\pm$15 & 154 \\
        \hyperlink{RLF}{RLF} & 19 & 22 & 60 & 64 & 54$\pm$8 & 111 \\
        \hyperlink{SI}{SI} & 19 & 19 & 68 & 77 & 64$\pm$9 & 114 \\
        \midrule
        \multicolumn{7}{c}{Fermionic-grouping} \\
        \midrule
        \hyperlink{LR}{LR} & 10 & 11 & 22 & 22 & 20$\pm$2 & 33 \\
        \hyperlink{FFF-LR}{FFF-LR} & 10 & 11 & 22 & 22 & 20$\pm$2 & 33 \\
        \midrule
        \multicolumn{7}{c}{This work} \\
        \midrule
        Scenario I & 9 & 9 & 9 & 15 & 45$\pm$38 & 6 \\
        Scenario II & 9 & 9 & 9 & 15 & 12$\pm$20 & 6 \\
        \bottomrule
    \end{tabular}
    \label{measurement_groups}
\end{table*}

\begin{table*}[h]
\centering
    \caption{Full results number of measurement (formula from \ref{Number of measurements section}) needed for different reduction methods, from \ref{Num_of_meas}. Every number is multiplied by $\times10^4$. Free H$_6$ refers to randomized molecule geometries and the values correspond to mean and standard deviation of the sample.}
    \begin{tabular}{ccccccc}
        \toprule
        \midrule
        Method & Linear H$_4$ & Square H$_4$ & Linear H$_6$ & Circular H$_6$ & Free H$_6$ & Linear H$_8$ \\
        \midrule
        \multicolumn{7}{c}{Pauli-grouping} \\
        \midrule
        \hyperlink{LF}{LF} & 11.89 & 6.41 & 19.10 & 35.10 & 16.06$\pm$6.36 & 26.88 \\
        \hyperlink{RLF}{RLF} & 4.99 & 9.94 & 22.24 & 27.08 & 16.60$\pm$6.65 & 25.89 \\
        \hyperlink{SI}{SI} & 3.48 & 2.77 & 3.58 & 4.27 & 4.60$\pm$2.15 & 3.60 \\
        \midrule
        \multicolumn{7}{c}{Fermionic-grouping} \\
        \midrule
        \hyperlink{LR}{LR} & 119.96 & 167.26 & 164.59 & 268.19 & 253.73$\pm$64.54 & 237.93 \\
        \hyperlink{FFF-LR}{FFF-LR} & 3.89 & 155.24 & 3.74 & 258.04 & 16.13$\pm$7.57 & 3.17 \\
        \midrule
        \multicolumn{7}{c}{This work} \\
        \midrule
        Scenario I & 2.28 & 2.44 & 2.55 & 3.17 & 5.69$\pm$4.57 & 2.60 \\
        Scenario II & 2.19 & 2.44 & 2.35 & 3.17 & 4.94$\pm$4.78 & 2.48 \\
        \bottomrule
    \end{tabular}
    \label{number_of_measurements}
\end{table*}

\subsection{Close-up visualization} \label{Close-up visualization}
\begin{figure*}
    \centering
    \subfigure[Linear H$_4$ Scenario I]{\includegraphics[width=0.4\textwidth]{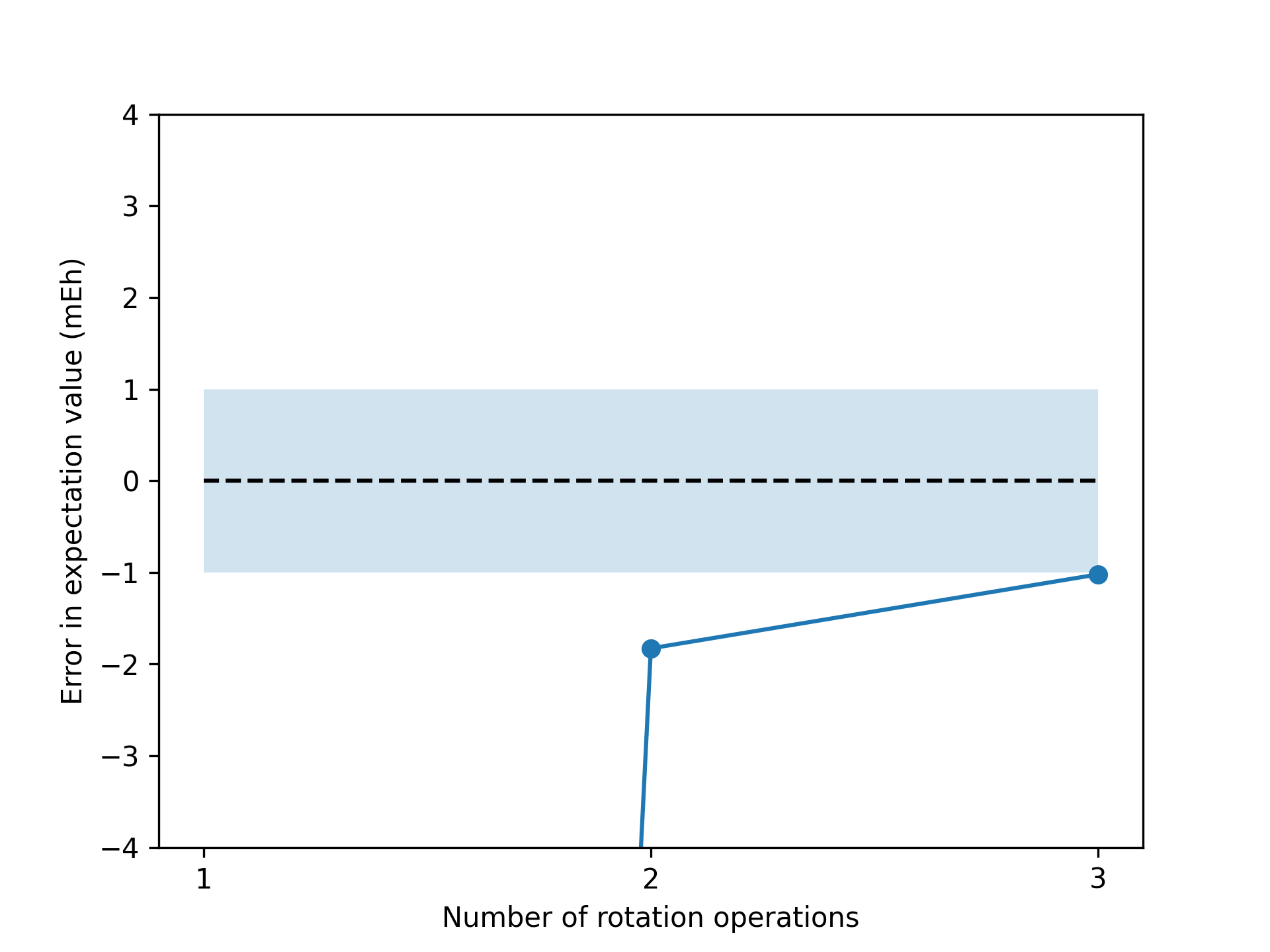}}
    \subfigure[Linear H$_4$ Scenario II]{\includegraphics[width=0.4\textwidth]{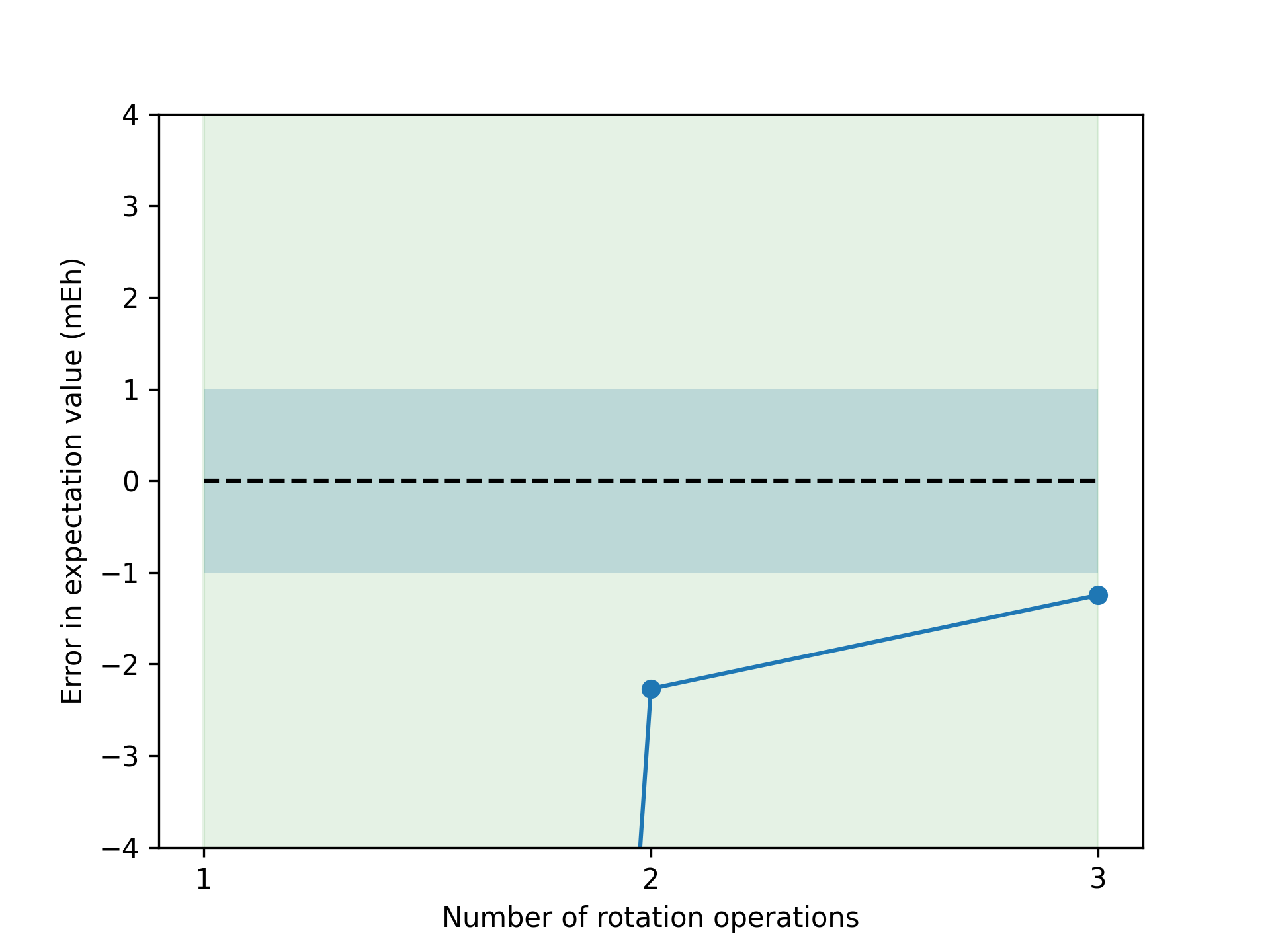}}
    \subfigure[Square H$_4$ Scenario I]{\includegraphics[width=0.4\textwidth]{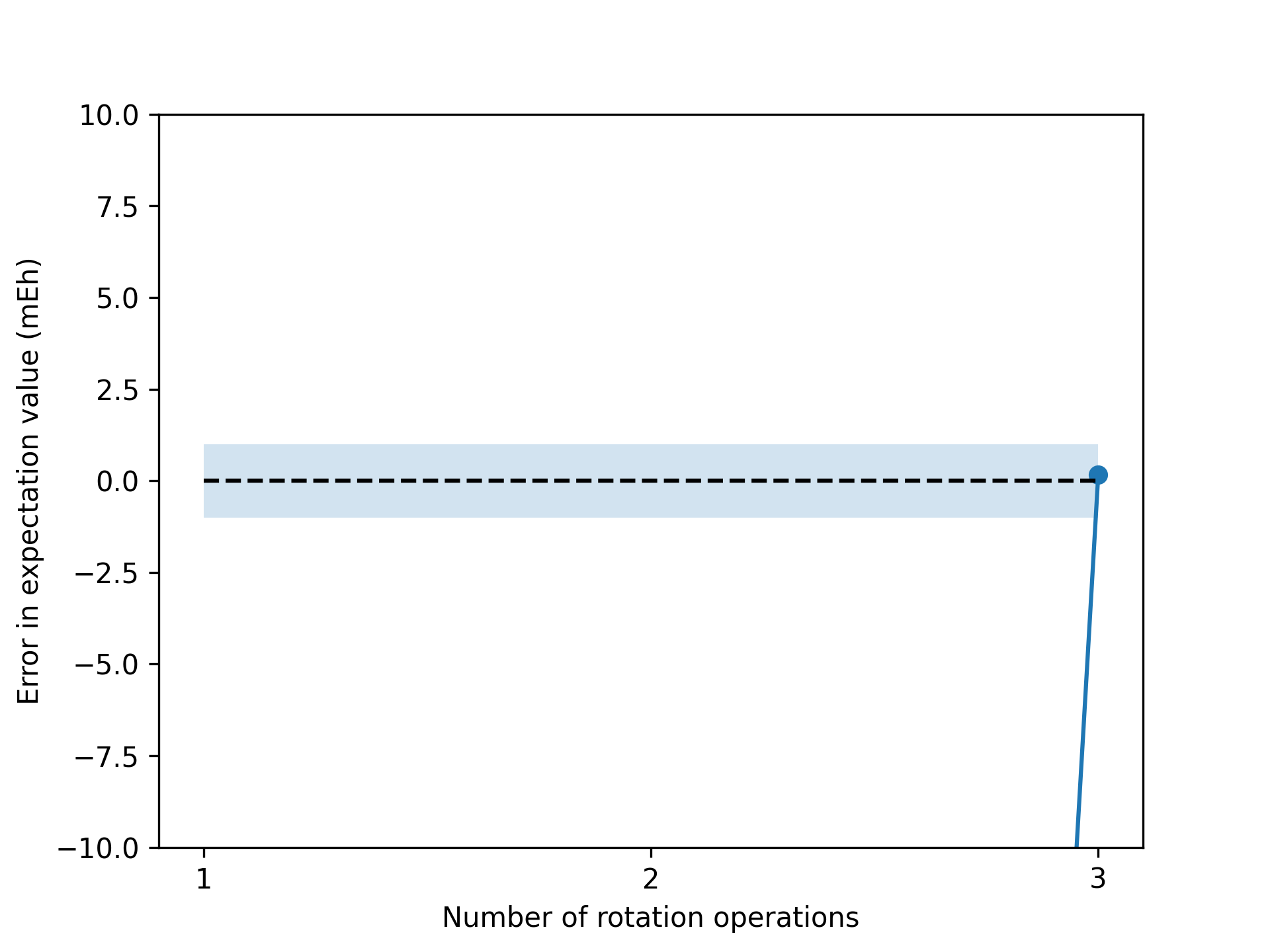}}
    \subfigure[Square H$_4$ Scenario II]{\includegraphics[width=0.4\textwidth]{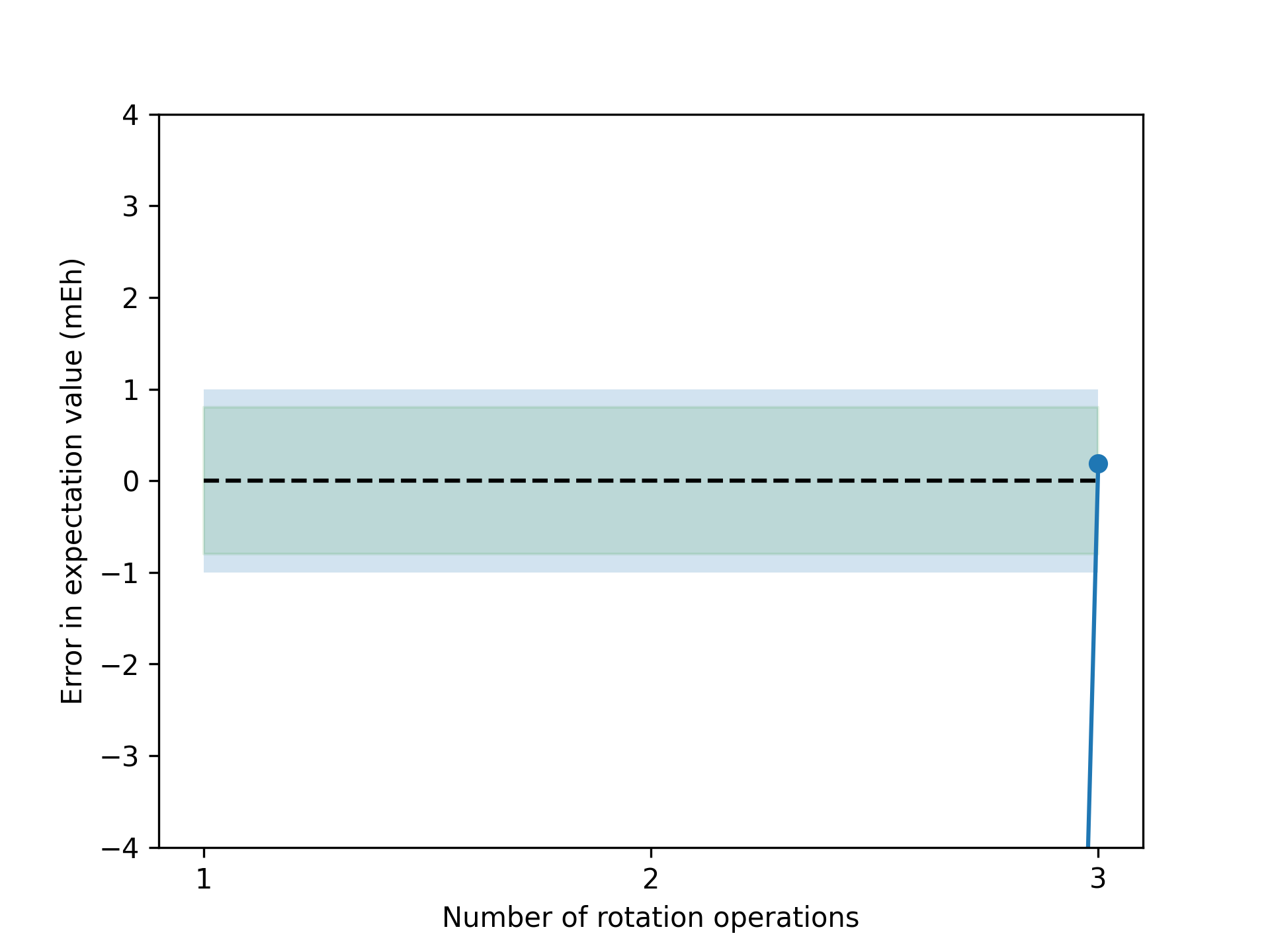}}
    \caption{Close-up visualization of Figure \ref{h4_result}.}
    \label{A3-1}
\end{figure*}

\begin{figure*}
    \centering
    \subfigure[Linear H$_6$ Scenario I]{\includegraphics[width=0.4\textwidth]{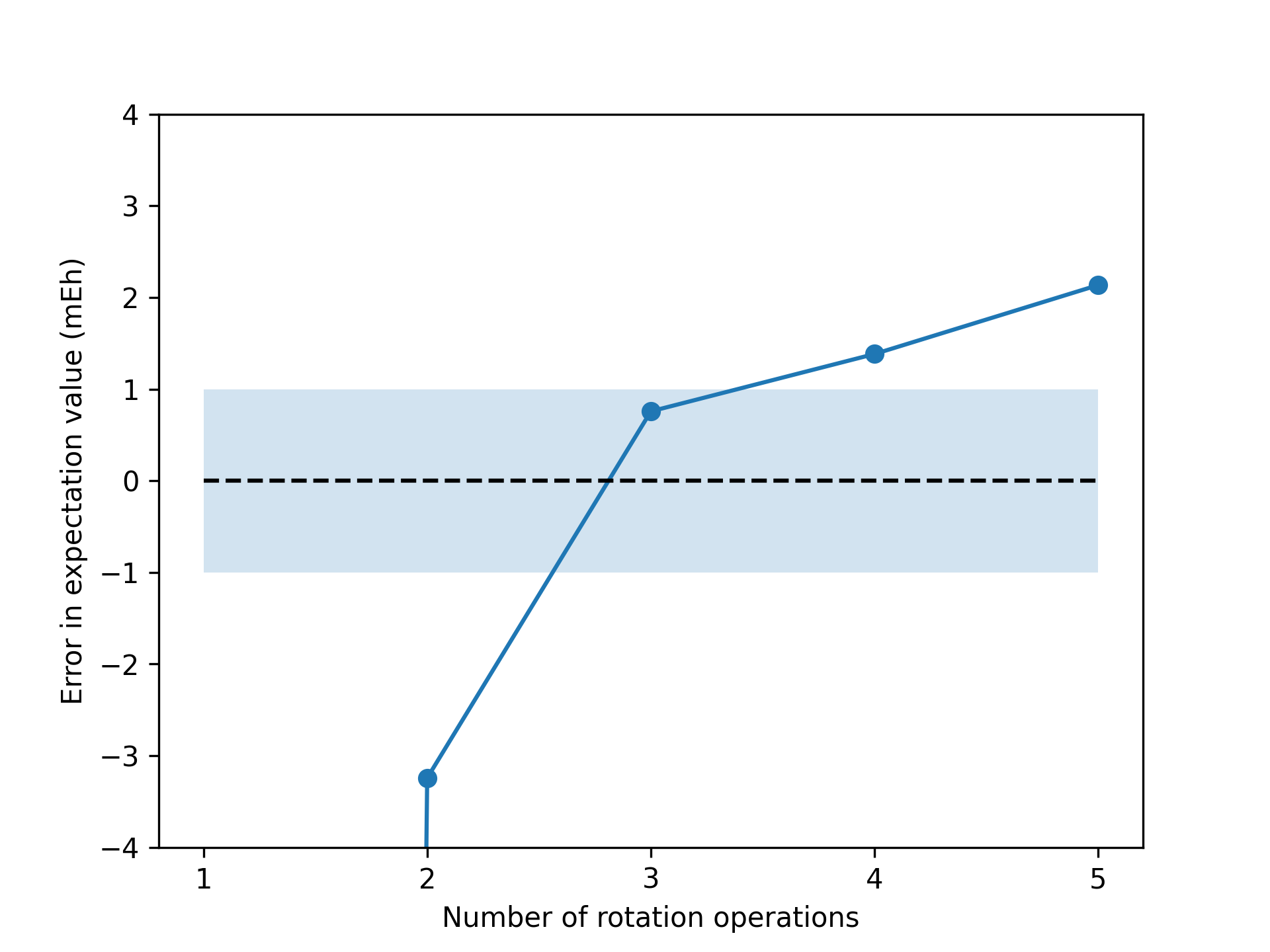}}
    \subfigure[Linear H$_6$ Scenario II]{\includegraphics[width=0.4\textwidth]{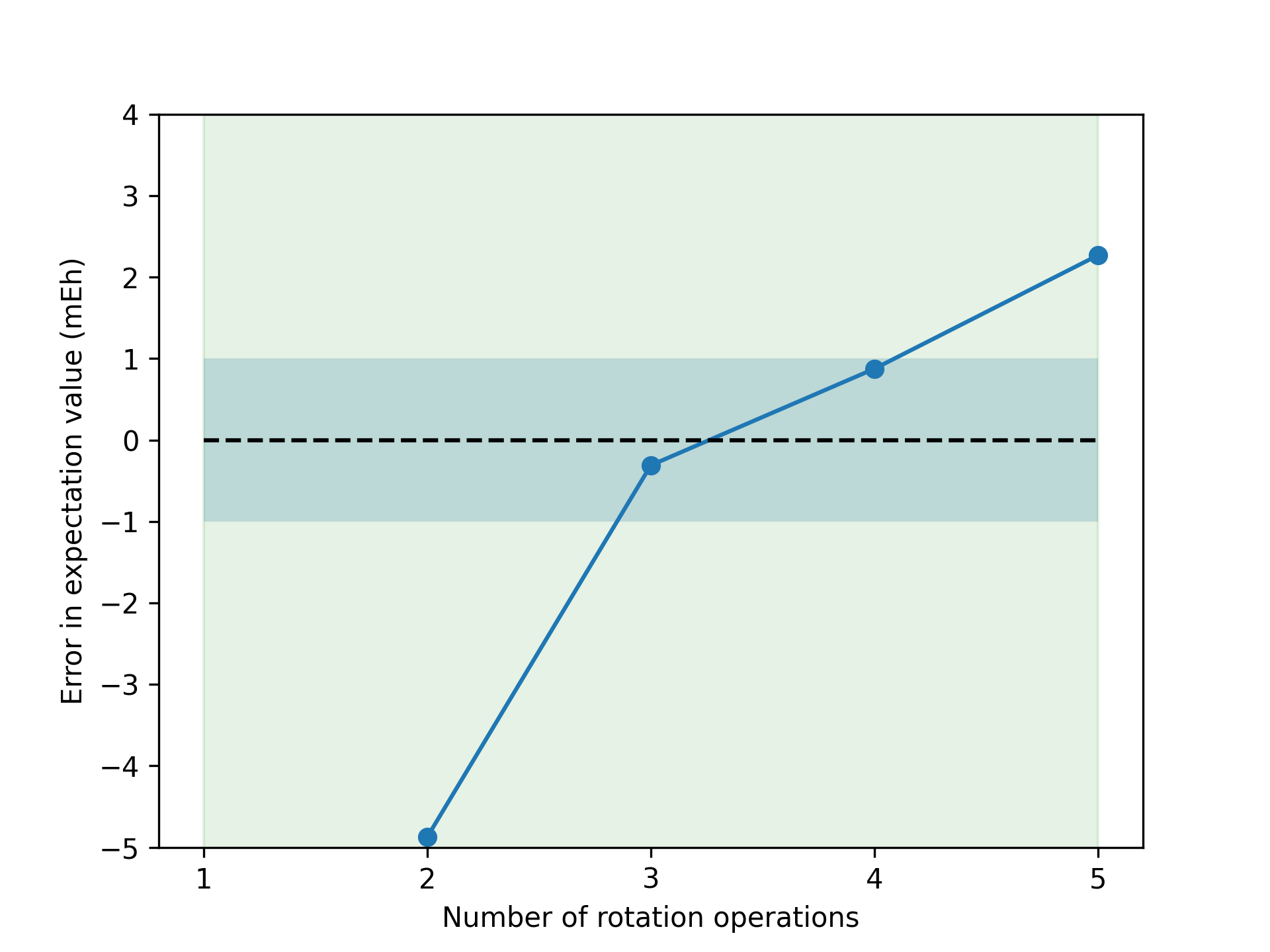}}
    \subfigure[Circular H$_6$ Scenario I]{\includegraphics[width=0.4\textwidth]{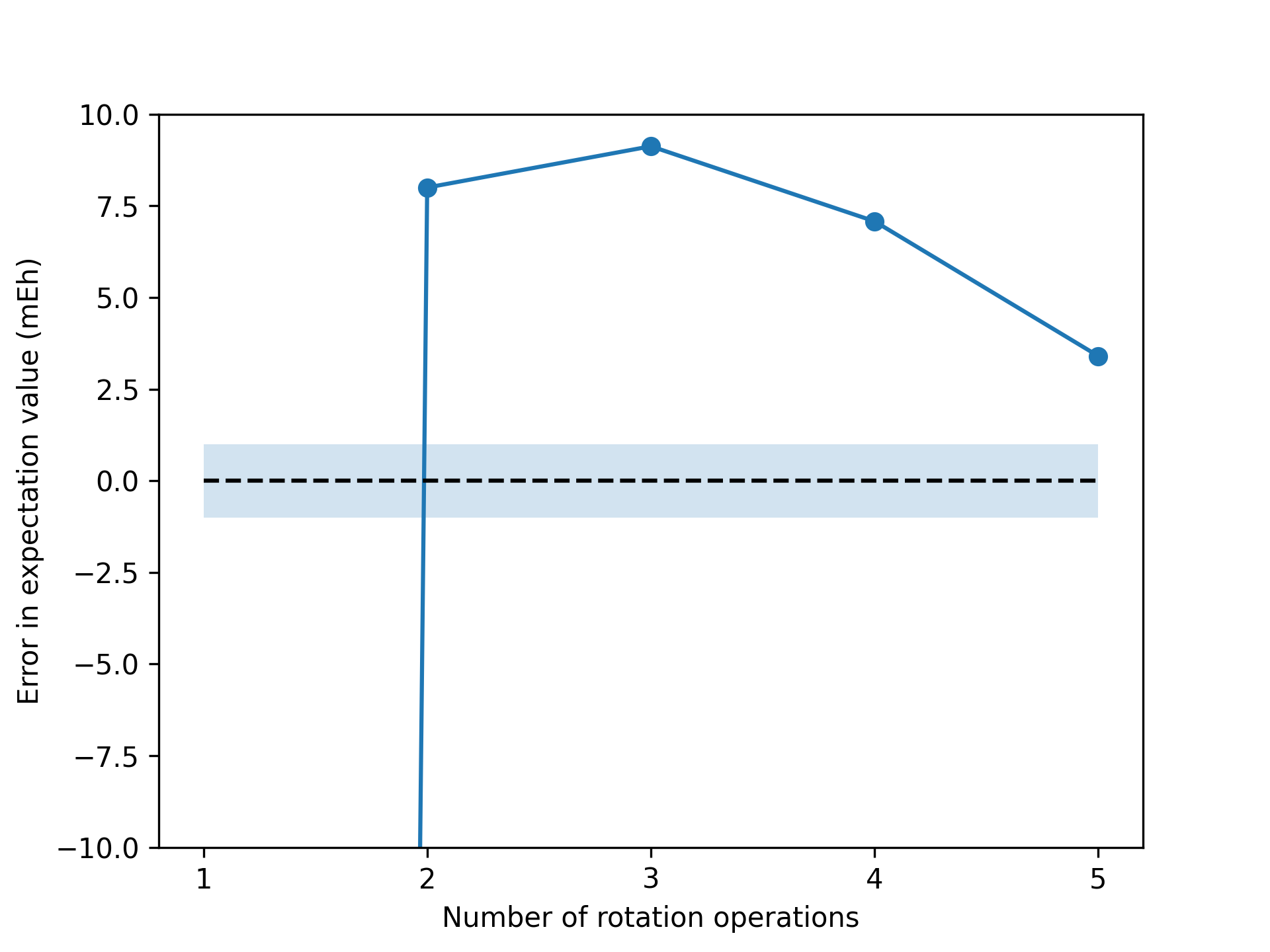}}
    \subfigure[Circular H$_6$ Scenario II]{\includegraphics[width=0.4\textwidth]{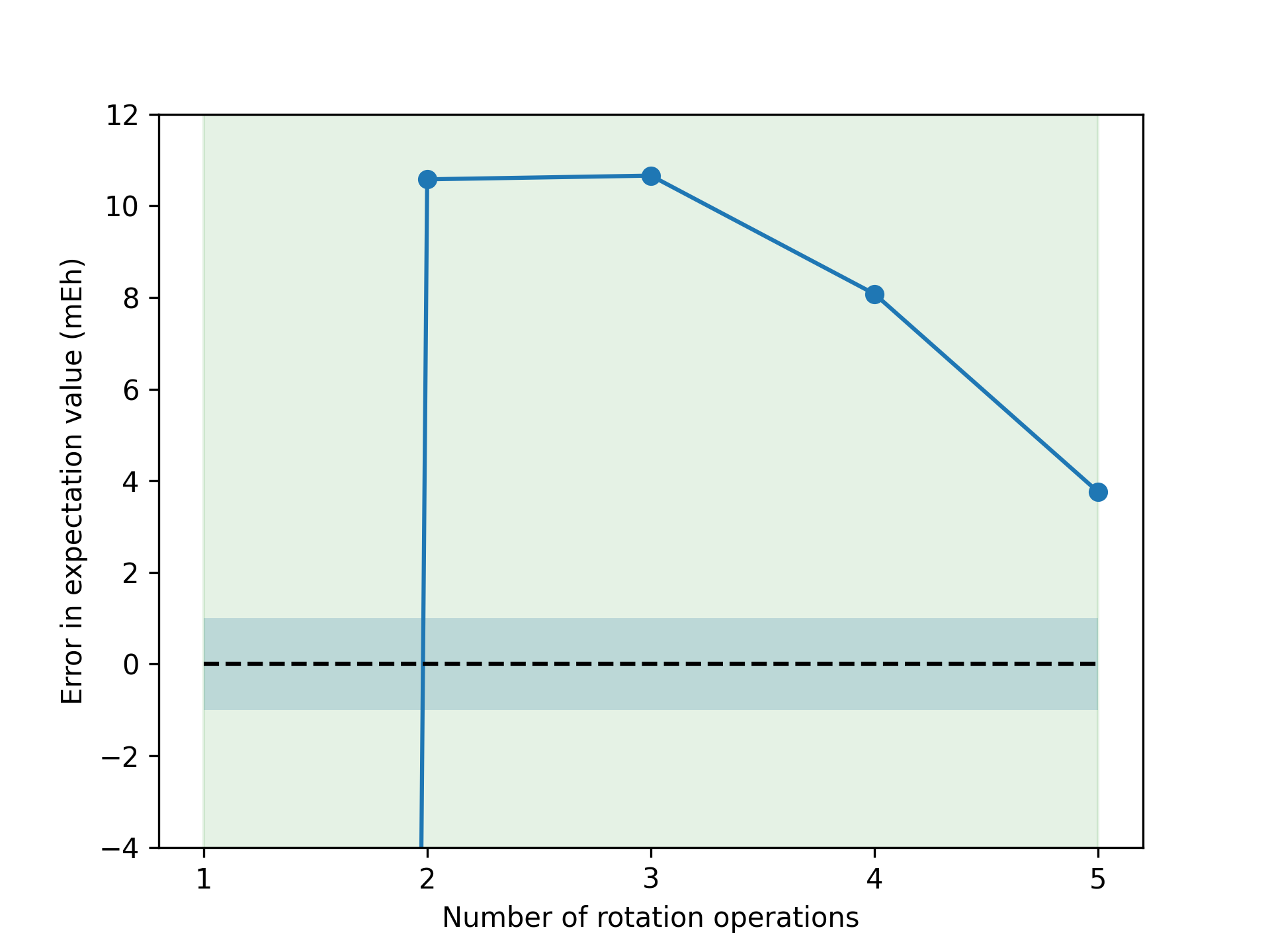}}
    \subfigure[Free H$_6$ Scenario I]{\includegraphics[width=0.4\textwidth]{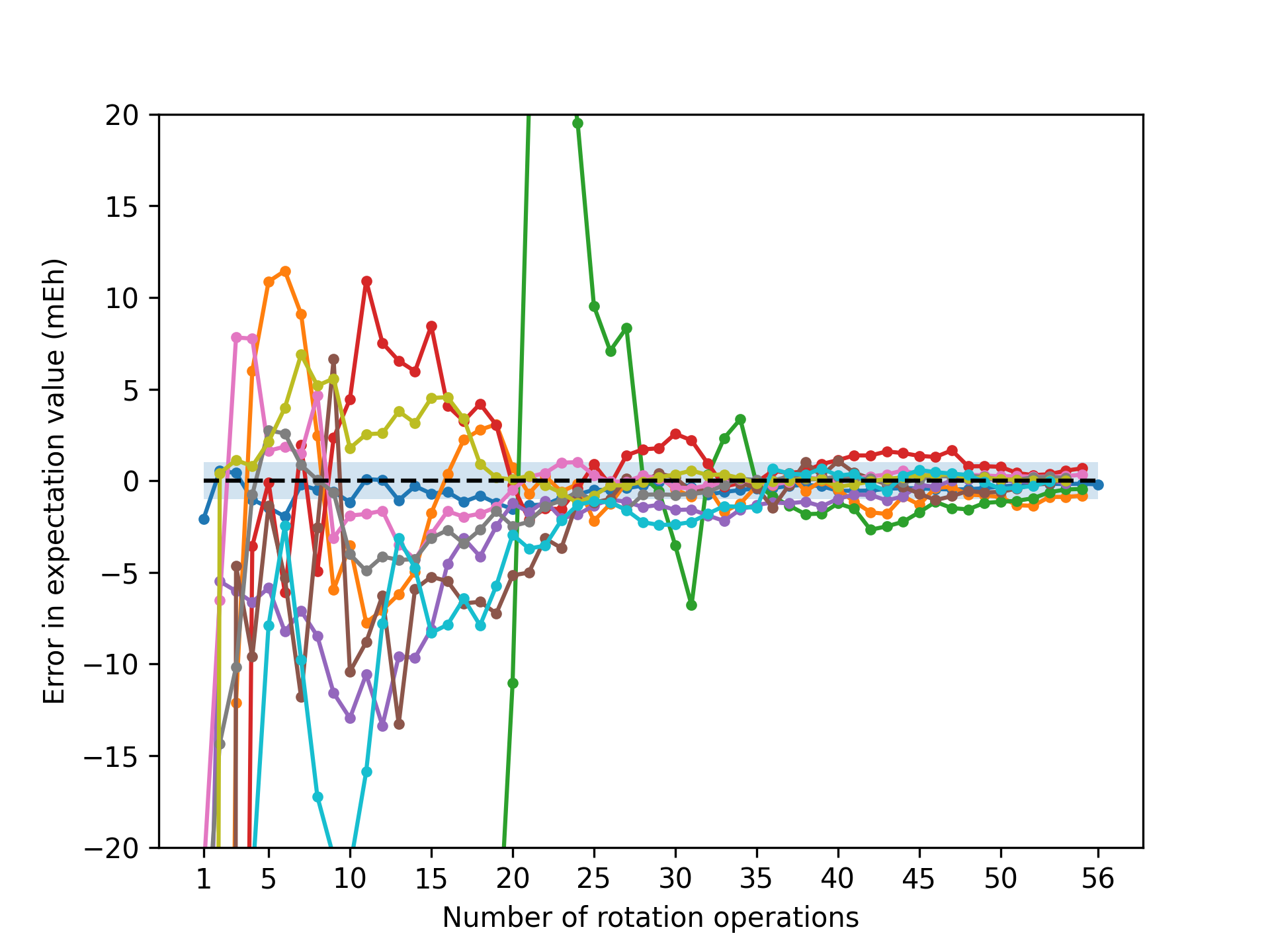}}
    \subfigure[Free H$_6$ Scenario II]{\includegraphics[width=0.4\textwidth]{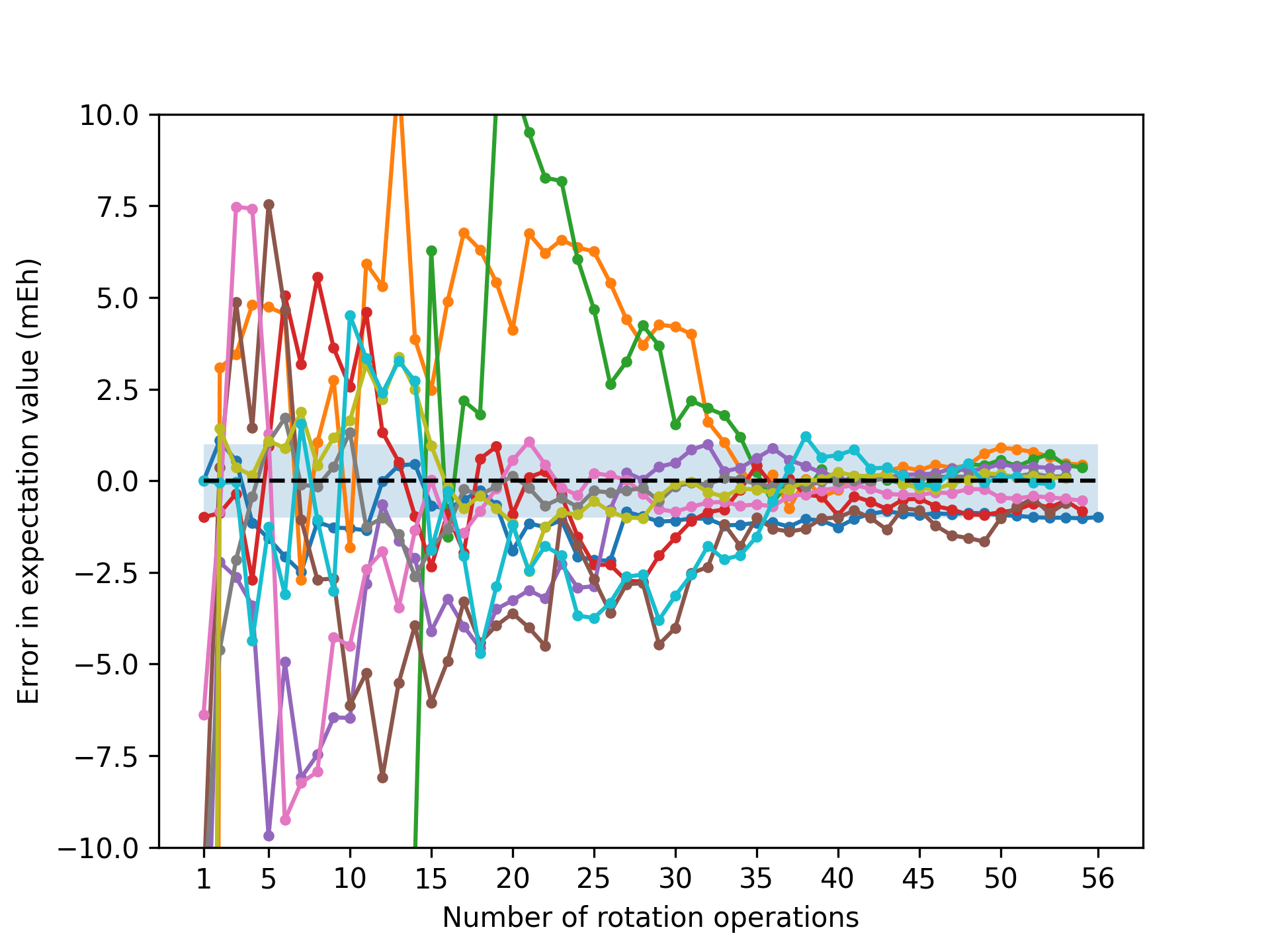}}
    \caption{Close-up visualization of Figure \ref{h6_result} and 10 molecules from Figure \ref{Random_H6_result}.}
    \label{A3-2}
\end{figure*}

\begin{figure*}
    \centering
    \subfigure[Linear H$_8$ Scenario I]{\includegraphics[width=0.4\textwidth]{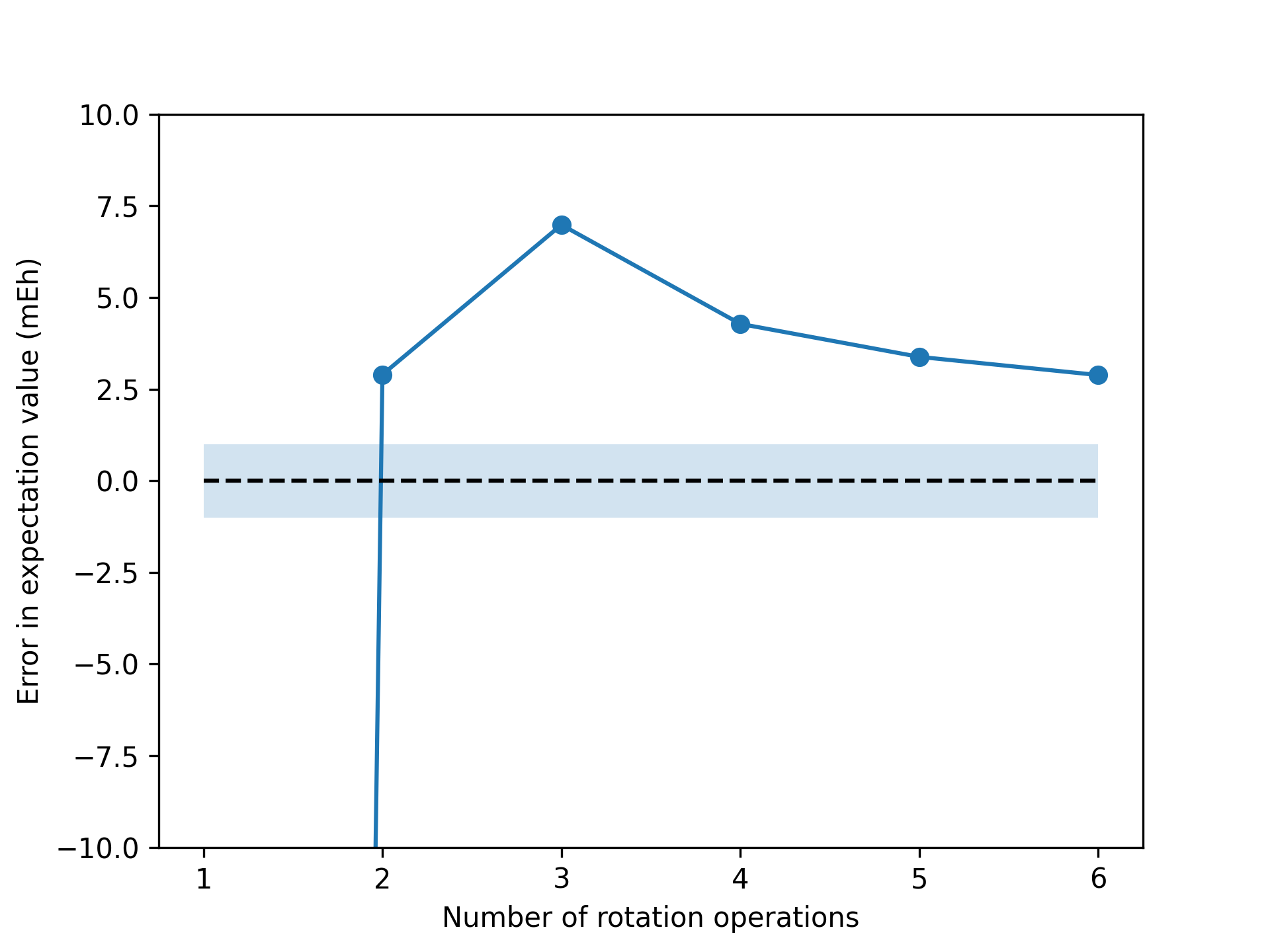}}
    \subfigure[Linear H$_8$ Scenario II]{\includegraphics[width=0.4\textwidth]{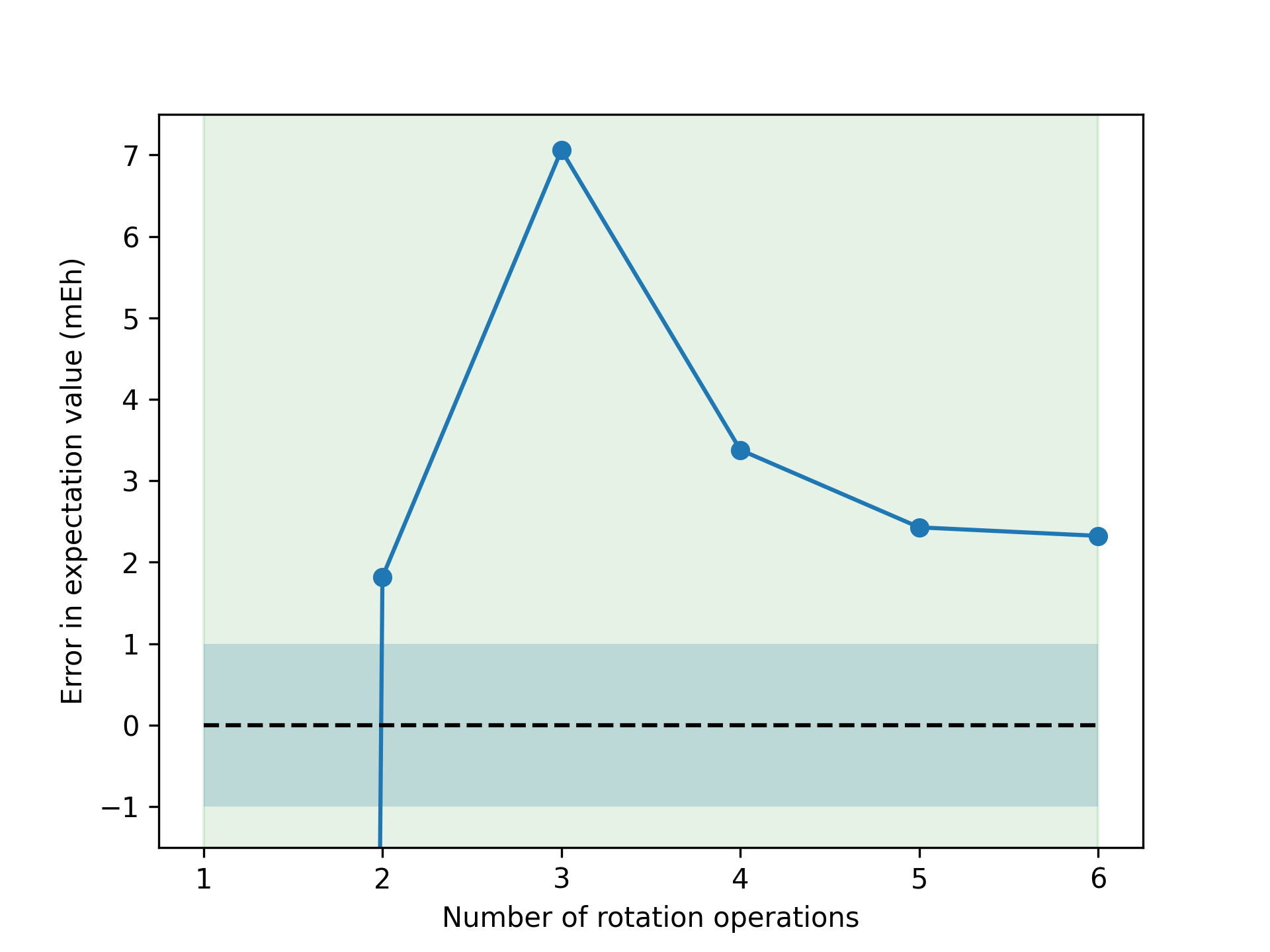}}
    \caption{Close-up visualization of Figure \ref{h8_result}.}
    \label{A3-3}
\end{figure*}

Figures \ref{A3-1}, \ref{A3-2} and \ref{A3-3} display close-up visualizations of Figures \ref{h4_result} and \ref{h6_result} and 10 molecules from Figure \ref{Random_H6_result}, which represent errors in approximating the molecular Hamiltonian operators of the selected examples.

\subsection{Distribution of number of measurements groups and number of measurements for free H$_6$ samples} \label{num_of_meas_freeH6}

\begin{figure*}
    \centering
    \subfigure[SI]{\includegraphics[width=0.3\textwidth]{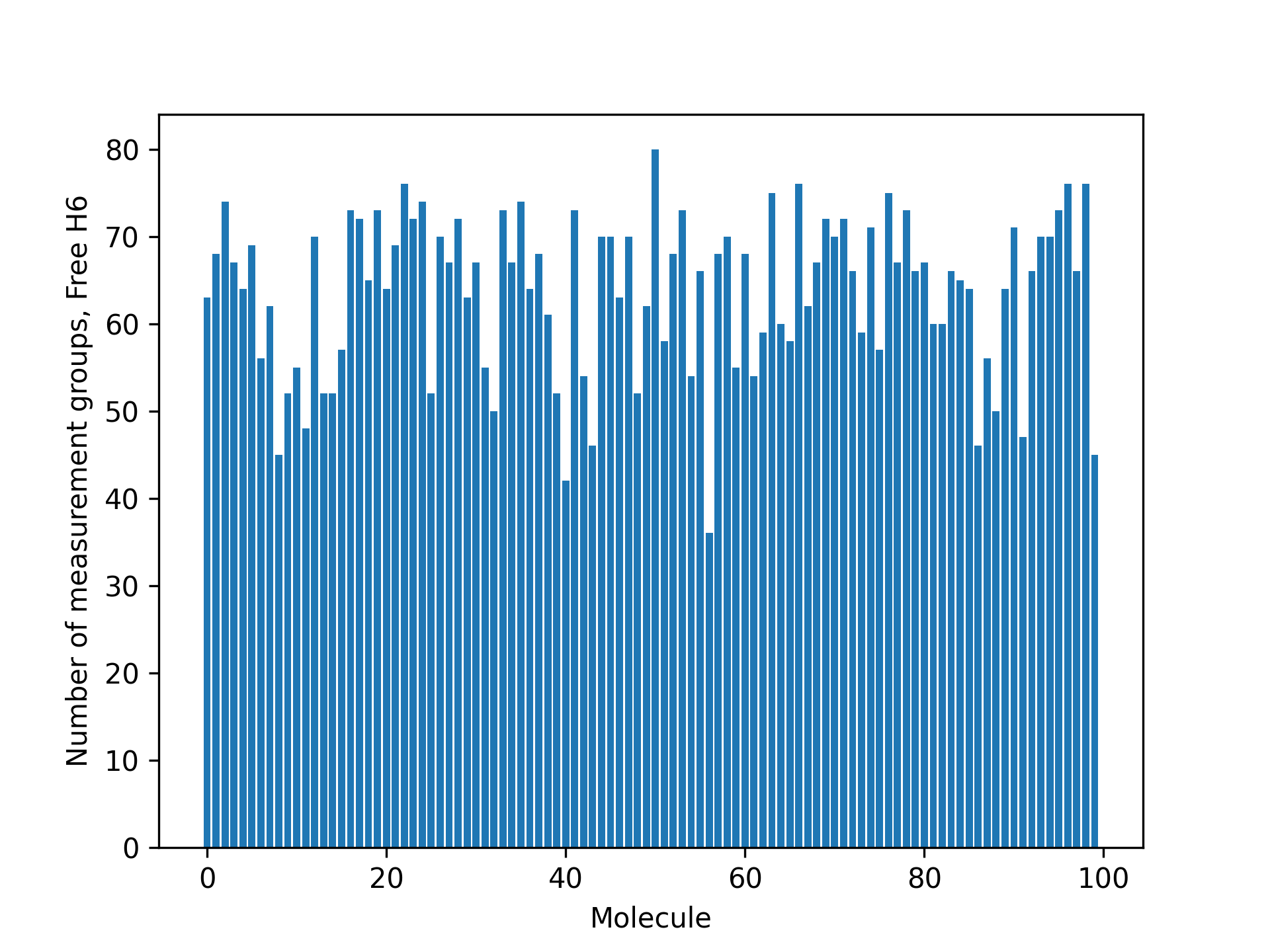}}
    \subfigure[Scenario I]{\includegraphics[width=0.3\textwidth]{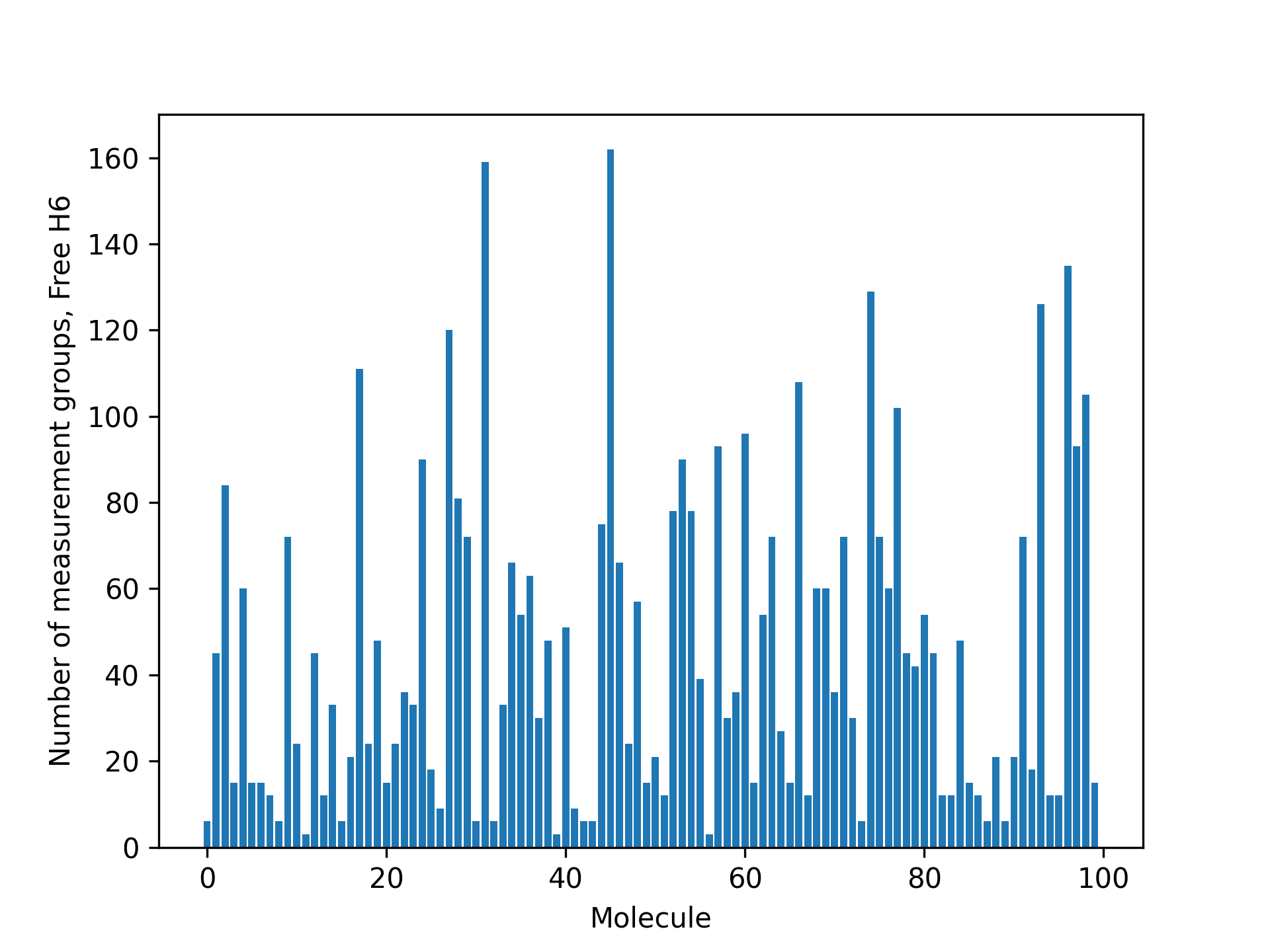}}
    \subfigure[Scenario II]{\includegraphics[width=0.3\textwidth]{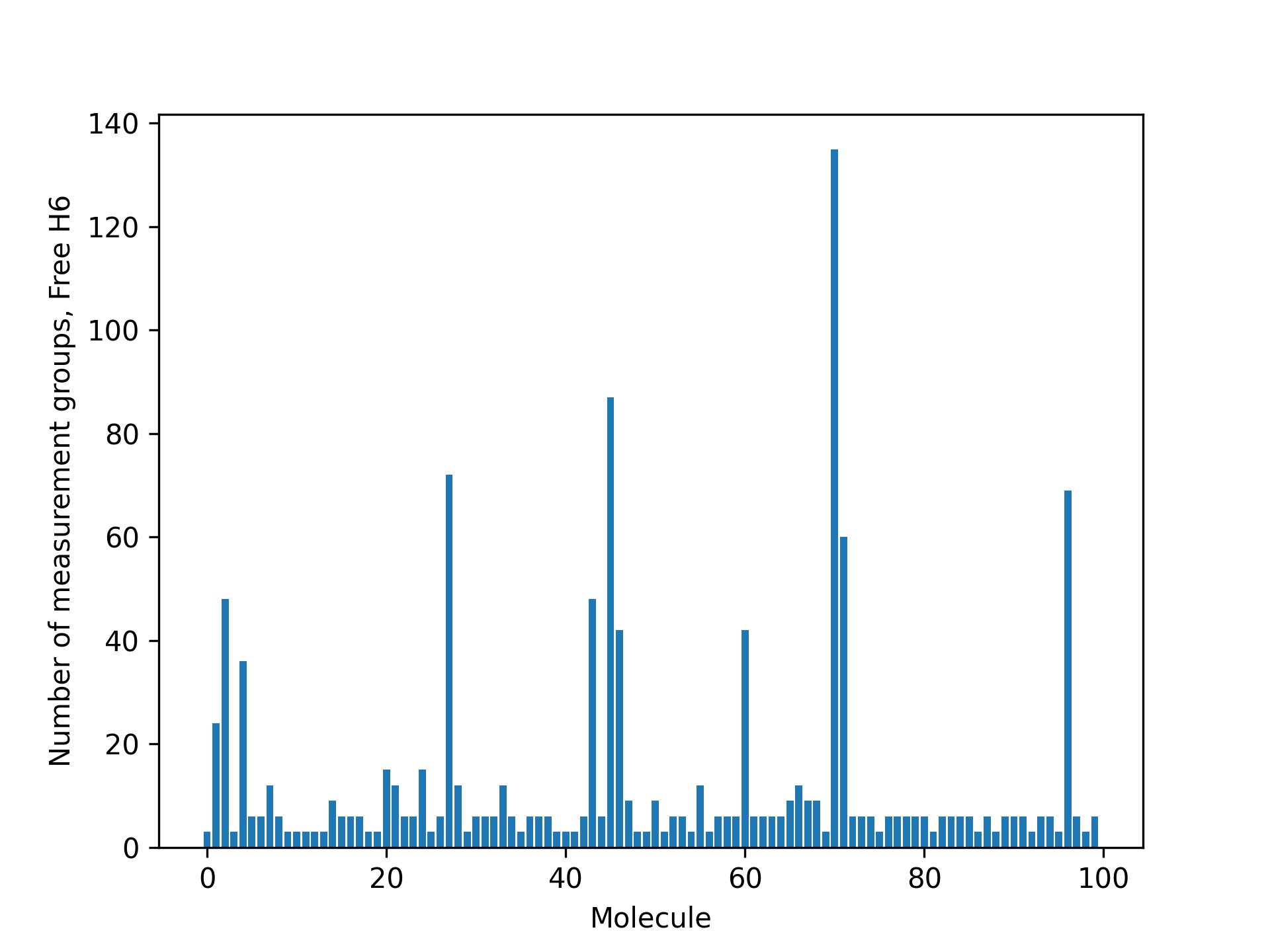}}
    \subfigure[SI]{\includegraphics[width=0.3\textwidth]{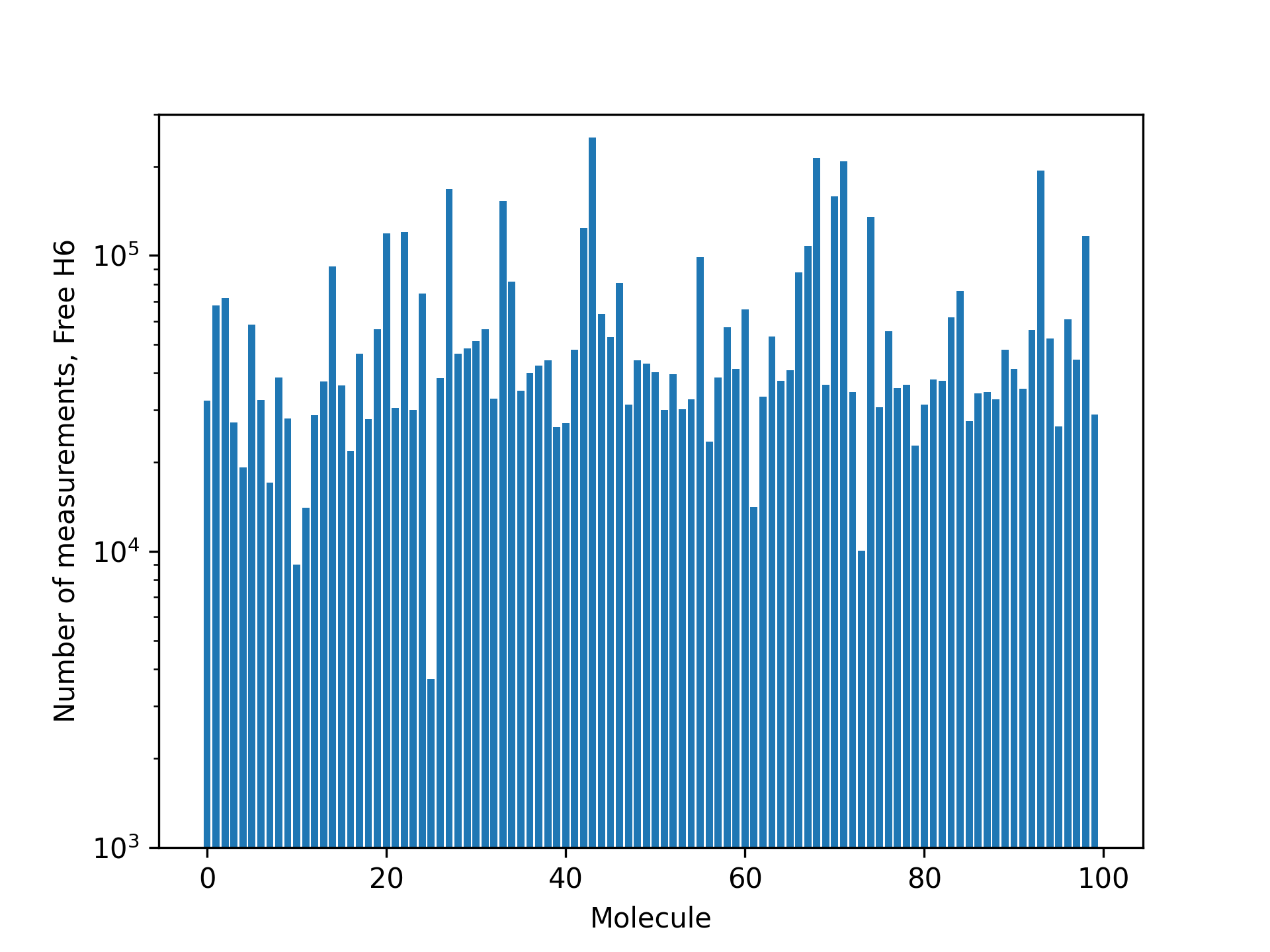}}
    \subfigure[Scenario I]{\includegraphics[width=0.265\textwidth]{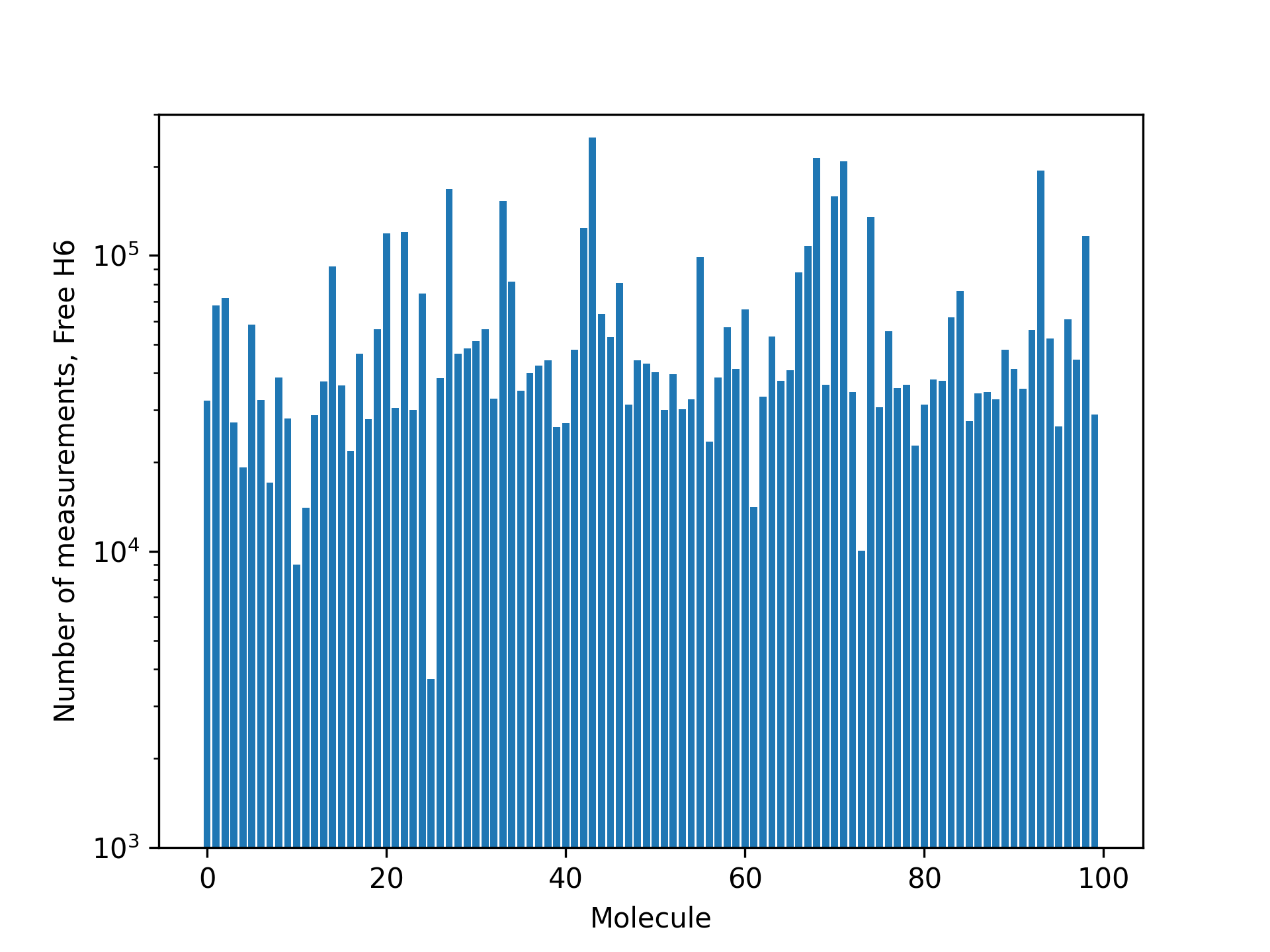}}
    \subfigure[Scenario II]{\includegraphics[width=0.3\textwidth]{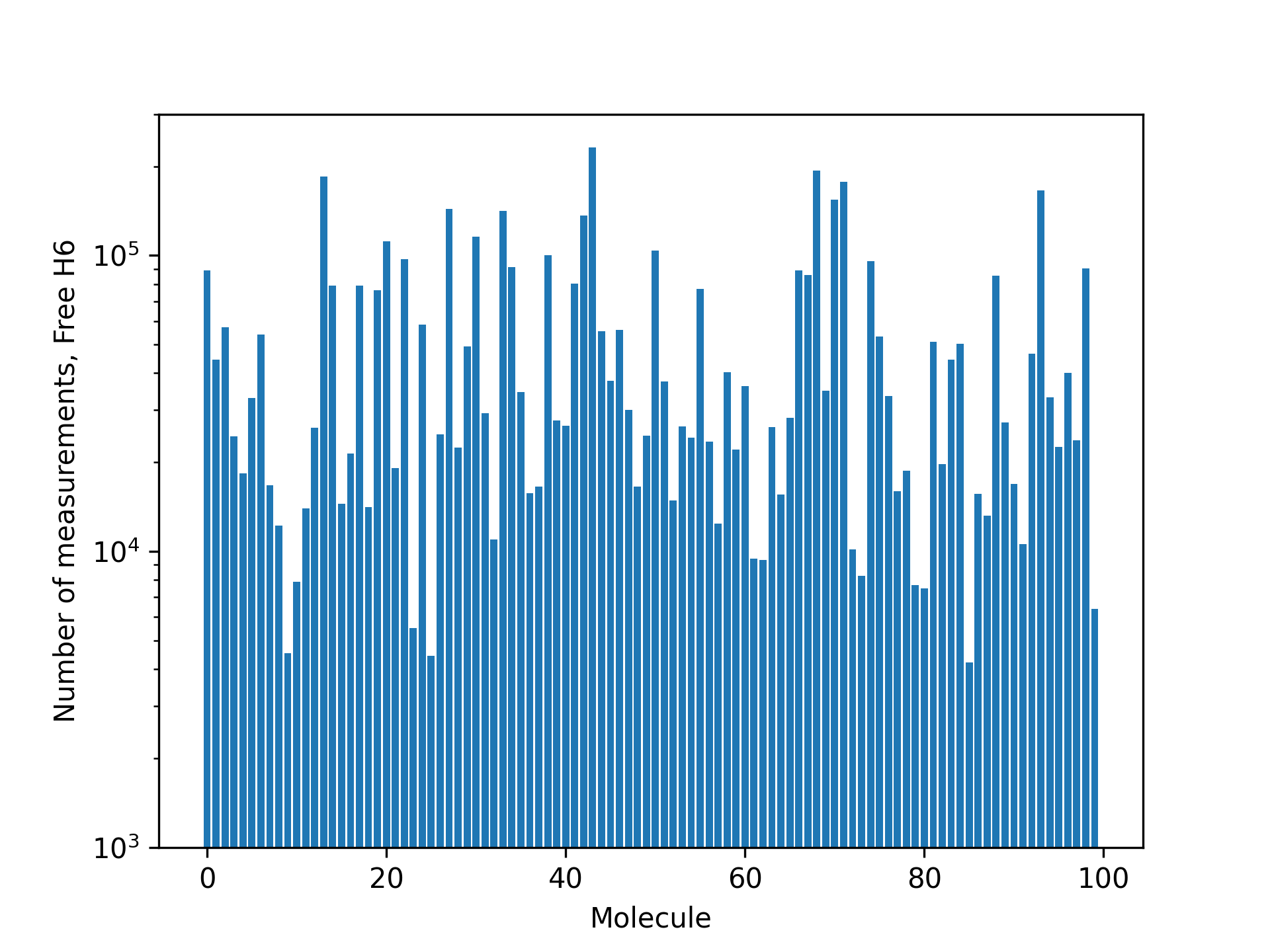}}
    \caption{(a-b-c) Distribution of number of measurement groups and (d-e-f) number of measurements for 100 samples of free geometry H$_6$ for SI, Scenario I and Scenario II.}
    \label{A4-1}
\end{figure*}

Figure \ref{A4-1} presents the full distribution of the number of measurement groups and number of measurements for the 100 samples of free geometry H$_6$ that achieves an error below 2 $\mathrm{mE_h}$, as shown in Figure \ref{Num_of_meas_groups} and Figure \ref{Num_of_meas}.

\clearpage

\section*{Glossary} \label{Glossary}
\hypertarget{LF}{\textbf{LF}: Large First \cite{verteletskyi2020measurement}} \\
\hypertarget{RLF}{\textbf{RLF}: Recursive Largest First \cite{verteletskyi2020measurement}} \\
\hypertarget{SI}{\textbf{SI}: Sorted Insertion \cite{crawford2019efficient, bansinghFidelity2022}} \\
\hypertarget{LR}{\textbf{LR}: Low-rank decomposition \cite{hugginsEfficient2021, yenCartan2021}} \\
\hypertarget{FFF-LR}{\textbf{FFF-LR}: Fluid Fermionic Fragments \cite{choiFluid2023}}

\end{document}